\documentclass[%
reprint,
superscriptaddress,
bibnotes,
amsmath,amssymb,
aps,
prl,
nolongbibliography,
]{revtex4-2}

\usepackage{graphicx}
\usepackage{dcolumn}
\usepackage{bm}
\usepackage{chemformula}
\usepackage{hyperref}

\usepackage{tabularx}
\usepackage{amsmath}
\newcommand{\tabincell}[2]{\begin{tabular}{@{}#1@{}}#2\end{tabular}}
\usepackage{float}
\usepackage{booktabs}
\usepackage{animate}

\begin{document}


\title{Giant magnetic in-plane anisotropy and competing instabilities in \ch{Na_3Co_2SbO_6}}

\author{Xintong~Li}
\email{xt.li@pku.edu.cn}
\thanks{These authors contributed equally to this study.}
\affiliation{International Center for Quantum Materials, School of Physics, Peking University, Beijing 100871, China}
\author{Yuchen~Gu}
\thanks{These authors contributed equally to this study.}
\affiliation{International Center for Quantum Materials, School of Physics, Peking University, Beijing 100871, China}
\author{Yue~Chen}
\thanks{These authors contributed equally to this study.}
\affiliation{International Center for Quantum Materials, School of Physics, Peking University, Beijing 100871, China}
\author{V.~Ovidiu~Garlea}
\affiliation{Neutron Scattering Division, Oak Ridge National Laboratory, Oak Ridge, Tennessee 37831, USA}
\author{Kazuki~Iida}
\affiliation{Neutron Science and Technology Center, Comprehensive Research Organization for Science and Society, Tokai, Ibaraki 319-1106, Japan}
\author{Kazuya~Kamazawa}
\affiliation{Neutron Science and Technology Center, Comprehensive Research Organization for Science and Society, Tokai, Ibaraki 319-1106, Japan}
\author{Yangmu~Li}
\affiliation{Beijing National Laboratory for Condensed Matter Physics, and Institute of Physics, Chinese Academy of Sciences, Beijing 100190, China}
\affiliation{Condensed Matter Physics and Materials Science Division, Brookhaven National Laboratory, Upton, New York 11973, USA}
\affiliation{University of Chinese Academy of Sciences, Beijing 100049, China}
\author{Guochu~Deng}
\affiliation{Australian Centre for Neutron Scattering, Australian Nuclear Science and Technology Organisation, Lucas Heights NSW-2234, Australia}
\author{Qian~Xiao}
\affiliation{International Center for Quantum Materials, School of Physics, Peking University, Beijing 100871, China}
\author{Xiquan~Zheng}
\affiliation{International Center for Quantum Materials, School of Physics, Peking University, Beijing 100871, China}
\author{Zirong~Ye}
\affiliation{International Center for Quantum Materials, School of Physics, Peking University, Beijing 100871, China}
\author{Yingying~Peng}
\affiliation{International Center for Quantum Materials, School of Physics, Peking University, Beijing 100871, China}
\author{I.~A.~Zaliznyak}
\email{zaliznyak@bnl.gov}
\affiliation{Condensed Matter Physics and Materials Science Division, Brookhaven National Laboratory, Upton, New York 11973, USA}
\author{J.~M.~Tranquada}
\affiliation{Condensed Matter Physics and Materials Science Division, Brookhaven National Laboratory, Upton, New York 11973, USA}
\author{Yuan~Li}
\email{yuan.li@pku.edu.cn}
\affiliation{International Center for Quantum Materials, School of Physics, Peking University, Beijing 100871, China}
\date{\today}

\begin{abstract}
We report magnetometry data obtained on twin-free single crystals of \ch{Na_3Co_2SbO_6}, which is considered a candidate material for realizing the Kitaev honeycomb model for quantum spin liquids. Contrary to a common belief that such materials can be modeled with the symmetries of an ideal honeycomb lattice, our data reveal a pronounced two-fold symmetry and in-plane anisotropy of over 200\%, despite the honeycomb layer's tiny orthorhombic distortion of less than 0.2\%. We further use magnetic neutron diffraction to elucidate a rich variety of field-induced phases observed in the magnetometry. These phases manifest themselves in the paramagnetic state as diffuse scattering signals associated with competing ferro- and antiferromagnetic instabilities, consistent with a theory that also predicts a quantum spin liquid phase nearby. Our results call for theoretical understanding of the observed in-plane anisotropy, and render \ch{Na_3Co_2SbO_6} a promising ground for finding exotic quantum phases by targeted external tuning.

\end{abstract}

\maketitle

Frustrated magnetic systems have the potential to realize exotic quantum spin liquids (QSLs) \cite{BalentsNature2010,ZhouRMP2017,BroholmScience2020}. The exactly solvable Kitaev model \cite{KitaevAP2006}, which features bond-dependent Ising interactions between effective spin-1/2 nearest neighbors on a honeycomb lattice, has motivated intensive QSL research in recent years. As a guiding principle, it is believed that such interactions can be realized in spin-orbit coupled Mott insulators \cite{JackeliPRL2009,TakagiNRP2019,MotomeJPCM2020,TrebstPRRSP2022}. Solid-state platforms for realizing the Kitaev model have evolved over the years from $5d$ iridium \cite{ChaloupkaPRL2010} to $4d$ ruthenium \cite{PlumbPRB2014} compounds, and most recently to $3d$ cobaltates \cite{LiuPRB2018,SanoPRB2018,LiuPRL2020,Kim_2021,LiuIJMPB2021}. Despite a potential drawback of weaker spin-orbit coupling, 
the cobaltates are believed to have relatively weak non-Kitaev and further-neighbor interactions compared to their $4d$ and $5d$ counterparts \cite{LiuPRB2018,SanoPRB2018,LiuIJMPB2021}.

As a reality of nature, essentially all candidate Kitaev magnets have long-range order at low temperatures \cite{ViciuJSSC2007,SinghPRB2010,LiuPRB2011,JohnsonPRB2015,ZhongSA2020,TakagiNRP2019}. This has been attributed to the presence of interactions beyond the Kitaev model \cite{ChaloupkaPRL2010,KimchiPRB2011,ChaloupkaPRL2013,RauPRL2014,SizyukPRB2014,YamajiPRL2014,KatukuriNJP2014,RousochatzakisPRX2015,ChaloupkaPRB2016,ChaloupkaPRB2015,WinterPRB2016}, such that additional tuning is needed to overcome the ordering tendency, \textit{e.g.}, by using thermal disorder and external fields  \cite{DoNP2017,WangPRB2018,WinterPRL2018,BanerjeeNPJ2018,GordonNC2019,HickeyNC2019,LiuPRL2020,YaoPRB2020}, in order to recover QSL behaviors. To this end, it is important to know how close the microscopic model of a given system is to an anticipated QSL phase. The cobaltate \ch{Na_3Co_2SbO_6} is promising in this regard, as its model is inferred to situate near boundaries between ferromagnetic (FM), antiferromagnetic (AFM), and QSL phases \cite{LiuPRL2020}. This understanding is supported by the relatively low N\'{e}el temperature ($T_\mathrm{N}$) and small saturation fields of the system compared to its sister compound \ch{Na_2Co_2TeO_6} \cite{ViciuJSSC2007,WongJSSC2016,YanPRM2019,StratanNJC2019}.

Notably, while non-Kitaev and further-neighbor terms are widely considered in theoretical constructions \cite{KimchiPRB2011,RauPRL2014,YamajiPRL2014,KatukuriNJP2014,SizyukPRB2014,RousochatzakisPRX2015,ChaloupkaPRB2016,ChaloupkaPRB2015,WinterPRB2016,WangPRB2017,JanssenPRB2017,RusnackoPRB2019,MaksimovPRR2020,LaurellNPJQM2020}, the low, monoclinic symmetry of many candidate materials, including \ch{Na_2IrO_3} \cite{SinghPRB2010,LiuPRB2011}, $\alpha$-\ch{RuCl_3} \cite{JohnsonPRB2015,CaoPRB2016}, and \ch{Na_3Co_2SbO_6} \cite{ViciuJSSC2007}, is often neglected. Even though originating from inter-layer stacking, the monoclinicity also means lack of $C_3$ rotational symmetry of the crystal field and a transition-metal ion's interactions with its neighbors in the same layer. Approximating the interactions with their bond-averaged values \cite{WinterPRB2016} is an assumption commonly taken but rarely checked. Two cobaltates, \ch{Na_2Co_2TeO_6} \cite{ViciuJSSC2007} and \ch{BaCo_2(AsO_4)_2} \cite{ZhongSA2020}, have the $C_3$ symmetry, but their zero-field ground states are reported to be dissimilar to the monoclinic systems \cite{ChenPRB2021,LeePRB2021,ZhongSA2020}, and no consensus has been reached concerning the microscopic models \cite{SongvilayPRB2020,LinNC2021,KimJPCM2021,SamarakoonPRB2021,SandersArxiv2021,YaoArxiv2022}. The lack of $C_3$ symmetry should result in magnetic in-plane anisotropy, as has been observed in $\alpha$-\ch{RuCl_3} \cite{LampenPRB2018}. However, the anisotropy is found to vary considerably \cite{LampenPRB2018,BalzPRB2021,KocsisArxiv2022}, possibly due to sample-dependent monoclinic domain population.

\begin{figure*}[!ht]
\includegraphics[width=6.5in]{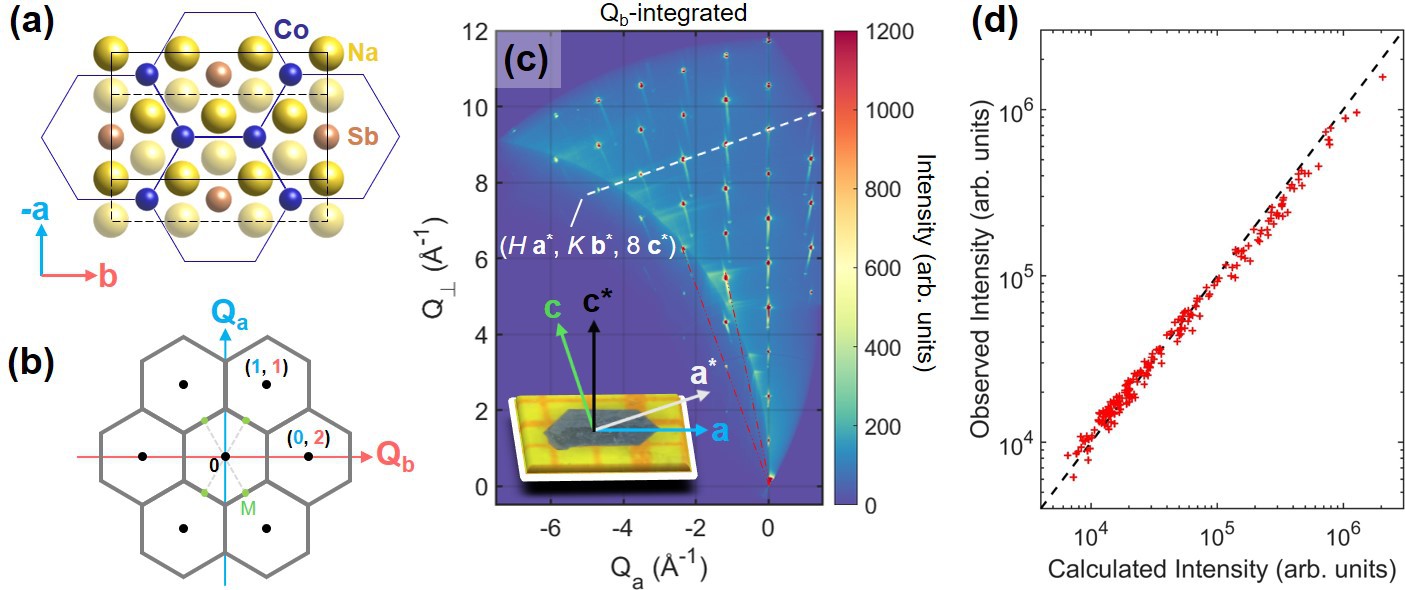}
\caption{(a) Unit cell of \ch{Na_3Co_2SbO_6} viewed from top perpendicular to the $ab$-plane, omitting oxygen atoms. Solid (dashed) rectangle indicates cell boundary in the top (bottom) Na layer. Hexagons indicate the Co sub-lattice in the middle layer. (b) Two-dimensional (2D) structural Brillouin zones, indexed in units of $a^*$ and $b^*$ projected into the $ab$-plane. Zone centers ($\mathrm{\Gamma}$-points) are $Q_a+Q_b=$~even, and M-points are mid-points between $\mathrm{\Gamma}$. (c) $X$-ray diffraction in the $a^* c^*$-plane, obtained on a twin-free crystal after integration along $\mathbf{b^*}$. The data can be indexed in a monoclinic setting without twinning. Radial Bragg tails are due to energy spread of the monochromated $X$-rays. Inset is a photo of a crystal with pertinent axes indicated. (d) Observed $X$-ray diffraction intensities from 204 indexed Bragg peaks compared to calculation from the ideal crystal structure.}
\label{fig1}
\end{figure*}

Here, we report a systematic study of \ch{Na_3Co_2SbO_6} aided by the use of twin-free crystals. Magnetometry reveals at low temperatures a strong $C_2$ in-plane anisotropy, in both the low-field susceptibility and the critical fields for switching toward a series of field-induced states. The magnitude of the anisotropy is unprecedented, yet the field-induced transitions resemble other systems to some extent. We further use neutron diffraction to determine the wave vectors of the field-induced states. They signify a series of AFM and FM instabilities, which closely compete and produce distinct diffuse scattering above $T_\mathrm{N}$ in zero field. These results render \ch{Na_3Co_2SbO_6} a highly intriguing system with the potential to realize exotic phases under targeted tuning.

\begin{figure*}
\includegraphics[width=4.8in]{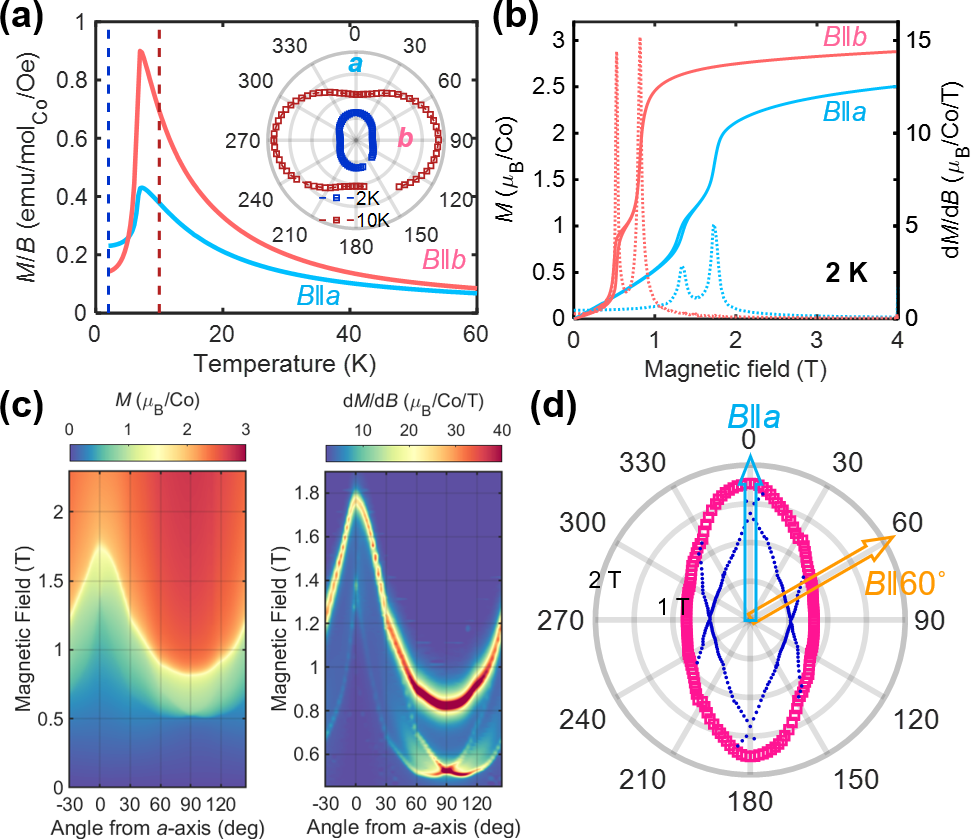}
\caption{(a) DC magnetic susceptibility measured in fields of 0.1 T along the $\mathbf{a}$ and $\mathbf{b}$ axes. Inset displays the susceptibility versus in-plane angle at 2 K and 10 K (dashed lines in the main panel), showing highly pronounced $C_2$ profiles which reverse the long and short axes across $T_\mathrm{N}$. (b) Magnetization versus field data (solid lines) reveal two transitions along both $\mathbf{a}$ and $\mathbf{b}$ at 2 K. The critical fields ($B_{\mathrm{c}1}$, $B_{\mathrm{c}2}$) are determined from the derivative (dashed lines) as (0.82 T, 1.76 T) for $\mathbf{a}$ and (0.52 T, 1.37 T) for $\mathbf{b}$, with an uncertainty of $\pm0.02$ T. Data are displayed for both field-up and -down sweeping directions, which are nearly identical except near $B_{\mathrm{c}1}$, indicating a hysteretic nature of the transition. (c) In-plane angle dependence of magnetization versus field (left) and the field-derivative (right) at 2 K. Measurement at each angle is performed over a field-up sweep, and the field is decreased to zero before moving to the next angle. It is seen that $B_{\mathrm{c}1}$ splits away from $\mathbf{a}$ and $\mathbf{b}$. (d) Summary of the result in (c) after $180^\circ$ symmetrization. Empty arrows are a reference for the field directions in Fig.~\ref{fig4}.}
\label{fig2}
\end{figure*}

\section{Magnetometry on twin-free crystals}

Figure~\ref{fig1}(a-b) presents the crystal and reciprocal-space structure of \ch{Na_3Co_2SbO_6}, which has the same space group ($C2/m$) as $\alpha$-\ch{RuCl_3} \cite{JohnsonPRB2015,CaoPRB2016}. A peculiarity common to both structures is in the stacking: adjacent honeycomb layers are offset from each other by $-\mathbf{a}/3$, hence we have $a/c\approx-3\cos\beta$, with $[a,\,b,\,c]=[5.371,\,9.289,\,5.653]\,\mathrm{\AA}$ and $\beta=108.6^\circ$ in \ch{Na_3Co_2SbO_6} \cite{YanPRM2019}. Similar to $\alpha$-\ch{RuCl_3} \cite{CaoPRB2016}, the orthorhombic distortion in the honeycomb layer of \ch{Na_3Co_2SbO_6} is tiny: the distortion is measured as $\sqrt{3}a/b-1 < 0.002$ \cite{YanPRM2019,StratanNJC2019}. Yet, we find that such a small distortion removes the overall $S_6$ and $C_3$ symmetries. We will next show the far-reaching consequences on the magnetism using a rare growth product: twin-free single crystals. Such crystals can be found by screening with Raman spectroscopy (Fig.~\ref{figS1} in \cite{SM}), and ultimately verified with $X$-ray diffraction [Fig.~\ref{fig1}(c)]. They have a well-defined $T_\mathrm{N}$ of about 6.6~K (Fig.~\ref{figS2} in \cite{SM}) with sample-dependent variation of no more than 1 K possibly caused by structural imperfections. The variation is considerably smaller than in the literature \cite{ViciuJSSC2007,WongJSSC2016,YanPRM2019,StratanNJC2019}. This is in line with the facts that our best crystals have very few stacking faults [Fig.~\ref{fig1}(c)] compared to a previous report \cite{YanPRM2019}, and that the observed Bragg intensities agree well with calculation based on the ideal crystal structure [Fig.~\ref{fig1}(d)]. Our further refinement attempts suggest that the agreement cannot be improved by introducing anti-site disorder between Co and Sb \cite{YanPRM2019}. While the data do not allow us to rule out disorder in the Na layers \cite{YanPRM2019}, we consider its role to be minor because such disorder is expected to cause stacking faults which are rare in our crystals.

\begin{figure*}[!ht]
\includegraphics[width=4.8in]{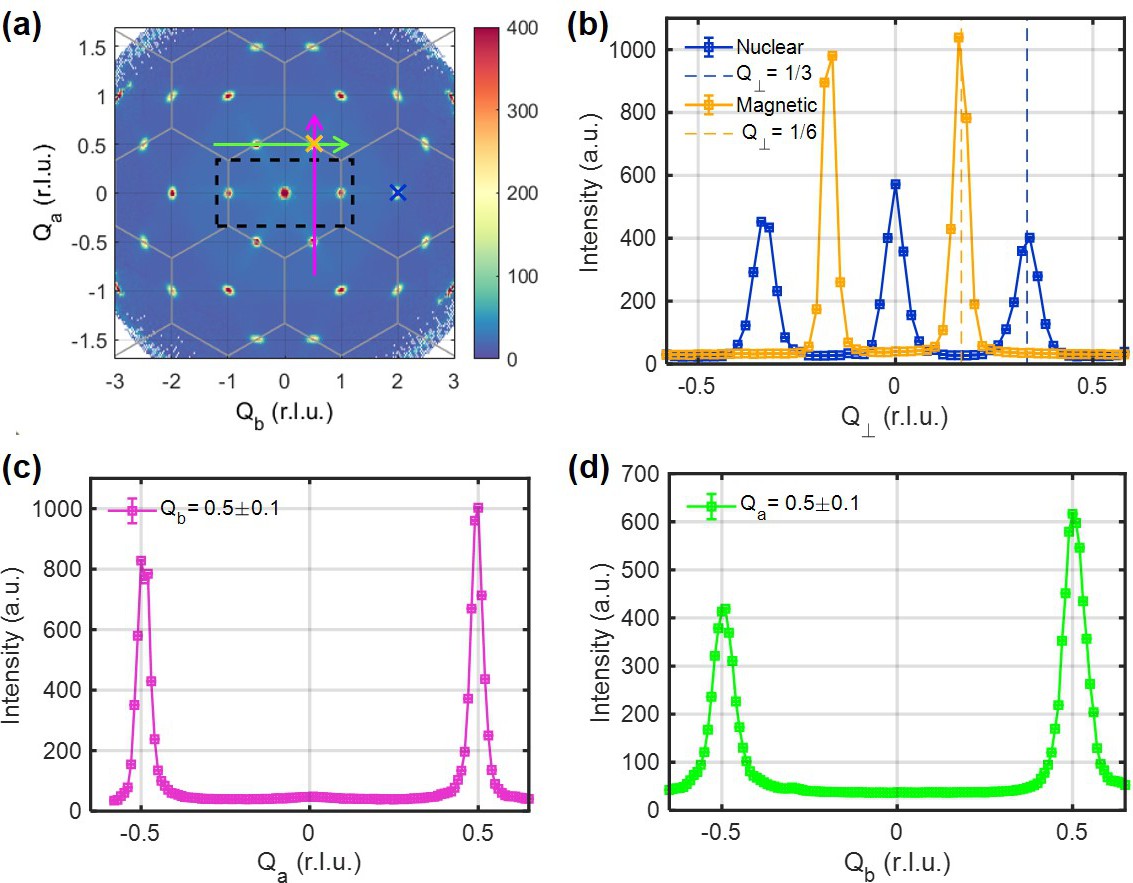}
\caption{\textcolor{black}{(a) $Q_{\perp}$-integrated diffraction data measured at 6$K$ and in zero field. Gray hexagons indicate 2D Brillouin zones. Color-coded crosses and arrows indicate locations of line cuts in (b-d). Dashed box indicates the restricted data coverage in Fig.~\ref{fig4}. (b) Line cuts along $Q_{\perp}$ through nuclear and magnetic Bragg peaks. The nuclear peaks at $Q_{\perp}=0$ and $\pm1/3$ are contributed by physical reflections $(0,\,2\mathbf{b^*},\,0)$ of $\mathrm{S}_a$ and $(\mathbf{a^*},\,\mathbf{b^*},\,0)$ of $\mathrm{S}_{60}$, respectively. The magnetic peaks at $Q_{\perp}=\pm1/6$ are contributed by reflections $(\pm\mathbf{a^*}/2,\,\pm\mathbf{b^*}/2,\,0)$ of both domains. (c-d) Line cuts along $Q_a$ and $Q_b$ through the magnetic reflections. The map in (a) has been symmetrized, whereas the line cuts in (b-d) are not symmetrized.} }
\label{fig3}
\end{figure*}

Figure~\ref{fig2}(a) shows how a pronounced $\mathbf{a}$-$\mathbf{b}$ anisotropy develops in the magnetic susceptibility upon cooling. Far above $T_\mathrm{N}$, we observe a $\sim10\%$ anisotropy consistent with the anisotropy of the $g$-factor, $g_b > g_a$ (Figs.~\ref{figS3}-\ref{figS4} and Table~\ref{tabS1} in \cite{SM}), as is also seen from high-field magnetization where moments are nearly polarized [Fig.~\ref{fig2}(b)]. The anisotropy drastically increases to over $200\%$ near $T_\mathrm{N}$ (see Fig.~\ref{figS3} in \cite{SM} for out-of-plane anisotropy), which signifies the role of the fluctuations -- the moments respond much more strongly in the easy direction nearly parallel to the developing order parameter \cite{YanPRM2019}. This understanding also explains why the anisotropy is reversed below $T_\mathrm{N}$. The reversal is no longer observed in $B=2$ T (Fig.~\ref{figS3} in \cite{SM}) which is large enough to overcome the AFM order. We make two remarks here to relate to previous works: (1) The $\mathbf{a}$-axis response clearly drops below $T_\mathrm{N}$ [Fig.~\ref{fig2}(a)], suggesting that the ordered moments are not entirely along $\mathbf{b}$ \cite{YanPRM2019}. (2) No reversal is observed below $T_\mathrm{N}$ in $\alpha$-\ch{RuCl_3} \cite{LampenPRB2018}, where the anisotropy also appears to be much weaker.

The competition between anisotropic interactions and the applied field is more clearly seen in the magnetization at 2~K [Fig.~\ref{fig2}(b-d)], where our twin-free sample reveals a wealth of remarkable features unnoticed in previous works \cite{ViciuJSSC2007,WongJSSC2016,YanPRM2019,StratanNJC2019}. Two well-separated transitions are observed along both $\mathbf{a}$ and $\mathbf{b}$, at critical fields [$B_{\mathrm{c}1}$ and $B_{\mathrm{c}2}$, Fig.~\ref{fig2}(b)] that again differ strongly between the two directions. The lower-field transition is clearly hysteretic, as indicated by the magnetization's dependence on the field-sweeping direction. It further splits into two hysteretic transitions, the critical fields of which we refer to as $B_{\mathrm{c1,low}}$ and $B_{\mathrm{c1,high}}$, when the field is applied in-plane but away from the high-symmetry $\mathbf{a}$ and $\mathbf{b}$ axes [Fig.~\ref{fig2}(c-d)]. The lowest $B_{\mathrm{c1,low}}$ value is found at about $15^\circ$ away from $\mathbf{b}$. The highest $B_{\mathrm{c1,high}}$ can approach $B_{\mathrm{c}2}$ and become no longer visible from the data, over a range of field directions between $10^\circ$ and $30^\circ$ away from $\mathbf{a}$. Hence, very unexpectedly, there is nearly no 6-fold symmetry in the results\textcolor{black}{, including in the nearly field-polarized state at 2 T (Fig.~\ref{figS3} of \cite{SM})}. The large magnitude of $\mathbf{a}$-$\mathbf{b}$ anisotropy sharply contrasts with the $C_3$-symmetric sister compound \ch{Na_2Co_2TeO_6}, where the magnetic responses along $\mathbf{a}$ and $\mathbf{a^*}$ are reasonably similar \cite{YaoPRB2020,LinNC2021}.
\textcolor{black}{We note that quenched disorders may play a role in the experimentally observable anisotropy in \ch{Na_3Co_2SbO_6}. By heating up a twin-free crystal to 600 $^\circ$C at 20 $^\circ$C/min, staying for 1 hour, and quenching the crystal in liquid nitrogen, we found the susceptibility anisotropy ratio [$\chi_b/\chi_a$, see Fig.~\ref{fig2}(a)] to change from 1.81 to 1.78 at 10 K, and from 0.53 to 0.75 at 2 K. The two field-induced transitions [Fig.~\ref{fig2}(b)] at 2 K also became considerably smeared out. According to $X$-ray diffraction, the crystal remained twin-free after the quenching.}

\begin{figure*}[!ht]
\includegraphics[width=\textwidth]{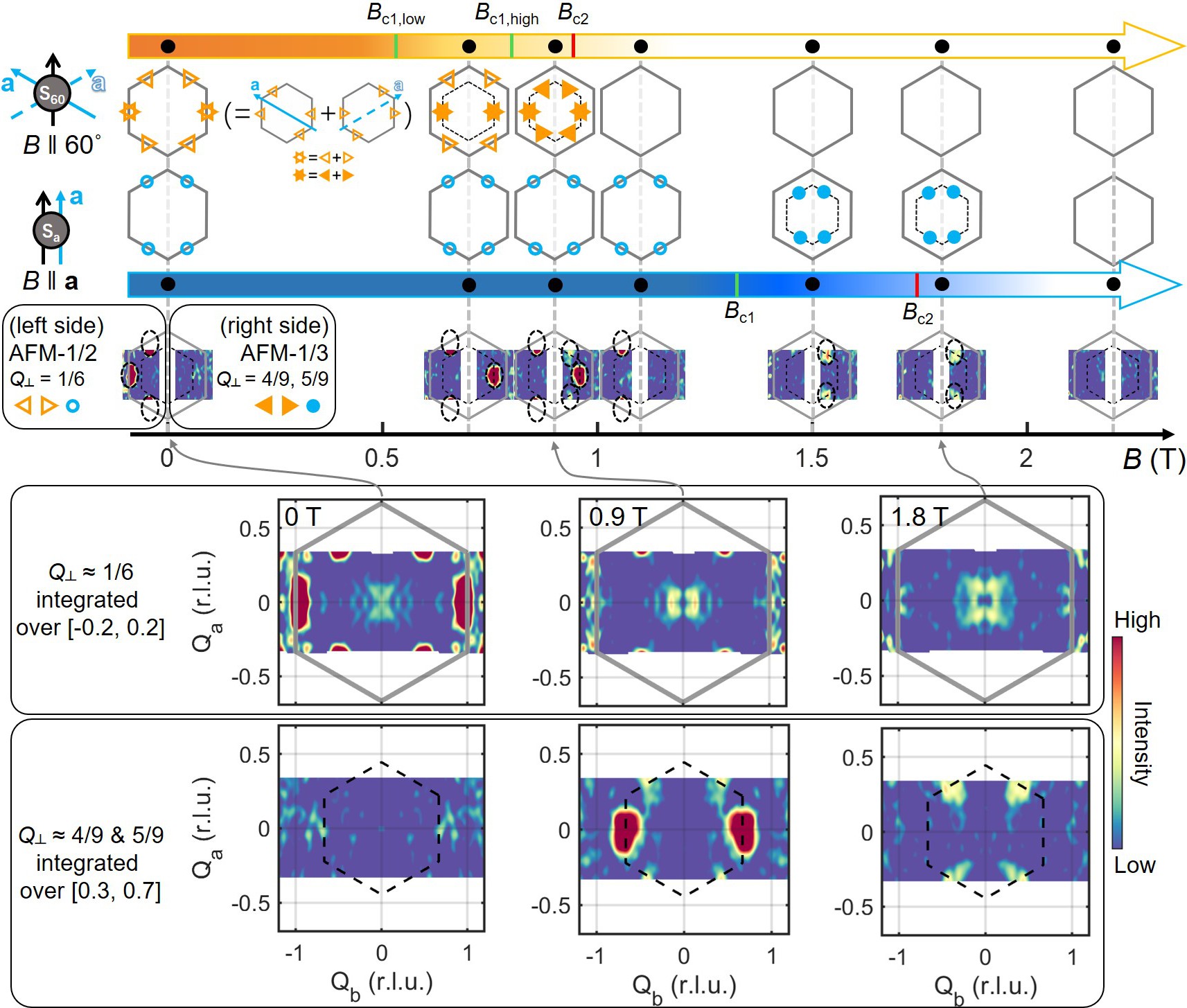}
\caption{The upper half illustrates the behavior of two sample parts ($\mathrm{S}_a$ or $\mathrm{S}_{60}$, see legends on the left and text) as the field passes through their respective phase boundaries [Fig.~\ref{fig2}(d)]. $\mathrm{S}_a$ is uniquely defined as per the in-plane orientation, and it contributes diffractions indicated by blue circles. $\mathrm{S}_{60}$ further contains two parts that are related by $180^\circ$ rotation about the field, which contribute diffractions indicated by left- and right-pointing orange triangles. Empty and filled symbols indicate AFM${\frac{1}{2}}$ and AFM${\frac{1}{3}}$ wave vectors (see text), respectively, which are measured by restricting $Q_\perp$ to $[-0.2,\,0.2]$ and $[0.3,\,0.7]$, respectively. Miniatures of diffraction data (measured at $T=0.25$~K) are displayed in the bottom row of the illustration in a left-right split fashion, where the observed diffraction peaks (encircled by dashed ellipses \textcolor{black}{centered at their expected locations, some of which fall beyond the data coverage}) are fully consistent with the ``$\mathrm{S}_a + \mathrm{S}_{60}$'' combination of the cartoons. Solid and dashed hexagons are the first Brillouin zone and the 2/3 of it, respectively. In the lower half, we display diffraction data measured at three selected fields, where the solid and dashed hexagons have the same meaning as in the upper-half illustration. \textcolor{black}{Note that the vertical ($\parallel Q_a$) data coverage is limited and the resolution is relatively poor compared to those in Fig.~\ref{fig3}. An animated view of the full variable-field data is presented in Fig.~\ref{figS10} of \cite{SM}.}}
\label{fig4}
\end{figure*}

\section{Magnetic neutron diffraction}

We next turn to the intermediate state(s) between $B_{\mathrm{c}1}$ and $B_{\mathrm{c}2}$. While the step-like and hysteretic (near $B_{\mathrm{c}1}$) behaviors hinted at a spin-flop origin \cite{ViciuJSSC2007,WongJSSC2016,YanPRM2019,StratanNJC2019}, our observation of the transitions along both $\mathbf{a}$ and $\mathbf{b}$ (and everywhere in between) defies such an interpretation. The result in Fig.~\ref{fig2}(c) is furthermore independent of field or temperature history, precluding the relevance of magnetic domain repopulation \cite{SearsPRB2017,BanerjeeNPJ2018}. Motivated by the fact that the magnetization above $B_{\mathrm{c}1}$ resembles ``plateau'' phases found in low-dimensional frustrated magnets \cite{HardyPRB2004,UedaPRL2005,CaoPRB2007,JoPRB2009}, \textit{i.e.}, it reaches about 1/3 and 1/2 of saturation [Fig.~\ref{fig2}(b)] for $B_{\parallel b}$ and $B_{\parallel a}$, respectively, we have performed neutron diffraction in magnetic fields to explore this possibility. The experiment was done on a co-aligned but twinned array of crystals with their $\mathbf{c^*}$ axis horizontal, such that the vertical field was along $\mathbf{a}$ for 1/3 of the sample ($\mathrm{S}_a$), and at 60$^\circ$ from $\mathbf{a}$ for the rest ($\mathrm{S}_{60}$), see illustrations in the upper-left corner of Fig.~\ref{fig4}. In spite of the twinning, there is no ambiguity in the domain origin ($\mathrm{S}_a$ or $\mathrm{S}_{60}$) of the field-evolving signals, \textcolor{black}{under the assumption that magnetization and diffraction see the same transitions (Fig.~\ref{figS5} in \cite{SM}). According to magnetization [Fig.~\ref{fig2}(d)], all transitions occur below (above) 1 Tesla for $\mathrm{S}_{60}$ ($\mathrm{S}_a$). The difference is illustrated by the thick horizontal arrow diagrams in the upper half of Fig.~\ref{fig4}.}

We use here a ``hybrid'' orthogonal coordinate system for the reciprocal space, illustrated in Fig.~\ref{fig1}(b). Wave vectors are denoted as $(Q_a,\,Q_b,\,Q_\perp)$, with $Q_b$ and $Q_\perp$ in units of $\mathbf{b^*}$ and $\mathbf{c^*}$, respectively. $Q_a$ is in units of $\mathbf{a^*}$ projected onto the real-space $a$-axis, and it is parallel to the vertical field. This coordinate system is convenient for describing a twinned sample, because the twinning features $C_6$ rotations within the $ab$-plane, and mixes up $Q_a$ and $Q_b$ while leaving $Q_\perp$ intact. We will write $\mathbf{a^*}$, $\mathbf{b^*}$, and $\mathbf{c^*}$ explicitly when we refer to the (physical) monoclinic indices. A table reference for transforming between the two indexing systems can be found in Table \ref{tabS3} \cite{SM}. \textcolor{black}{To give some examples, nuclear Bragg peaks at $(0,\,2,\,\pm1/3)$ in the hybrid notation [Fig.~\ref{fig3}(a-b)] are associated with physical indices $(\pm\mathbf{a^*},\,\pm\mathbf{b^*},\,0)$ of $\mathrm{S}_{60}$. In zero field, the AFM wave vectors $(\pm\mathbf{a^*}/2,\,\pm\mathbf{b^*}/2,\,0)$ \cite{YanPRM2019} of $\mathrm{S}_a$ transform into $(\pm0.5,\,\pm0.5,\,\pm1/6)$ in the hybrid notation [Fig.~\ref{fig3}(b-d)],} producing diffractions at four $(Q_a,\,Q_b)$ locations, whereas the same diffractions from the two copies of $\mathrm{S}_{60}$ (Fig.~\ref{fig4}) are expected at six $(Q_a,\,Q_b)$ locations. All of these AFM wave vectors have $\left|Q_\perp\right|=1/6$ as indicated by empty symbols in Fig.~\ref{fig4}. \textcolor{black}{While the data coverage in Fig.~\ref{fig4} along the $Q_a$ direction is limited compared to that in Fig.~\ref{fig3}), magnetic diffractions above and below the $Q_a=0$ (horizontal) plane are partly observed. This is enabled by vertical focusing optics \cite{SM}, which relaxes the momentum resolution and elongates diffraction features in the $Q_a$ direction.}

The results in Fig.~\ref{fig4} can be summarized as follows: the AFM wave vectors switch from $(\pm\mathbf{a^*}/2,\,\pm\mathbf{b^*}/2,\,0)$ at $B=0$, to $(\pm\mathbf{a^*}/3,\,\pm\mathbf{b^*}/3,\,\pm\mathbf{c^*}/3)$ above $B_{\mathrm{c}1}$, and eventually no AFM is left above 2.2 T (see Methods and Figs.~\ref{figS5}-\ref{figS7} in \cite{SM} for additional evidence for the peak indexing). We therefore refer to the zero-field and the intermediate states as AFM${\frac{1}{2}}$ and AFM${\frac{1}{3}}$, respectively. The AFM${\frac{1}{3}}$ wave vectors all have $Q_\perp=4/9$ or $5/9$ \cite{SM}, which allows the diffraction peaks to be observed separately from the AFM${\frac{1}{2}}$ ones by restricting $Q_\perp$ in the experiment (Fig.~\ref{fig4}). Notably, due to the low-symmetry field direction for $\mathrm{S}_{60}$, the wave vectors in this part of the sample do not switch together. Instead, the switching occurs in two steps for the diffraction peaks situated on different $\Gamma$-M lines [Fig.~\ref{fig1}(b)] relative to the field, see the comparison of the 0, 0.7, and 0.9 T illustrations for $\mathrm{S}_{60}$ and the associated data in the upper half of Fig.~\ref{fig4}. This two-step switching behavior is fully consistent with the two transitions at $B_\mathrm{c1,low}$ and $B_\mathrm{c1,high}$ for the same field direction, which we have established with magnetometry [Fig.~\ref{fig2}(d)]. We note that at each transition, the switching wave vectors remain on the same $\Gamma$-M lines, without intermixing between the 2D momentum directions.

The wave-vector switch supports the idea that AFM${\frac{1}{3}}$ is a ferrimagnetic phase with an enlarged 2D cell compared to AFM${\frac{1}{2}}$, such as ``$\uparrow\uparrow\downarrow$'' compared to ``$\uparrow\downarrow$''. With the understanding that AFM${\frac{1}{2}}$ features zigzag order \cite{YanPRM2019}, which consists of alternating FM chains running along zigzag lines of the honeycomb lattice, AFM${\frac{1}{3}}$ could feature alternating wide and narrow FM ribbons and chains. An illustration of such FM chains without the alternating correlation can be found in Fig.~\ref{fig6}(a). \textcolor{black}{The two-step transitions of $B_{\mathrm{c}1}$ and the single-step transition of $B_{\mathrm{c}2}$ introduce some restrictions on the magnetic structure, which we discuss in \cite{SM}.} We further note that $B_{\mathrm{c}2}$ does \textit{not} necessarily mark entrance into a field-polarized state, certainly not for $\mathrm{S}_{a}$, since the AFM${\frac{1}{3}}$ diffraction peaks persist above $B_{\mathrm{c}2}$ (Fig.~\ref{fig4}). The nature of $B_{\mathrm{c}2}$ will be reported elsewhere. Further above $B_{\mathrm{c}2}$, all magnetic diffraction eventually coincide with nuclear Bragg peaks, as expected for a field-polarized state.

\begin{figure}[!ht]
\includegraphics[width=3.375in]{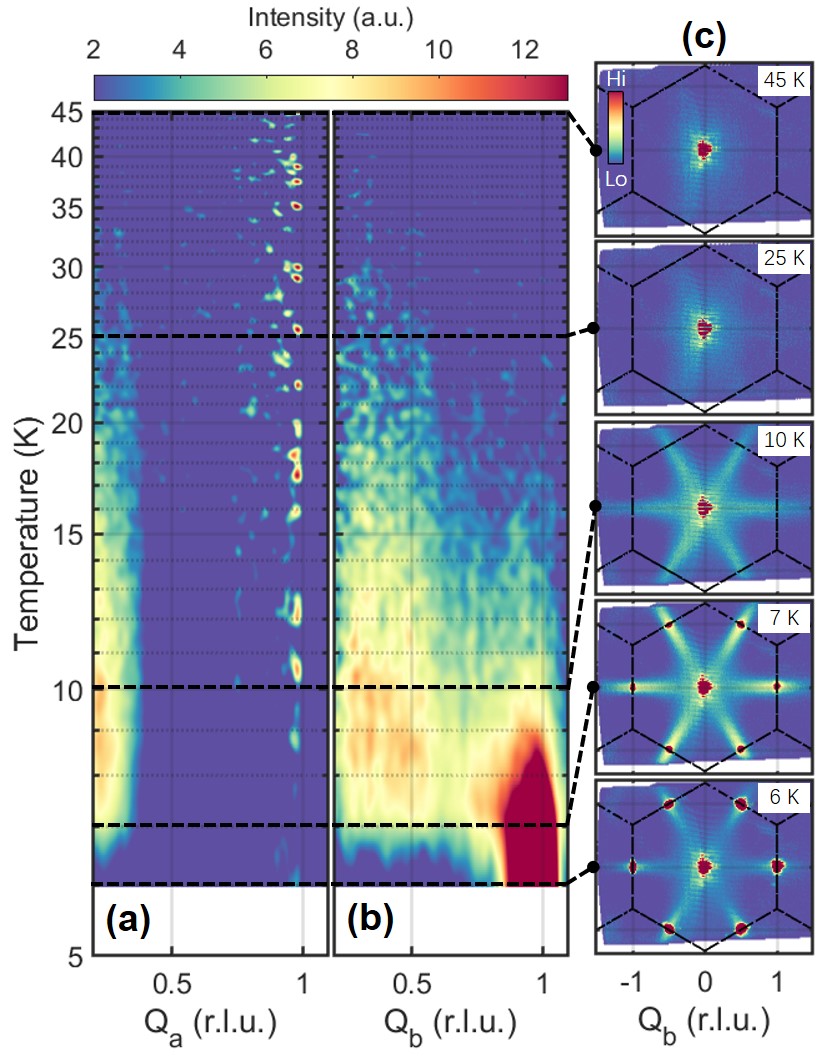}
\caption{ (a-b) Variable-$T$ diffuse magnetic scattering in zero field, viewed along $Q_a$ and $Q_b$ starting from near the $\mathrm{\Gamma}$-point, after subtracting the $T=45$~K data as background. (c) Scattering in the 2D Brillouin zone at selected temperatures. Unlike the nuclear ($\mathrm{\Gamma}$-point) and magnetic (M-point) Bragg peaks, the diffuse scattering does not show noticeable $Q_\perp$ dependence, and the displayed data are $Q_\perp$ integrated and restricted to a small energy window of $\pm0.2$ meV.}
\label{fig5}
\end{figure}

\begin{figure*}[!ht]
\includegraphics[width=\textwidth]{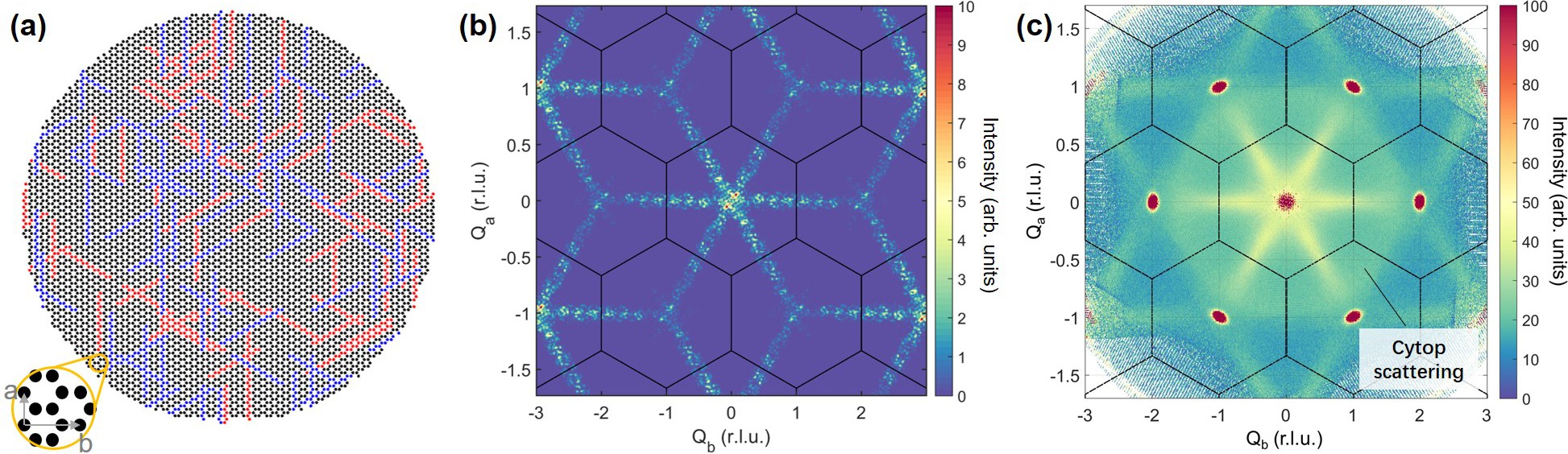}
\caption{(a) A random placement of ferromagnetic zigzag chain segments on a honeycomb lattice. Blue, red, and black circles indicate spin-up, spin-down, and spin-less sites, respectively, which contribute positive, negative, and zero neutron-scattering amplitudes in the simulation. (b) Fourier component squared of the scattering amplitudes in the field of view in (a), computed on a fine 2D momentum grid. (c) Neutron diffraction intensities at 10 K, similar to those in Fig.~\ref{fig5}(c) but acquired with a higher incident energy (Table~\ref{figS2} in \cite{SM}). Data are $Q_\perp$-integrated, and have been $C_6$-symmetrized for better comparison with (b). Black hexagons indicate 2D Brillouin zones. Sharp diffraction spots are nuclear Bragg peaks. Halo-like diffuse intensities in the first and part of the second zones are $T$-independent scattering from the sample holder and glue (Cytop). }
\label{fig6}
\end{figure*}

Taken together, the results show that at very low $T$ and in external in-plane fields, \ch{Na_3Co_2SbO_6} sequentially goes through magnetic states characterized by the M-point, the ``$\frac{2}{3}$M''-point, and eventually the zone-center $\mathrm{\Gamma}$-point, forming an evolution along the $\mathrm{\Gamma}$-M lines [Fig.~\ref{fig1}(b)]. The direction of the field affects only when, but not whether, the transitions would occur. It is therefore tempting to think that the system possesses competing AFM-FM instabilities with wave vectors lined up along $\mathrm{\Gamma}$-M. In Fig.~\ref{fig5}, we use variable-$T$ neutron diffraction to show that this is indeed the case. The experiment was performed on a twinned sample, in \textit{zero} magnetic field. The most remarkable observation is found at 10 K [Fig.~\ref{fig5}(c)]: we see distinct hexagonal-star-shaped diffuse scattering, which ``flows'' into the long-range magnetic Bragg peaks at the M-points upon further cooling [Fig.~\ref{fig5}(b)]. The observed star consists of six narrow streaks which precisely cover the $\mathrm{\Gamma}$-M lines. In a twin-free sample, the number of streaks would likely be four [Fig.~\ref{fig1}(b)], which would help explain the giant in-plane magnetic anisotropy, and it warrants further experimental confirmation. The streaks are, in fact, quasi-2D objects in reciprocal space with only weak dependence on $Q_\perp$ (Fig.~\ref{figS8} in \cite{SM}). They correspond to quasi-1D correlations in real space (Fig.~\ref{fig6}, further discussed below), and can be viewed as a counterpart of rod-like diffuse scattering in \ch{Yb_2Ti_2O_7} \cite{RossPRL2009,ThompsonPRL2011}, which has been attributed to coexisting FM and AFM correlations \cite{ScheiePNAS2020,ScheieArxiv2022}. Below $T_\mathrm{N}$, the body of the star is depleted, including the FM-like diffuse scattering near $\mathrm{\Gamma}$ [Fig.~\ref{fig5}(a)]. Such a temperature-evolution, together with the field-evolution at low $T$ (Fig.~\ref{fig4}), signifies a close competition between a variety of AFM and FM instabilities, with or without thermal disorder. Indeed, the M-point AFM order might be energetically favored in zero field by only a small margin. We have further found evidence for a weak field-trainable net moment in a twin-free sample (Fig.~\ref{figS9} in \cite{SM}), which supports an incipience of the ferromagnetism.

\section{Discussion}

Our results motivate further exploration of exotic quantum phases in \ch{Na_3Co_2SbO_6}, as well as in extended Kitaev and related theoretical models. Magnetic field-induced phases in candidate Kitaev materials have been under intense research in recent years \cite{BalzPRB2021,ZhongSA2020,YaoPRB2020,LinNC2021,ZhengPRL2017,WolterPRB2017,LeahyPRL2017,BaekPRL2017,YuPRL2018,WellmPRB2018,KasaharaNature2018,BanerjeeNPJ2018,SahasrabudhePRB2020,YokoiScience2021}, and the magnetization's step-like transitions into and out of the intermediate states in Fig.~\ref{fig2} resemble some of the reports \cite{BalzPRB2021,ZhongSA2020,YaoPRB2020}, even though the wave-vector switching behavior might not be the same \cite{BalzPRB2021,ZhongSA2020}. These results suggest that the candidate materials commonly possess multiple magnetic instabilities -- a hallmark of frustration. Our findings are consistent with the view that \ch{Na_3Co_2SbO_6} is close to a trisecting point of FM, AFM, and QSL phases \cite{LiuPRL2020}, yet the pronounced in-plane anisotropy clearly adds complexity and novelty to the previous understanding. Specifically, given the anisotropy, uniaxial strains both perpendicular \cite{LiuPRL2020} and parallel to the honeycomb layers might help promote QSL physics.

Meanwhile, the $\mathrm{\Gamma}$-M characteristics of the ordering wave vectors and diffuse scattering imply a particular combination of competing instabilities, which are not commonly seen in model systems \cite{LaurellNPJQM2020}. In Fig.~\ref{fig6}, we show that the star-shaped diffuse scattering pattern can be well-simulated by FM zigzag chains randomly placed on a honeycomb lattice. Each star streak in $\mathbf{Q}$ space is contributed by chains in real space that run perpendicular to the streak. Because neutron scattering probes magnetic moments perpendicular to $\mathbf{Q}$, we infer that the magnetic moments in the FM chains point largely parallel to the chains -- similar to those in a typical zigzag magnetic structure \cite{LiuPRB2011,ChenPRB2021}. Given that the AFM order below $T_\mathrm{N}$ is preceded by the short-range FM chains above $T_\mathrm{N}$, a plausible scenario is that the system's leading magnetic interactions are strongly in favor of individual FM-chain formation, yet at the same time, they are weakly in favor of an alternating side-by-side arrangement of the chains, \textit{i.e.}, \textit{against} the formation of 2D FM order. Together with our inference of the moment direction above, the scenario echoes with the idea of bond-dependent anisotropic interactions, which is at the core of the Kitaev and related models.

We further notice that, among three types of parameters that are commonly considered for explaining the zigzag order \cite{JanssenPRB2017}, our result appears to be consistent with the expected behaviors of models with a leading nearest-neighbor symmetric off-diagonal interaction term, $\Gamma_1 > 0$. This is because for a given nearest-neighbor pair, the $\Gamma_1 > 0$ term favors FM alignment of the spin component parallel to the bond, but AFM alignment of the component perpendicular to both the bond and the Ising axis of the Kitaev term. Together with the geometry of the honeycomb lattice, the $\Gamma_1 > 0$ term can thus explain both the FM chains' formation tendency and their resistance to form 2D FM order. Indeed, in the Appendix of Ref.~\onlinecite{JanssenPRB2017}, we find a discussion of such models' several similar behaviors compared to our observations, including the formation of AFM${\frac{1}{3}}$ order and competing instabilities at a star-shaped set of wave vectors. Another major feature of such models is their demonstrated ability to produce large magnetic-response anisotropy without a highly anisotropic $g$ tensor \cite{JanssenPRB2017}. We thus expect our results to motivate further theoretical research of frustrated magnetism in the off-diagonal models \cite{SamarakoonPRB2018,SahaPRL2019,LuoNPJ2021}, some of which may have a QSL ground state \cite{LuoNPJ2021}.

To conclude, we have elucidated the field-induced phases and competing instabilities in the quantum magnet \ch{Na_3Co_2SbO_6}, and uncovered an unexpectedly large magnetic anisotropy. The results indicate exotic magnetic phases and render this system highly promising for further explorations using targeted external tuning. The results also stimulate future theoretical research of spin-orbit-coupled quantum magnets.

\begin{acknowledgments}

We are grateful for technical support by Qizhi Li and Jianping Sun, and for discussions with Wenjie Chen, Ji Feng, L. Janssen, G. Khaliullin, V. Kocsis, Huimei Liu, Qiang Luo, A.~U.~B. Wolter, and Shilong Zhang. The work at Peking University was supported by the National Basic Research Program of China (Grant Nos. 2021YFA1401900 and 2018YFA0305602) and the NSF of China (Grant Nos. 12061131004, 11874069, and 11888101). The work at Brookhaven National Laboratory was supported by Office of Basic Energy Sciences (BES), Division of Materials Sciences and Engineering, U.S. Department of Energy (DOE), under contract DE-SC0012704. X.L. further acknowledges support from China Postdoctoral Science Foundation (Grant No. 2020M680179). A portion of this research used resources at Spallation Neutron Source, a DOE Office of Science User Facility operated by the Oak Ridge National Laboratory. One of the neutron scattering experiments was performed at the MLF, J-PARC, Japan, under a user program (No. 2020B0148).

\end{acknowledgments}

\nocite{apsrev42Control}
\bibliographystyle{apsrev4-2}

\bibliography{NCSO_aniso_ref_reformat}

\begin{thebibliography}{84}%
\makeatletter
\providecommand \@ifxundefined [1]{%
 \@ifx{#1\undefined}
}%
\providecommand \@ifnum [1]{%
 \ifnum #1\expandafter \@firstoftwo
 \else \expandafter \@secondoftwo
 \fi
}%
\providecommand \@ifx [1]{%
 \ifx #1\expandafter \@firstoftwo
 \else \expandafter \@secondoftwo
 \fi
}%
\providecommand \natexlab [1]{#1}%
\providecommand \enquote  [1]{``#1''}%
\providecommand \bibnamefont  [1]{#1}%
\providecommand \bibfnamefont [1]{#1}%
\providecommand \citenamefont [1]{#1}%
\providecommand \href@noop [0]{\@secondoftwo}%
\providecommand \href [0]{\begingroup \@sanitize@url \@href}%
\providecommand \@href[1]{\@@startlink{#1}\@@href}%
\providecommand \@@href[1]{\endgroup#1\@@endlink}%
\providecommand \@sanitize@url [0]{\catcode `\\12\catcode `\$12\catcode
  `\&12\catcode `\#12\catcode `\^12\catcode `\_12\catcode `\%12\relax}%
\providecommand \@@startlink[1]{}%
\providecommand \@@endlink[0]{}%
\providecommand \url  [0]{\begingroup\@sanitize@url \@url }%
\providecommand \@url [1]{\endgroup\@href {#1}{\urlprefix }}%
\providecommand \urlprefix  [0]{URL }%
\providecommand \Eprint [0]{\href }%
\providecommand \doibase [0]{https://doi.org/}%
\providecommand \selectlanguage [0]{\@gobble}%
\providecommand \bibinfo  [0]{\@secondoftwo}%
\providecommand \bibfield  [0]{\@secondoftwo}%
\providecommand \translation [1]{[#1]}%
\providecommand \BibitemOpen [0]{}%
\providecommand \bibitemStop [0]{}%
\providecommand \bibitemNoStop [0]{.\EOS\space}%
\providecommand \EOS [0]{\spacefactor3000\relax}%
\providecommand \BibitemShut  [1]{\csname bibitem#1\endcsname}%
\let\auto@bib@innerbib\@empty
\bibitem [{\citenamefont {Balents}(2010)}]{BalentsNature2010}%
  \BibitemOpen
  \bibfield  {author} {\bibinfo {author} {\bibfnamefont {L.}~\bibnamefont
  {Balents}},\ }\href {https://doi.org/10.1038/nature08917} {\bibfield
  {journal} {\bibinfo  {journal} {Nature}\ }\textbf {\bibinfo {volume} {464}},\
  \bibinfo {pages} {199} (\bibinfo {year} {2010})}\BibitemShut {NoStop}%
\bibitem [{\citenamefont {Zhou}\ \emph {et~al.}(2017)\citenamefont {Zhou},
  \citenamefont {Kanoda},\ and\ \citenamefont {Ng}}]{ZhouRMP2017}%
  \BibitemOpen
  \bibfield  {author} {\bibinfo {author} {\bibfnamefont {Y.}~\bibnamefont
  {Zhou}}, \bibinfo {author} {\bibfnamefont {K.}~\bibnamefont {Kanoda}},\ and\
  \bibinfo {author} {\bibfnamefont {T.-K.}\ \bibnamefont {Ng}},\ }\href
  {https://doi.org/10.1103/RevModPhys.89.025003} {\bibfield  {journal}
  {\bibinfo  {journal} {Rev. Mod. Phys.}\ }\textbf {\bibinfo {volume} {89}},\
  \bibinfo {pages} {025003} (\bibinfo {year} {2017})}\BibitemShut {NoStop}%
\bibitem [{\citenamefont {Broholm}\ \emph {et~al.}(2020)\citenamefont
  {Broholm}, \citenamefont {Cava}, \citenamefont {Kivelson}, \citenamefont
  {Nocera}, \citenamefont {Norman},\ and\ \citenamefont
  {Senthil}}]{BroholmScience2020}%
  \BibitemOpen
  \bibfield  {author} {\bibinfo {author} {\bibfnamefont {C.}~\bibnamefont
  {Broholm}}, \bibinfo {author} {\bibfnamefont {R.}~\bibnamefont {Cava}},
  \bibinfo {author} {\bibfnamefont {S.}~\bibnamefont {Kivelson}}, \bibinfo
  {author} {\bibfnamefont {D.}~\bibnamefont {Nocera}}, \bibinfo {author}
  {\bibfnamefont {M.}~\bibnamefont {Norman}},\ and\ \bibinfo {author}
  {\bibfnamefont {T.}~\bibnamefont {Senthil}},\ }\href
  {https://doi.org/10.1126/science.aay0668} {\bibfield  {journal} {\bibinfo
  {journal} {Science}\ }\textbf {\bibinfo {volume} {367}},\ \bibinfo {pages}
  {eaay0668} (\bibinfo {year} {2020})}\BibitemShut {NoStop}%
\bibitem [{\citenamefont {Kitaev}(2006)}]{KitaevAP2006}%
  \BibitemOpen
  \bibfield  {author} {\bibinfo {author} {\bibfnamefont {A.}~\bibnamefont
  {Kitaev}},\ }\href
  {https://doi.org/https://doi.org/10.1016/j.aop.2005.10.005} {\bibfield
  {journal} {\bibinfo  {journal} {Ann. Phys.}\ }\textbf {\bibinfo {volume}
  {321}},\ \bibinfo {pages} {2 } (\bibinfo {year} {2006})}\BibitemShut
  {NoStop}%
\bibitem [{\citenamefont {Jackeli}\ and\ \citenamefont
  {Khaliullin}(2009)}]{JackeliPRL2009}%
  \BibitemOpen
  \bibfield  {author} {\bibinfo {author} {\bibfnamefont {G.}~\bibnamefont
  {Jackeli}}\ and\ \bibinfo {author} {\bibfnamefont {G.}~\bibnamefont
  {Khaliullin}},\ }\href {https://doi.org/10.1103/PhysRevLett.102.017205}
  {\bibfield  {journal} {\bibinfo  {journal} {Phys. Rev. Lett.}\ }\textbf
  {\bibinfo {volume} {102}},\ \bibinfo {pages} {017205} (\bibinfo {year}
  {2009})}\BibitemShut {NoStop}%
\bibitem [{\citenamefont {Takagi}\ \emph {et~al.}(2019)\citenamefont {Takagi},
  \citenamefont {Takayama}, \citenamefont {Jackeli}, \citenamefont
  {Khaliullin},\ and\ \citenamefont {Nagler}}]{TakagiNRP2019}%
  \BibitemOpen
  \bibfield  {author} {\bibinfo {author} {\bibfnamefont {H.}~\bibnamefont
  {Takagi}}, \bibinfo {author} {\bibfnamefont {T.}~\bibnamefont {Takayama}},
  \bibinfo {author} {\bibfnamefont {G.}~\bibnamefont {Jackeli}}, \bibinfo
  {author} {\bibfnamefont {G.}~\bibnamefont {Khaliullin}},\ and\ \bibinfo
  {author} {\bibfnamefont {S.~E.}\ \bibnamefont {Nagler}},\ }\href
  {https://doi.org/10.1038/s42254-019-0038-2} {\bibfield  {journal} {\bibinfo
  {journal} {Nat. Rev. Phys.}\ }\textbf {\bibinfo {volume} {1}},\ \bibinfo
  {pages} {264} (\bibinfo {year} {2019})}\BibitemShut {NoStop}%
\bibitem [{\citenamefont {Motome}\ \emph {et~al.}(2020)\citenamefont {Motome},
  \citenamefont {Sano}, \citenamefont {Jang}, \citenamefont {Sugita},\ and\
  \citenamefont {Kato}}]{MotomeJPCM2020}%
  \BibitemOpen
  \bibfield  {author} {\bibinfo {author} {\bibfnamefont {Y.}~\bibnamefont
  {Motome}}, \bibinfo {author} {\bibfnamefont {R.}~\bibnamefont {Sano}},
  \bibinfo {author} {\bibfnamefont {S.}~\bibnamefont {Jang}}, \bibinfo {author}
  {\bibfnamefont {Y.}~\bibnamefont {Sugita}},\ and\ \bibinfo {author}
  {\bibfnamefont {Y.}~\bibnamefont {Kato}},\ }\href
  {https://doi.org/10.1088/1361-648x/ab8525} {\bibfield  {journal} {\bibinfo
  {journal} {Journal of Physics: Condensed Matter}\ }\textbf {\bibinfo {volume}
  {32}},\ \bibinfo {pages} {404001} (\bibinfo {year} {2020})}\BibitemShut
  {NoStop}%
\bibitem [{\citenamefont {Trebst}\ and\ \citenamefont
  {Hickey}(2022)}]{TrebstPRRSP2022}%
  \BibitemOpen
  \bibfield  {author} {\bibinfo {author} {\bibfnamefont {S.}~\bibnamefont
  {Trebst}}\ and\ \bibinfo {author} {\bibfnamefont {C.}~\bibnamefont
  {Hickey}},\ }\href
  {https://doi.org/https://doi.org/10.1016/j.physrep.2021.11.003} {\bibfield
  {journal} {\bibinfo  {journal} {Physics Reports}\ }\textbf {\bibinfo {volume}
  {950}},\ \bibinfo {pages} {1} (\bibinfo {year} {2022})}\BibitemShut {NoStop}%
\bibitem [{\citenamefont {Chaloupka}\ \emph {et~al.}(2010)\citenamefont
  {Chaloupka}, \citenamefont {Jackeli},\ and\ \citenamefont
  {Khaliullin}}]{ChaloupkaPRL2010}%
  \BibitemOpen
  \bibfield  {author} {\bibinfo {author} {\bibfnamefont {J.}~\bibnamefont
  {Chaloupka}}, \bibinfo {author} {\bibfnamefont {G.}~\bibnamefont {Jackeli}},\
  and\ \bibinfo {author} {\bibfnamefont {G.}~\bibnamefont {Khaliullin}},\
  }\href {https://doi.org/10.1103/PhysRevLett.105.027204} {\bibfield  {journal}
  {\bibinfo  {journal} {Phys. Rev. Lett.}\ }\textbf {\bibinfo {volume} {105}},\
  \bibinfo {pages} {027204} (\bibinfo {year} {2010})}\BibitemShut {NoStop}%
\bibitem [{\citenamefont {Plumb}\ \emph {et~al.}(2014)\citenamefont {Plumb},
  \citenamefont {Clancy}, \citenamefont {Sandilands}, \citenamefont {Shankar},
  \citenamefont {Hu}, \citenamefont {Burch}, \citenamefont {Kee},\ and\
  \citenamefont {Kim}}]{PlumbPRB2014}%
  \BibitemOpen
  \bibfield  {author} {\bibinfo {author} {\bibfnamefont {K.~W.}\ \bibnamefont
  {Plumb}}, \bibinfo {author} {\bibfnamefont {J.~P.}\ \bibnamefont {Clancy}},
  \bibinfo {author} {\bibfnamefont {L.~J.}\ \bibnamefont {Sandilands}},
  \bibinfo {author} {\bibfnamefont {V.~V.}\ \bibnamefont {Shankar}}, \bibinfo
  {author} {\bibfnamefont {Y.~F.}\ \bibnamefont {Hu}}, \bibinfo {author}
  {\bibfnamefont {K.~S.}\ \bibnamefont {Burch}}, \bibinfo {author}
  {\bibfnamefont {H.-Y.}\ \bibnamefont {Kee}},\ and\ \bibinfo {author}
  {\bibfnamefont {Y.-J.}\ \bibnamefont {Kim}},\ }\href
  {https://doi.org/10.1103/PhysRevB.90.041112} {\bibfield  {journal} {\bibinfo
  {journal} {Phys. Rev. B}\ }\textbf {\bibinfo {volume} {90}},\ \bibinfo
  {pages} {041112} (\bibinfo {year} {2014})}\BibitemShut {NoStop}%
\bibitem [{\citenamefont {Liu}\ and\ \citenamefont
  {Khaliullin}(2018)}]{LiuPRB2018}%
  \BibitemOpen
  \bibfield  {author} {\bibinfo {author} {\bibfnamefont {H.}~\bibnamefont
  {Liu}}\ and\ \bibinfo {author} {\bibfnamefont {G.}~\bibnamefont
  {Khaliullin}},\ }\href {https://doi.org/10.1103/PhysRevB.97.014407}
  {\bibfield  {journal} {\bibinfo  {journal} {Phys. Rev. B}\ }\textbf {\bibinfo
  {volume} {97}},\ \bibinfo {pages} {014407} (\bibinfo {year}
  {2018})}\BibitemShut {NoStop}%
\bibitem [{\citenamefont {Sano}\ \emph {et~al.}(2018)\citenamefont {Sano},
  \citenamefont {Kato},\ and\ \citenamefont {Motome}}]{SanoPRB2018}%
  \BibitemOpen
  \bibfield  {author} {\bibinfo {author} {\bibfnamefont {R.}~\bibnamefont
  {Sano}}, \bibinfo {author} {\bibfnamefont {Y.}~\bibnamefont {Kato}},\ and\
  \bibinfo {author} {\bibfnamefont {Y.}~\bibnamefont {Motome}},\ }\href
  {https://doi.org/10.1103/PhysRevB.97.014408} {\bibfield  {journal} {\bibinfo
  {journal} {Phys. Rev. B}\ }\textbf {\bibinfo {volume} {97}},\ \bibinfo
  {pages} {014408} (\bibinfo {year} {2018})}\BibitemShut {NoStop}%
\bibitem [{\citenamefont {Liu}\ \emph {et~al.}(2020)\citenamefont {Liu},
  \citenamefont {Chaloupka},\ and\ \citenamefont {Khaliullin}}]{LiuPRL2020}%
  \BibitemOpen
  \bibfield  {author} {\bibinfo {author} {\bibfnamefont {H.}~\bibnamefont
  {Liu}}, \bibinfo {author} {\bibfnamefont {J.}~\bibnamefont {Chaloupka}},\
  and\ \bibinfo {author} {\bibfnamefont {G.}~\bibnamefont {Khaliullin}},\
  }\href {https://doi.org/10.1103/PhysRevLett.125.047201} {\bibfield  {journal}
  {\bibinfo  {journal} {Phys. Rev. Lett.}\ }\textbf {\bibinfo {volume} {125}},\
  \bibinfo {pages} {047201} (\bibinfo {year} {2020})}\BibitemShut {NoStop}%
\bibitem [{\citenamefont {Kim}\ \emph {et~al.}(2021{\natexlab{a}})\citenamefont
  {Kim}, \citenamefont {Kim},\ and\ \citenamefont {Park}}]{Kim_2021}%
  \BibitemOpen
  \bibfield  {author} {\bibinfo {author} {\bibfnamefont {C.}~\bibnamefont
  {Kim}}, \bibinfo {author} {\bibfnamefont {H.-S.}\ \bibnamefont {Kim}},\ and\
  \bibinfo {author} {\bibfnamefont {J.-G.}\ \bibnamefont {Park}},\ }\href
  {https://doi.org/10.1088/1361-648x/ac2d5d} {\bibfield  {journal} {\bibinfo
  {journal} {Journal of Physics: Condensed Matter}\ }\textbf {\bibinfo {volume}
  {34}},\ \bibinfo {pages} {023001} (\bibinfo {year}
  {2021}{\natexlab{a}})}\BibitemShut {NoStop}%
\bibitem [{\citenamefont {Liu}(2021)}]{LiuIJMPB2021}%
  \BibitemOpen
  \bibfield  {author} {\bibinfo {author} {\bibfnamefont {H.}~\bibnamefont
  {Liu}},\ }\href {https://doi.org/10.1142/S0217979221300061} {\bibfield
  {journal} {\bibinfo  {journal} {Int. J. Mod. Phys. B}\ }\textbf {\bibinfo
  {volume} {35}},\ \bibinfo {pages} {2130006} (\bibinfo {year}
  {2021})}\BibitemShut {NoStop}%
\bibitem [{\citenamefont {Viciu}\ \emph {et~al.}(2007)\citenamefont {Viciu},
  \citenamefont {Huang}, \citenamefont {Morosan}, \citenamefont {Zandbergen},
  \citenamefont {Greenbaum}, \citenamefont {McQueen},\ and\ \citenamefont
  {Cava}}]{ViciuJSSC2007}%
  \BibitemOpen
  \bibfield  {author} {\bibinfo {author} {\bibfnamefont {L.}~\bibnamefont
  {Viciu}}, \bibinfo {author} {\bibfnamefont {Q.}~\bibnamefont {Huang}},
  \bibinfo {author} {\bibfnamefont {E.}~\bibnamefont {Morosan}}, \bibinfo
  {author} {\bibfnamefont {H.}~\bibnamefont {Zandbergen}}, \bibinfo {author}
  {\bibfnamefont {N.}~\bibnamefont {Greenbaum}}, \bibinfo {author}
  {\bibfnamefont {T.}~\bibnamefont {McQueen}},\ and\ \bibinfo {author}
  {\bibfnamefont {R.}~\bibnamefont {Cava}},\ }\href
  {https://dx.doi.org/10.1016/j.jssc.2007.01.002} {\bibfield  {journal}
  {\bibinfo  {journal} {J. Solid State Chem.}\ }\textbf {\bibinfo {volume}
  {180}},\ \bibinfo {pages} {1060} (\bibinfo {year} {2007})}\BibitemShut
  {NoStop}%
\bibitem [{\citenamefont {Singh}\ and\ \citenamefont
  {Gegenwart}(2010)}]{SinghPRB2010}%
  \BibitemOpen
  \bibfield  {author} {\bibinfo {author} {\bibfnamefont {Y.}~\bibnamefont
  {Singh}}\ and\ \bibinfo {author} {\bibfnamefont {P.}~\bibnamefont
  {Gegenwart}},\ }\href {https://doi.org/10.1103/PhysRevB.82.064412} {\bibfield
   {journal} {\bibinfo  {journal} {Phys. Rev. B}\ }\textbf {\bibinfo {volume}
  {82}},\ \bibinfo {pages} {064412} (\bibinfo {year} {2010})}\BibitemShut
  {NoStop}%
\bibitem [{\citenamefont {Liu}\ \emph {et~al.}(2011)\citenamefont {Liu},
  \citenamefont {Berlijn}, \citenamefont {Yin}, \citenamefont {Ku},
  \citenamefont {Tsvelik}, \citenamefont {Kim}, \citenamefont {Gretarsson},
  \citenamefont {Singh}, \citenamefont {Gegenwart},\ and\ \citenamefont
  {Hill}}]{LiuPRB2011}%
  \BibitemOpen
  \bibfield  {author} {\bibinfo {author} {\bibfnamefont {X.}~\bibnamefont
  {Liu}}, \bibinfo {author} {\bibfnamefont {T.}~\bibnamefont {Berlijn}},
  \bibinfo {author} {\bibfnamefont {W.-G.}\ \bibnamefont {Yin}}, \bibinfo
  {author} {\bibfnamefont {W.}~\bibnamefont {Ku}}, \bibinfo {author}
  {\bibfnamefont {A.}~\bibnamefont {Tsvelik}}, \bibinfo {author} {\bibfnamefont
  {Y.-J.}\ \bibnamefont {Kim}}, \bibinfo {author} {\bibfnamefont
  {H.}~\bibnamefont {Gretarsson}}, \bibinfo {author} {\bibfnamefont
  {Y.}~\bibnamefont {Singh}}, \bibinfo {author} {\bibfnamefont
  {P.}~\bibnamefont {Gegenwart}},\ and\ \bibinfo {author} {\bibfnamefont
  {J.~P.}\ \bibnamefont {Hill}},\ }\href
  {https://doi.org/10.1103/PhysRevB.83.220403} {\bibfield  {journal} {\bibinfo
  {journal} {Phys. Rev. B}\ }\textbf {\bibinfo {volume} {83}},\ \bibinfo
  {pages} {220403} (\bibinfo {year} {2011})}\BibitemShut {NoStop}%
\bibitem [{\citenamefont {Johnson}\ \emph {et~al.}(2015)\citenamefont
  {Johnson}, \citenamefont {Williams}, \citenamefont {Haghighirad},
  \citenamefont {Singleton}, \citenamefont {Zapf}, \citenamefont {Manuel},
  \citenamefont {Mazin}, \citenamefont {Li}, \citenamefont {Jeschke},
  \citenamefont {Valent\'{\i}},\ and\ \citenamefont {Coldea}}]{JohnsonPRB2015}%
  \BibitemOpen
  \bibfield  {author} {\bibinfo {author} {\bibfnamefont {R.~D.}\ \bibnamefont
  {Johnson}}, \bibinfo {author} {\bibfnamefont {S.~C.}\ \bibnamefont
  {Williams}}, \bibinfo {author} {\bibfnamefont {A.~A.}\ \bibnamefont
  {Haghighirad}}, \bibinfo {author} {\bibfnamefont {J.}~\bibnamefont
  {Singleton}}, \bibinfo {author} {\bibfnamefont {V.}~\bibnamefont {Zapf}},
  \bibinfo {author} {\bibfnamefont {P.}~\bibnamefont {Manuel}}, \bibinfo
  {author} {\bibfnamefont {I.~I.}\ \bibnamefont {Mazin}}, \bibinfo {author}
  {\bibfnamefont {Y.}~\bibnamefont {Li}}, \bibinfo {author} {\bibfnamefont
  {H.~O.}\ \bibnamefont {Jeschke}}, \bibinfo {author} {\bibfnamefont
  {R.}~\bibnamefont {Valent\'{\i}}},\ and\ \bibinfo {author} {\bibfnamefont
  {R.}~\bibnamefont {Coldea}},\ }\href
  {https://doi.org/10.1103/PhysRevB.92.235119} {\bibfield  {journal} {\bibinfo
  {journal} {Phys. Rev. B}\ }\textbf {\bibinfo {volume} {92}},\ \bibinfo
  {pages} {235119} (\bibinfo {year} {2015})}\BibitemShut {NoStop}%
\bibitem [{\citenamefont {Zhong}\ \emph {et~al.}(2020)\citenamefont {Zhong},
  \citenamefont {Gao}, \citenamefont {Ong},\ and\ \citenamefont
  {Cava}}]{ZhongSA2020}%
  \BibitemOpen
  \bibfield  {author} {\bibinfo {author} {\bibfnamefont {R.}~\bibnamefont
  {Zhong}}, \bibinfo {author} {\bibfnamefont {T.}~\bibnamefont {Gao}}, \bibinfo
  {author} {\bibfnamefont {N.~P.}\ \bibnamefont {Ong}},\ and\ \bibinfo {author}
  {\bibfnamefont {R.~J.}\ \bibnamefont {Cava}},\ }\href
  {https://doi.org/10.1126/sciadv.aay6953} {\bibfield  {journal} {\bibinfo
  {journal} {Science Advances}\ }\textbf {\bibinfo {volume} {6}},\ \bibinfo
  {pages} {eaay6953} (\bibinfo {year} {2020})}\BibitemShut {NoStop}%
\bibitem [{\citenamefont {Kimchi}\ and\ \citenamefont
  {You}(2011)}]{KimchiPRB2011}%
  \BibitemOpen
  \bibfield  {author} {\bibinfo {author} {\bibfnamefont {I.}~\bibnamefont
  {Kimchi}}\ and\ \bibinfo {author} {\bibfnamefont {Y.-Z.}\ \bibnamefont
  {You}},\ }\href {https://doi.org/10.1103/PhysRevB.84.180407} {\bibfield
  {journal} {\bibinfo  {journal} {Phys. Rev. B}\ }\textbf {\bibinfo {volume}
  {84}},\ \bibinfo {pages} {180407} (\bibinfo {year} {2011})}\BibitemShut
  {NoStop}%
\bibitem [{\citenamefont {Chaloupka}\ \emph {et~al.}(2013)\citenamefont
  {Chaloupka}, \citenamefont {Jackeli},\ and\ \citenamefont
  {Khaliullin}}]{ChaloupkaPRL2013}%
  \BibitemOpen
  \bibfield  {author} {\bibinfo {author} {\bibfnamefont {J.}~\bibnamefont
  {Chaloupka}}, \bibinfo {author} {\bibfnamefont {G.}~\bibnamefont {Jackeli}},\
  and\ \bibinfo {author} {\bibfnamefont {G.}~\bibnamefont {Khaliullin}},\
  }\href {https://doi.org/10.1103/PhysRevLett.110.097204} {\bibfield  {journal}
  {\bibinfo  {journal} {Phys. Rev. Lett.}\ }\textbf {\bibinfo {volume} {110}},\
  \bibinfo {pages} {097204} (\bibinfo {year} {2013})}\BibitemShut {NoStop}%
\bibitem [{\citenamefont {Rau}\ \emph {et~al.}(2014)\citenamefont {Rau},
  \citenamefont {Lee},\ and\ \citenamefont {Kee}}]{RauPRL2014}%
  \BibitemOpen
  \bibfield  {author} {\bibinfo {author} {\bibfnamefont {J.~G.}\ \bibnamefont
  {Rau}}, \bibinfo {author} {\bibfnamefont {E.~K.-H.}\ \bibnamefont {Lee}},\
  and\ \bibinfo {author} {\bibfnamefont {H.-Y.}\ \bibnamefont {Kee}},\ }\href
  {https://doi.org/10.1103/PhysRevLett.112.077204} {\bibfield  {journal}
  {\bibinfo  {journal} {Phys. Rev. Lett.}\ }\textbf {\bibinfo {volume} {112}},\
  \bibinfo {pages} {077204} (\bibinfo {year} {2014})}\BibitemShut {NoStop}%
\bibitem [{\citenamefont {Sizyuk}\ \emph {et~al.}(2014)\citenamefont {Sizyuk},
  \citenamefont {Price}, \citenamefont {W\"olfle},\ and\ \citenamefont
  {Perkins}}]{SizyukPRB2014}%
  \BibitemOpen
  \bibfield  {author} {\bibinfo {author} {\bibfnamefont {Y.}~\bibnamefont
  {Sizyuk}}, \bibinfo {author} {\bibfnamefont {C.}~\bibnamefont {Price}},
  \bibinfo {author} {\bibfnamefont {P.}~\bibnamefont {W\"olfle}},\ and\
  \bibinfo {author} {\bibfnamefont {N.~B.}\ \bibnamefont {Perkins}},\ }\href
  {https://doi.org/10.1103/PhysRevB.90.155126} {\bibfield  {journal} {\bibinfo
  {journal} {Phys. Rev. B}\ }\textbf {\bibinfo {volume} {90}},\ \bibinfo
  {pages} {155126} (\bibinfo {year} {2014})}\BibitemShut {NoStop}%
\bibitem [{\citenamefont {Yamaji}\ \emph {et~al.}(2014)\citenamefont {Yamaji},
  \citenamefont {Nomura}, \citenamefont {Kurita}, \citenamefont {Arita},\ and\
  \citenamefont {Imada}}]{YamajiPRL2014}%
  \BibitemOpen
  \bibfield  {author} {\bibinfo {author} {\bibfnamefont {Y.}~\bibnamefont
  {Yamaji}}, \bibinfo {author} {\bibfnamefont {Y.}~\bibnamefont {Nomura}},
  \bibinfo {author} {\bibfnamefont {M.}~\bibnamefont {Kurita}}, \bibinfo
  {author} {\bibfnamefont {R.}~\bibnamefont {Arita}},\ and\ \bibinfo {author}
  {\bibfnamefont {M.}~\bibnamefont {Imada}},\ }\href
  {https://doi.org/10.1103/PhysRevLett.113.107201} {\bibfield  {journal}
  {\bibinfo  {journal} {Phys. Rev. Lett.}\ }\textbf {\bibinfo {volume} {113}},\
  \bibinfo {pages} {107201} (\bibinfo {year} {2014})}\BibitemShut {NoStop}%
\bibitem [{\citenamefont {Katukuri}\ \emph {et~al.}(2014)\citenamefont
  {Katukuri}, \citenamefont {Nishimoto}, \citenamefont {Yushankhai},
  \citenamefont {Stoyanova}, \citenamefont {Kandpal}, \citenamefont {Choi},
  \citenamefont {Coldea}, \citenamefont {Rousochatzakis}, \citenamefont
  {Hozoi},\ and\ \citenamefont {van~den Brink}}]{KatukuriNJP2014}%
  \BibitemOpen
  \bibfield  {author} {\bibinfo {author} {\bibfnamefont {V.~M.}\ \bibnamefont
  {Katukuri}}, \bibinfo {author} {\bibfnamefont {S.}~\bibnamefont {Nishimoto}},
  \bibinfo {author} {\bibfnamefont {V.}~\bibnamefont {Yushankhai}}, \bibinfo
  {author} {\bibfnamefont {A.}~\bibnamefont {Stoyanova}}, \bibinfo {author}
  {\bibfnamefont {H.}~\bibnamefont {Kandpal}}, \bibinfo {author} {\bibfnamefont
  {S.}~\bibnamefont {Choi}}, \bibinfo {author} {\bibfnamefont {R.}~\bibnamefont
  {Coldea}}, \bibinfo {author} {\bibfnamefont {I.}~\bibnamefont
  {Rousochatzakis}}, \bibinfo {author} {\bibfnamefont {L.}~\bibnamefont
  {Hozoi}},\ and\ \bibinfo {author} {\bibfnamefont {J.}~\bibnamefont {van~den
  Brink}},\ }\href {https://doi.org/10.1088/1367-2630/16/1/013056} {\bibfield
  {journal} {\bibinfo  {journal} {New J. Phys.}\ }\textbf {\bibinfo {volume}
  {16}},\ \bibinfo {pages} {013056} (\bibinfo {year} {2014})}\BibitemShut
  {NoStop}%
\bibitem [{\citenamefont {Rousochatzakis}\ \emph {et~al.}(2015)\citenamefont
  {Rousochatzakis}, \citenamefont {Reuther}, \citenamefont {Thomale},
  \citenamefont {Rachel},\ and\ \citenamefont
  {Perkins}}]{RousochatzakisPRX2015}%
  \BibitemOpen
  \bibfield  {author} {\bibinfo {author} {\bibfnamefont {I.}~\bibnamefont
  {Rousochatzakis}}, \bibinfo {author} {\bibfnamefont {J.}~\bibnamefont
  {Reuther}}, \bibinfo {author} {\bibfnamefont {R.}~\bibnamefont {Thomale}},
  \bibinfo {author} {\bibfnamefont {S.}~\bibnamefont {Rachel}},\ and\ \bibinfo
  {author} {\bibfnamefont {N.~B.}\ \bibnamefont {Perkins}},\ }\href
  {https://doi.org/10.1103/PhysRevX.5.041035} {\bibfield  {journal} {\bibinfo
  {journal} {Phys. Rev. X}\ }\textbf {\bibinfo {volume} {5}},\ \bibinfo {pages}
  {041035} (\bibinfo {year} {2015})}\BibitemShut {NoStop}%
\bibitem [{\citenamefont {Chaloupka}\ and\ \citenamefont
  {Khaliullin}(2016)}]{ChaloupkaPRB2016}%
  \BibitemOpen
  \bibfield  {author} {\bibinfo {author} {\bibfnamefont {J.}~\bibnamefont
  {Chaloupka}}\ and\ \bibinfo {author} {\bibfnamefont {G.}~\bibnamefont
  {Khaliullin}},\ }\href {https://doi.org/10.1103/PhysRevB.94.064435}
  {\bibfield  {journal} {\bibinfo  {journal} {Phys. Rev. B}\ }\textbf {\bibinfo
  {volume} {94}},\ \bibinfo {pages} {064435} (\bibinfo {year}
  {2016})}\BibitemShut {NoStop}%
\bibitem [{\citenamefont {Chaloupka}\ and\ \citenamefont
  {Khaliullin}(2015)}]{ChaloupkaPRB2015}%
  \BibitemOpen
  \bibfield  {author} {\bibinfo {author} {\bibfnamefont {J.}~\bibnamefont
  {Chaloupka}}\ and\ \bibinfo {author} {\bibfnamefont {G.}~\bibnamefont
  {Khaliullin}},\ }\href {https://doi.org/10.1103/PhysRevB.92.024413}
  {\bibfield  {journal} {\bibinfo  {journal} {Phys. Rev. B}\ }\textbf {\bibinfo
  {volume} {92}},\ \bibinfo {pages} {024413} (\bibinfo {year}
  {2015})}\BibitemShut {NoStop}%
\bibitem [{\citenamefont {Winter}\ \emph {et~al.}(2016)\citenamefont {Winter},
  \citenamefont {Li}, \citenamefont {Jeschke},\ and\ \citenamefont
  {Valent\'{\i}}}]{WinterPRB2016}%
  \BibitemOpen
  \bibfield  {author} {\bibinfo {author} {\bibfnamefont {S.~M.}\ \bibnamefont
  {Winter}}, \bibinfo {author} {\bibfnamefont {Y.}~\bibnamefont {Li}}, \bibinfo
  {author} {\bibfnamefont {H.~O.}\ \bibnamefont {Jeschke}},\ and\ \bibinfo
  {author} {\bibfnamefont {R.}~\bibnamefont {Valent\'{\i}}},\ }\href
  {https://doi.org/10.1103/PhysRevB.93.214431} {\bibfield  {journal} {\bibinfo
  {journal} {Phys. Rev. B}\ }\textbf {\bibinfo {volume} {93}},\ \bibinfo
  {pages} {214431} (\bibinfo {year} {2016})}\BibitemShut {NoStop}%
\bibitem [{\citenamefont {Do}\ \emph {et~al.}(2017)\citenamefont {Do},
  \citenamefont {Park}, \citenamefont {Yoshitake}, \citenamefont {Nasu},
  \citenamefont {Motome}, \citenamefont {Kwon}, \citenamefont {Adroja},
  \citenamefont {Voneshen}, \citenamefont {Kim}, \citenamefont {Jang},
  \citenamefont {Park}, \citenamefont {Choi},\ and\ \citenamefont
  {Ji}}]{DoNP2017}%
  \BibitemOpen
  \bibfield  {author} {\bibinfo {author} {\bibfnamefont {S.-H.}\ \bibnamefont
  {Do}}, \bibinfo {author} {\bibfnamefont {S.-Y.}\ \bibnamefont {Park}},
  \bibinfo {author} {\bibfnamefont {J.}~\bibnamefont {Yoshitake}}, \bibinfo
  {author} {\bibfnamefont {J.}~\bibnamefont {Nasu}}, \bibinfo {author}
  {\bibfnamefont {Y.}~\bibnamefont {Motome}}, \bibinfo {author} {\bibfnamefont
  {Y.~S.}\ \bibnamefont {Kwon}}, \bibinfo {author} {\bibfnamefont
  {D.}~\bibnamefont {Adroja}}, \bibinfo {author} {\bibfnamefont
  {D.}~\bibnamefont {Voneshen}}, \bibinfo {author} {\bibfnamefont
  {K.}~\bibnamefont {Kim}}, \bibinfo {author} {\bibfnamefont {T.-H.}\
  \bibnamefont {Jang}}, \bibinfo {author} {\bibfnamefont {J.-H.}\ \bibnamefont
  {Park}}, \bibinfo {author} {\bibfnamefont {K.-Y.}\ \bibnamefont {Choi}},\
  and\ \bibinfo {author} {\bibfnamefont {S.}~\bibnamefont {Ji}},\ }\href
  {https://doi.org/10.1038/nphys4264} {\bibfield  {journal} {\bibinfo
  {journal} {Nat. Phys.}\ }\textbf {\bibinfo {volume} {13}},\ \bibinfo {pages}
  {1079} (\bibinfo {year} {2017})}\BibitemShut {NoStop}%
\bibitem [{\citenamefont {Wang}\ \emph {et~al.}(2018)\citenamefont {Wang},
  \citenamefont {Guo}, \citenamefont {Tafti}, \citenamefont {Hegg},
  \citenamefont {Sen}, \citenamefont {Sidorov}, \citenamefont {Wang},
  \citenamefont {Cai}, \citenamefont {Yi}, \citenamefont {Zhou}, \citenamefont
  {Wang}, \citenamefont {Zhang}, \citenamefont {Yang}, \citenamefont {Li},
  \citenamefont {Li}, \citenamefont {Li}, \citenamefont {Liu}, \citenamefont
  {Shi}, \citenamefont {Ku}, \citenamefont {Wu}, \citenamefont {Cava},\ and\
  \citenamefont {Sun}}]{WangPRB2018}%
  \BibitemOpen
  \bibfield  {author} {\bibinfo {author} {\bibfnamefont {Z.}~\bibnamefont
  {Wang}}, \bibinfo {author} {\bibfnamefont {J.}~\bibnamefont {Guo}}, \bibinfo
  {author} {\bibfnamefont {F.~F.}\ \bibnamefont {Tafti}}, \bibinfo {author}
  {\bibfnamefont {A.}~\bibnamefont {Hegg}}, \bibinfo {author} {\bibfnamefont
  {S.}~\bibnamefont {Sen}}, \bibinfo {author} {\bibfnamefont {V.~A.}\
  \bibnamefont {Sidorov}}, \bibinfo {author} {\bibfnamefont {L.}~\bibnamefont
  {Wang}}, \bibinfo {author} {\bibfnamefont {S.}~\bibnamefont {Cai}}, \bibinfo
  {author} {\bibfnamefont {W.}~\bibnamefont {Yi}}, \bibinfo {author}
  {\bibfnamefont {Y.}~\bibnamefont {Zhou}}, \bibinfo {author} {\bibfnamefont
  {H.}~\bibnamefont {Wang}}, \bibinfo {author} {\bibfnamefont {S.}~\bibnamefont
  {Zhang}}, \bibinfo {author} {\bibfnamefont {K.}~\bibnamefont {Yang}},
  \bibinfo {author} {\bibfnamefont {A.}~\bibnamefont {Li}}, \bibinfo {author}
  {\bibfnamefont {X.}~\bibnamefont {Li}}, \bibinfo {author} {\bibfnamefont
  {Y.}~\bibnamefont {Li}}, \bibinfo {author} {\bibfnamefont {J.}~\bibnamefont
  {Liu}}, \bibinfo {author} {\bibfnamefont {Y.}~\bibnamefont {Shi}}, \bibinfo
  {author} {\bibfnamefont {W.}~\bibnamefont {Ku}}, \bibinfo {author}
  {\bibfnamefont {Q.}~\bibnamefont {Wu}}, \bibinfo {author} {\bibfnamefont
  {R.~J.}\ \bibnamefont {Cava}},\ and\ \bibinfo {author} {\bibfnamefont
  {L.}~\bibnamefont {Sun}},\ }\href
  {https://doi.org/10.1103/PhysRevB.97.245149} {\bibfield  {journal} {\bibinfo
  {journal} {Phys. Rev. B}\ }\textbf {\bibinfo {volume} {97}},\ \bibinfo
  {pages} {245149} (\bibinfo {year} {2018})}\BibitemShut {NoStop}%
\bibitem [{\citenamefont {Winter}\ \emph {et~al.}(2018)\citenamefont {Winter},
  \citenamefont {Riedl}, \citenamefont {Kaib}, \citenamefont {Coldea},\ and\
  \citenamefont {Valent\'{\i}}}]{WinterPRL2018}%
  \BibitemOpen
  \bibfield  {author} {\bibinfo {author} {\bibfnamefont {S.~M.}\ \bibnamefont
  {Winter}}, \bibinfo {author} {\bibfnamefont {K.}~\bibnamefont {Riedl}},
  \bibinfo {author} {\bibfnamefont {D.}~\bibnamefont {Kaib}}, \bibinfo {author}
  {\bibfnamefont {R.}~\bibnamefont {Coldea}},\ and\ \bibinfo {author}
  {\bibfnamefont {R.}~\bibnamefont {Valent\'{\i}}},\ }\href
  {https://doi.org/10.1103/PhysRevLett.120.077203} {\bibfield  {journal}
  {\bibinfo  {journal} {Phys. Rev. Lett.}\ }\textbf {\bibinfo {volume} {120}},\
  \bibinfo {pages} {077203} (\bibinfo {year} {2018})}\BibitemShut {NoStop}%
\bibitem [{\citenamefont {Banerjee}\ \emph {et~al.}(2018)\citenamefont
  {Banerjee}, \citenamefont {Lampen-Kelley}, \citenamefont {Knolle},
  \citenamefont {Balz}, \citenamefont {Aczel}, \citenamefont {Winn},
  \citenamefont {Liu}, \citenamefont {Pajerowski}, \citenamefont {Yan},
  \citenamefont {Bridges}, \citenamefont {Savici}, \citenamefont {Chakoumakos},
  \citenamefont {Lumsden}, \citenamefont {Tennant}, \citenamefont {Moessner},
  \citenamefont {Mandrus},\ and\ \citenamefont {Nagler}}]{BanerjeeNPJ2018}%
  \BibitemOpen
  \bibfield  {author} {\bibinfo {author} {\bibfnamefont {A.}~\bibnamefont
  {Banerjee}}, \bibinfo {author} {\bibfnamefont {P.}~\bibnamefont
  {Lampen-Kelley}}, \bibinfo {author} {\bibfnamefont {J.}~\bibnamefont
  {Knolle}}, \bibinfo {author} {\bibfnamefont {C.}~\bibnamefont {Balz}},
  \bibinfo {author} {\bibfnamefont {A.~A.}\ \bibnamefont {Aczel}}, \bibinfo
  {author} {\bibfnamefont {B.}~\bibnamefont {Winn}}, \bibinfo {author}
  {\bibfnamefont {Y.}~\bibnamefont {Liu}}, \bibinfo {author} {\bibfnamefont
  {D.}~\bibnamefont {Pajerowski}}, \bibinfo {author} {\bibfnamefont
  {J.}~\bibnamefont {Yan}}, \bibinfo {author} {\bibfnamefont {C.~A.}\
  \bibnamefont {Bridges}}, \bibinfo {author} {\bibfnamefont {A.~T.}\
  \bibnamefont {Savici}}, \bibinfo {author} {\bibfnamefont {B.~C.}\
  \bibnamefont {Chakoumakos}}, \bibinfo {author} {\bibfnamefont {M.~D.}\
  \bibnamefont {Lumsden}}, \bibinfo {author} {\bibfnamefont {D.~A.}\
  \bibnamefont {Tennant}}, \bibinfo {author} {\bibfnamefont {R.}~\bibnamefont
  {Moessner}}, \bibinfo {author} {\bibfnamefont {D.~G.}\ \bibnamefont
  {Mandrus}},\ and\ \bibinfo {author} {\bibfnamefont {S.~E.}\ \bibnamefont
  {Nagler}},\ }\href {https://doi.org/10.1038/s41535-018-0079-2} {\bibfield
  {journal} {\bibinfo  {journal} {npj Quantum Materials}\ }\textbf {\bibinfo
  {volume} {3}},\ \bibinfo {pages} {8} (\bibinfo {year} {2018})}\BibitemShut
  {NoStop}%
\bibitem [{\citenamefont {Gordon}\ \emph {et~al.}(2019)\citenamefont {Gordon},
  \citenamefont {Catuneanu}, \citenamefont {S$\mathrm{\o}$rensen},\ and\
  \citenamefont {Kee}}]{GordonNC2019}%
  \BibitemOpen
  \bibfield  {author} {\bibinfo {author} {\bibfnamefont {J.~S.}\ \bibnamefont
  {Gordon}}, \bibinfo {author} {\bibfnamefont {A.}~\bibnamefont {Catuneanu}},
  \bibinfo {author} {\bibfnamefont {E.~S.}\ \bibnamefont
  {S$\mathrm{\o}$rensen}},\ and\ \bibinfo {author} {\bibfnamefont {H.-Y.}\
  \bibnamefont {Kee}},\ }\href {https://doi.org/10.1038/s41467-019-10405-8}
  {\bibfield  {journal} {\bibinfo  {journal} {Nat. Commun.}\ }\textbf {\bibinfo
  {volume} {10}},\ \bibinfo {pages} {2470} (\bibinfo {year}
  {2019})}\BibitemShut {NoStop}%
\bibitem [{\citenamefont {Hickey}\ and\ \citenamefont
  {Trebst}(2019)}]{HickeyNC2019}%
  \BibitemOpen
  \bibfield  {author} {\bibinfo {author} {\bibfnamefont {C.}~\bibnamefont
  {Hickey}}\ and\ \bibinfo {author} {\bibfnamefont {S.}~\bibnamefont
  {Trebst}},\ }\href {https://doi.org/10.1038/s41467-019-08459-9} {\bibfield
  {journal} {\bibinfo  {journal} {Nat. Commun.}\ }\textbf {\bibinfo {volume}
  {10}},\ \bibinfo {pages} {530} (\bibinfo {year} {2019})}\BibitemShut
  {NoStop}%
\bibitem [{\citenamefont {Yao}\ and\ \citenamefont {Li}(2020)}]{YaoPRB2020}%
  \BibitemOpen
  \bibfield  {author} {\bibinfo {author} {\bibfnamefont {W.}~\bibnamefont
  {Yao}}\ and\ \bibinfo {author} {\bibfnamefont {Y.}~\bibnamefont {Li}},\
  }\href {https://doi.org/10.1103/PhysRevB.101.085120} {\bibfield  {journal}
  {\bibinfo  {journal} {Phys. Rev. B}\ }\textbf {\bibinfo {volume} {101}},\
  \bibinfo {pages} {085120} (\bibinfo {year} {2020})}\BibitemShut {NoStop}%
\bibitem [{\citenamefont {Wong}\ \emph {et~al.}(2016)\citenamefont {Wong},
  \citenamefont {Avdeev},\ and\ \citenamefont {Ling}}]{WongJSSC2016}%
  \BibitemOpen
  \bibfield  {author} {\bibinfo {author} {\bibfnamefont {C.}~\bibnamefont
  {Wong}}, \bibinfo {author} {\bibfnamefont {M.}~\bibnamefont {Avdeev}},\ and\
  \bibinfo {author} {\bibfnamefont {C.~D.}\ \bibnamefont {Ling}},\ }\href
  {https://doi.org/10.1016/j.jssc.2016.07.032} {\bibfield  {journal} {\bibinfo
  {journal} {J. Solid State Chem.}\ }\textbf {\bibinfo {volume} {243}},\
  \bibinfo {pages} {18} (\bibinfo {year} {2016})}\BibitemShut {NoStop}%
\bibitem [{\citenamefont {Yan}\ \emph {et~al.}(2019)\citenamefont {Yan},
  \citenamefont {Okamoto}, \citenamefont {Wu}, \citenamefont {Zheng},
  \citenamefont {Zhou}, \citenamefont {Cao},\ and\ \citenamefont
  {McGuire}}]{YanPRM2019}%
  \BibitemOpen
  \bibfield  {author} {\bibinfo {author} {\bibfnamefont {J.-Q.}\ \bibnamefont
  {Yan}}, \bibinfo {author} {\bibfnamefont {S.}~\bibnamefont {Okamoto}},
  \bibinfo {author} {\bibfnamefont {Y.}~\bibnamefont {Wu}}, \bibinfo {author}
  {\bibfnamefont {Q.}~\bibnamefont {Zheng}}, \bibinfo {author} {\bibfnamefont
  {H.~D.}\ \bibnamefont {Zhou}}, \bibinfo {author} {\bibfnamefont {H.~B.}\
  \bibnamefont {Cao}},\ and\ \bibinfo {author} {\bibfnamefont {M.~A.}\
  \bibnamefont {McGuire}},\ }\href
  {https://doi.org/10.1103/PhysRevMaterials.3.074405} {\bibfield  {journal}
  {\bibinfo  {journal} {Phys. Rev. Materials}\ }\textbf {\bibinfo {volume}
  {3}},\ \bibinfo {pages} {074405} (\bibinfo {year} {2019})}\BibitemShut
  {NoStop}%
\bibitem [{\citenamefont {Stratan}\ \emph {et~al.}(2019)\citenamefont
  {Stratan}, \citenamefont {Shukaev}, \citenamefont {Vasilchikova},
  \citenamefont {Vasiliev}, \citenamefont {Korshunov}, \citenamefont
  {Kurbakov}, \citenamefont {Nalbandyan},\ and\ \citenamefont
  {Zvereva}}]{StratanNJC2019}%
  \BibitemOpen
  \bibfield  {author} {\bibinfo {author} {\bibfnamefont {M.~I.}\ \bibnamefont
  {Stratan}}, \bibinfo {author} {\bibfnamefont {I.~L.}\ \bibnamefont
  {Shukaev}}, \bibinfo {author} {\bibfnamefont {T.~M.}\ \bibnamefont
  {Vasilchikova}}, \bibinfo {author} {\bibfnamefont {A.~N.}\ \bibnamefont
  {Vasiliev}}, \bibinfo {author} {\bibfnamefont {A.~N.}\ \bibnamefont
  {Korshunov}}, \bibinfo {author} {\bibfnamefont {A.~I.}\ \bibnamefont
  {Kurbakov}}, \bibinfo {author} {\bibfnamefont {V.~B.}\ \bibnamefont
  {Nalbandyan}},\ and\ \bibinfo {author} {\bibfnamefont {E.~A.}\ \bibnamefont
  {Zvereva}},\ }\href {https://doi.org/10.1039/C9NJ03627J} {\bibfield
  {journal} {\bibinfo  {journal} {New J. Chem.}\ }\textbf {\bibinfo {volume}
  {43}},\ \bibinfo {pages} {13545} (\bibinfo {year} {2019})}\BibitemShut
  {NoStop}%
\bibitem [{\citenamefont {Wang}\ \emph {et~al.}(2017)\citenamefont {Wang},
  \citenamefont {Dong}, \citenamefont {Yu},\ and\ \citenamefont
  {Li}}]{WangPRB2017}%
  \BibitemOpen
  \bibfield  {author} {\bibinfo {author} {\bibfnamefont {W.}~\bibnamefont
  {Wang}}, \bibinfo {author} {\bibfnamefont {Z.-Y.}\ \bibnamefont {Dong}},
  \bibinfo {author} {\bibfnamefont {S.-L.}\ \bibnamefont {Yu}},\ and\ \bibinfo
  {author} {\bibfnamefont {J.-X.}\ \bibnamefont {Li}},\ }\href
  {https://doi.org/10.1103/PhysRevB.96.115103} {\bibfield  {journal} {\bibinfo
  {journal} {Phys. Rev. B}\ }\textbf {\bibinfo {volume} {96}},\ \bibinfo
  {pages} {115103} (\bibinfo {year} {2017})}\BibitemShut {NoStop}%
\bibitem [{\citenamefont {Janssen}\ \emph {et~al.}(2017)\citenamefont
  {Janssen}, \citenamefont {Andrade},\ and\ \citenamefont
  {Vojta}}]{JanssenPRB2017}%
  \BibitemOpen
  \bibfield  {author} {\bibinfo {author} {\bibfnamefont {L.}~\bibnamefont
  {Janssen}}, \bibinfo {author} {\bibfnamefont {E.~C.}\ \bibnamefont
  {Andrade}},\ and\ \bibinfo {author} {\bibfnamefont {M.}~\bibnamefont
  {Vojta}},\ }\href {https://doi.org/10.1103/PhysRevB.96.064430} {\bibfield
  {journal} {\bibinfo  {journal} {Phys. Rev. B}\ }\textbf {\bibinfo {volume}
  {96}},\ \bibinfo {pages} {064430} (\bibinfo {year} {2017})}\BibitemShut
  {NoStop}%
\bibitem [{\citenamefont {Rusna\ifmmode~\check{c}\else \v{c}\fi{}ko}\ \emph
  {et~al.}(2019)\citenamefont {Rusna\ifmmode~\check{c}\else \v{c}\fi{}ko},
  \citenamefont {Gotfryd},\ and\ \citenamefont {Chaloupka}}]{RusnackoPRB2019}%
  \BibitemOpen
  \bibfield  {author} {\bibinfo {author} {\bibfnamefont {J.}~\bibnamefont
  {Rusna\ifmmode~\check{c}\else \v{c}\fi{}ko}}, \bibinfo {author}
  {\bibfnamefont {D.}~\bibnamefont {Gotfryd}},\ and\ \bibinfo {author}
  {\bibfnamefont {J.}~\bibnamefont {Chaloupka}},\ }\href
  {https://doi.org/10.1103/PhysRevB.99.064425} {\bibfield  {journal} {\bibinfo
  {journal} {Phys. Rev. B}\ }\textbf {\bibinfo {volume} {99}},\ \bibinfo
  {pages} {064425} (\bibinfo {year} {2019})}\BibitemShut {NoStop}%
\bibitem [{\citenamefont {Maksimov}\ and\ \citenamefont
  {Chernyshev}(2020)}]{MaksimovPRR2020}%
  \BibitemOpen
  \bibfield  {author} {\bibinfo {author} {\bibfnamefont {P.~A.}\ \bibnamefont
  {Maksimov}}\ and\ \bibinfo {author} {\bibfnamefont {A.~L.}\ \bibnamefont
  {Chernyshev}},\ }\href {https://doi.org/10.1103/PhysRevResearch.2.033011}
  {\bibfield  {journal} {\bibinfo  {journal} {Phys. Rev. Research}\ }\textbf
  {\bibinfo {volume} {2}},\ \bibinfo {pages} {033011} (\bibinfo {year}
  {2020})}\BibitemShut {NoStop}%
\bibitem [{\citenamefont {Laurell}\ and\ \citenamefont
  {Okamoto}(2020)}]{LaurellNPJQM2020}%
  \BibitemOpen
  \bibfield  {author} {\bibinfo {author} {\bibfnamefont {P.}~\bibnamefont
  {Laurell}}\ and\ \bibinfo {author} {\bibfnamefont {S.}~\bibnamefont
  {Okamoto}},\ }\href {https://doi.org/10.1038/s41535-019-0203-y} {\bibfield
  {journal} {\bibinfo  {journal} {npj Quantum Materials}\ }\textbf {\bibinfo
  {volume} {5}},\ \bibinfo {pages} {1} (\bibinfo {year} {2020})}\BibitemShut
  {NoStop}%
\bibitem [{\citenamefont {Cao}\ \emph {et~al.}(2016)\citenamefont {Cao},
  \citenamefont {Banerjee}, \citenamefont {Yan}, \citenamefont {Bridges},
  \citenamefont {Lumsden}, \citenamefont {Mandrus}, \citenamefont {Tennant},
  \citenamefont {Chakoumakos},\ and\ \citenamefont {Nagler}}]{CaoPRB2016}%
  \BibitemOpen
  \bibfield  {author} {\bibinfo {author} {\bibfnamefont {H.~B.}\ \bibnamefont
  {Cao}}, \bibinfo {author} {\bibfnamefont {A.}~\bibnamefont {Banerjee}},
  \bibinfo {author} {\bibfnamefont {J.-Q.}\ \bibnamefont {Yan}}, \bibinfo
  {author} {\bibfnamefont {C.~A.}\ \bibnamefont {Bridges}}, \bibinfo {author}
  {\bibfnamefont {M.~D.}\ \bibnamefont {Lumsden}}, \bibinfo {author}
  {\bibfnamefont {D.~G.}\ \bibnamefont {Mandrus}}, \bibinfo {author}
  {\bibfnamefont {D.~A.}\ \bibnamefont {Tennant}}, \bibinfo {author}
  {\bibfnamefont {B.~C.}\ \bibnamefont {Chakoumakos}},\ and\ \bibinfo {author}
  {\bibfnamefont {S.~E.}\ \bibnamefont {Nagler}},\ }\href
  {https://doi.org/10.1103/PhysRevB.93.134423} {\bibfield  {journal} {\bibinfo
  {journal} {Phys. Rev. B}\ }\textbf {\bibinfo {volume} {93}},\ \bibinfo
  {pages} {134423} (\bibinfo {year} {2016})}\BibitemShut {NoStop}%
\bibitem [{\citenamefont {Chen}\ \emph {et~al.}(2021)\citenamefont {Chen},
  \citenamefont {Li}, \citenamefont {Hu}, \citenamefont {Hu}, \citenamefont
  {Yue}, \citenamefont {Sutarto}, \citenamefont {He}, \citenamefont {Iida},
  \citenamefont {Kamazawa}, \citenamefont {Yu}, \citenamefont {Lin},\ and\
  \citenamefont {Li}}]{ChenPRB2021}%
  \BibitemOpen
  \bibfield  {author} {\bibinfo {author} {\bibfnamefont {W.}~\bibnamefont
  {Chen}}, \bibinfo {author} {\bibfnamefont {X.}~\bibnamefont {Li}}, \bibinfo
  {author} {\bibfnamefont {Z.}~\bibnamefont {Hu}}, \bibinfo {author}
  {\bibfnamefont {Z.}~\bibnamefont {Hu}}, \bibinfo {author} {\bibfnamefont
  {L.}~\bibnamefont {Yue}}, \bibinfo {author} {\bibfnamefont {R.}~\bibnamefont
  {Sutarto}}, \bibinfo {author} {\bibfnamefont {F.}~\bibnamefont {He}},
  \bibinfo {author} {\bibfnamefont {K.}~\bibnamefont {Iida}}, \bibinfo {author}
  {\bibfnamefont {K.}~\bibnamefont {Kamazawa}}, \bibinfo {author}
  {\bibfnamefont {W.}~\bibnamefont {Yu}}, \bibinfo {author} {\bibfnamefont
  {X.}~\bibnamefont {Lin}},\ and\ \bibinfo {author} {\bibfnamefont
  {Y.}~\bibnamefont {Li}},\ }\href
  {https://doi.org/10.1103/PhysRevB.103.L180404} {\bibfield  {journal}
  {\bibinfo  {journal} {Phys. Rev. B}\ }\textbf {\bibinfo {volume} {103}},\
  \bibinfo {pages} {L180404} (\bibinfo {year} {2021})}\BibitemShut {NoStop}%
\bibitem [{\citenamefont {Lee}\ \emph {et~al.}(2021)\citenamefont {Lee},
  \citenamefont {Lee}, \citenamefont {Choi}, \citenamefont {Jang},
  \citenamefont {Kalaivanan}, \citenamefont {Sankar},\ and\ \citenamefont
  {Choi}}]{LeePRB2021}%
  \BibitemOpen
  \bibfield  {author} {\bibinfo {author} {\bibfnamefont {C.~H.}\ \bibnamefont
  {Lee}}, \bibinfo {author} {\bibfnamefont {S.}~\bibnamefont {Lee}}, \bibinfo
  {author} {\bibfnamefont {Y.~S.}\ \bibnamefont {Choi}}, \bibinfo {author}
  {\bibfnamefont {Z.~H.}\ \bibnamefont {Jang}}, \bibinfo {author}
  {\bibfnamefont {R.}~\bibnamefont {Kalaivanan}}, \bibinfo {author}
  {\bibfnamefont {R.}~\bibnamefont {Sankar}},\ and\ \bibinfo {author}
  {\bibfnamefont {K.-Y.}\ \bibnamefont {Choi}},\ }\href
  {https://doi.org/10.1103/PhysRevB.103.214447} {\bibfield  {journal} {\bibinfo
   {journal} {Phys. Rev. B}\ }\textbf {\bibinfo {volume} {103}},\ \bibinfo
  {pages} {214447} (\bibinfo {year} {2021})}\BibitemShut {NoStop}%
\bibitem [{\citenamefont {Songvilay}\ \emph {et~al.}(2020)\citenamefont
  {Songvilay}, \citenamefont {Robert}, \citenamefont {Petit}, \citenamefont
  {Rodriguez-Rivera}, \citenamefont {Ratcliff}, \citenamefont {Damay},
  \citenamefont {Bal\'edent}, \citenamefont {Jim\'enez-Ruiz}, \citenamefont
  {Lejay}, \citenamefont {Pachoud}, \citenamefont {Hadj-Azzem}, \citenamefont
  {Simonet},\ and\ \citenamefont {Stock}}]{SongvilayPRB2020}%
  \BibitemOpen
  \bibfield  {author} {\bibinfo {author} {\bibfnamefont {M.}~\bibnamefont
  {Songvilay}}, \bibinfo {author} {\bibfnamefont {J.}~\bibnamefont {Robert}},
  \bibinfo {author} {\bibfnamefont {S.}~\bibnamefont {Petit}}, \bibinfo
  {author} {\bibfnamefont {J.~A.}\ \bibnamefont {Rodriguez-Rivera}}, \bibinfo
  {author} {\bibfnamefont {W.~D.}\ \bibnamefont {Ratcliff}}, \bibinfo {author}
  {\bibfnamefont {F.}~\bibnamefont {Damay}}, \bibinfo {author} {\bibfnamefont
  {V.}~\bibnamefont {Bal\'edent}}, \bibinfo {author} {\bibfnamefont
  {M.}~\bibnamefont {Jim\'enez-Ruiz}}, \bibinfo {author} {\bibfnamefont
  {P.}~\bibnamefont {Lejay}}, \bibinfo {author} {\bibfnamefont
  {E.}~\bibnamefont {Pachoud}}, \bibinfo {author} {\bibfnamefont
  {A.}~\bibnamefont {Hadj-Azzem}}, \bibinfo {author} {\bibfnamefont
  {V.}~\bibnamefont {Simonet}},\ and\ \bibinfo {author} {\bibfnamefont
  {C.}~\bibnamefont {Stock}},\ }\href
  {https://doi.org/10.1103/PhysRevB.102.224429} {\bibfield  {journal} {\bibinfo
   {journal} {Phys. Rev. B}\ }\textbf {\bibinfo {volume} {102}},\ \bibinfo
  {pages} {224429} (\bibinfo {year} {2020})}\BibitemShut {NoStop}%
\bibitem [{\citenamefont {Lin}\ \emph {et~al.}(2021)\citenamefont {Lin},
  \citenamefont {Jeong}, \citenamefont {Kim}, \citenamefont {Wang},
  \citenamefont {Huang}, \citenamefont {Masuda}, \citenamefont {Asai},
  \citenamefont {Itoh}, \citenamefont {G{\"u}nther}, \citenamefont {Russina},
  \citenamefont {Lu}, \citenamefont {Sheng}, \citenamefont {Wang},
  \citenamefont {Wang}, \citenamefont {Wang}, \citenamefont {Ren},
  \citenamefont {Xi}, \citenamefont {Tong}, \citenamefont {Ling}, \citenamefont
  {Liu}, \citenamefont {Wu}, \citenamefont {Mei}, \citenamefont {Qu},
  \citenamefont {Zhou}, \citenamefont {Park},\ and\ \citenamefont
  {Ma}}]{LinNC2021}%
  \BibitemOpen
  \bibfield  {author} {\bibinfo {author} {\bibfnamefont {G.}~\bibnamefont
  {Lin}}, \bibinfo {author} {\bibfnamefont {J.}~\bibnamefont {Jeong}}, \bibinfo
  {author} {\bibfnamefont {C.}~\bibnamefont {Kim}}, \bibinfo {author}
  {\bibfnamefont {Y.}~\bibnamefont {Wang}}, \bibinfo {author} {\bibfnamefont
  {Q.}~\bibnamefont {Huang}}, \bibinfo {author} {\bibfnamefont
  {T.}~\bibnamefont {Masuda}}, \bibinfo {author} {\bibfnamefont
  {S.}~\bibnamefont {Asai}}, \bibinfo {author} {\bibfnamefont {S.}~\bibnamefont
  {Itoh}}, \bibinfo {author} {\bibfnamefont {G.}~\bibnamefont {G{\"u}nther}},
  \bibinfo {author} {\bibfnamefont {M.}~\bibnamefont {Russina}}, \bibinfo
  {author} {\bibfnamefont {Z.}~\bibnamefont {Lu}}, \bibinfo {author}
  {\bibfnamefont {J.}~\bibnamefont {Sheng}}, \bibinfo {author} {\bibfnamefont
  {L.}~\bibnamefont {Wang}}, \bibinfo {author} {\bibfnamefont {J.}~\bibnamefont
  {Wang}}, \bibinfo {author} {\bibfnamefont {G.}~\bibnamefont {Wang}}, \bibinfo
  {author} {\bibfnamefont {Q.}~\bibnamefont {Ren}}, \bibinfo {author}
  {\bibfnamefont {C.}~\bibnamefont {Xi}}, \bibinfo {author} {\bibfnamefont
  {W.}~\bibnamefont {Tong}}, \bibinfo {author} {\bibfnamefont {L.}~\bibnamefont
  {Ling}}, \bibinfo {author} {\bibfnamefont {Z.}~\bibnamefont {Liu}}, \bibinfo
  {author} {\bibfnamefont {L.}~\bibnamefont {Wu}}, \bibinfo {author}
  {\bibfnamefont {J.}~\bibnamefont {Mei}}, \bibinfo {author} {\bibfnamefont
  {Z.}~\bibnamefont {Qu}}, \bibinfo {author} {\bibfnamefont {H.}~\bibnamefont
  {Zhou}}, \bibinfo {author} {\bibfnamefont {J.-G.}\ \bibnamefont {Park}},\
  and\ \bibinfo {author} {\bibfnamefont {J.}~\bibnamefont {Ma}},\ }\href
  {https://doi.org/10.1038/s41467-021-25567-7} {\bibfield  {journal} {\bibinfo
  {journal} {Nat. Commun.}\ }\textbf {\bibinfo {volume} {12}},\ \bibinfo
  {pages} {5559} (\bibinfo {year} {2021})}\BibitemShut {NoStop}%
\bibitem [{\citenamefont {Kim}\ \emph {et~al.}(2021{\natexlab{b}})\citenamefont
  {Kim}, \citenamefont {Jeong}, \citenamefont {Lin}, \citenamefont {Park},
  \citenamefont {Masuda}, \citenamefont {Asai}, \citenamefont {Itoh},
  \citenamefont {Kim}, \citenamefont {Zhou}, \citenamefont {Ma},\ and\
  \citenamefont {Park}}]{KimJPCM2021}%
  \BibitemOpen
  \bibfield  {author} {\bibinfo {author} {\bibfnamefont {C.}~\bibnamefont
  {Kim}}, \bibinfo {author} {\bibfnamefont {J.}~\bibnamefont {Jeong}}, \bibinfo
  {author} {\bibfnamefont {G.}~\bibnamefont {Lin}}, \bibinfo {author}
  {\bibfnamefont {P.}~\bibnamefont {Park}}, \bibinfo {author} {\bibfnamefont
  {T.}~\bibnamefont {Masuda}}, \bibinfo {author} {\bibfnamefont
  {S.}~\bibnamefont {Asai}}, \bibinfo {author} {\bibfnamefont {S.}~\bibnamefont
  {Itoh}}, \bibinfo {author} {\bibfnamefont {H.-S.}\ \bibnamefont {Kim}},
  \bibinfo {author} {\bibfnamefont {H.}~\bibnamefont {Zhou}}, \bibinfo {author}
  {\bibfnamefont {J.}~\bibnamefont {Ma}},\ and\ \bibinfo {author}
  {\bibfnamefont {J.-G.}\ \bibnamefont {Park}},\ }\href
  {https://doi.org/10.1088/1361-648x/ac2644} {\bibfield  {journal} {\bibinfo
  {journal} {Journal of Physics: Condensed Matter}\ }\textbf {\bibinfo {volume}
  {34}},\ \bibinfo {pages} {045802} (\bibinfo {year}
  {2021}{\natexlab{b}})}\BibitemShut {NoStop}%
\bibitem [{\citenamefont {Samarakoon}\ \emph {et~al.}(2021)\citenamefont
  {Samarakoon}, \citenamefont {Chen}, \citenamefont {Zhou},\ and\ \citenamefont
  {Garlea}}]{SamarakoonPRB2021}%
  \BibitemOpen
  \bibfield  {author} {\bibinfo {author} {\bibfnamefont {A.~M.}\ \bibnamefont
  {Samarakoon}}, \bibinfo {author} {\bibfnamefont {Q.}~\bibnamefont {Chen}},
  \bibinfo {author} {\bibfnamefont {H.}~\bibnamefont {Zhou}},\ and\ \bibinfo
  {author} {\bibfnamefont {V.~O.}\ \bibnamefont {Garlea}},\ }\href
  {https://doi.org/10.1103/PhysRevB.104.184415} {\bibfield  {journal} {\bibinfo
   {journal} {Phys. Rev. B}\ }\textbf {\bibinfo {volume} {104}},\ \bibinfo
  {pages} {184415} (\bibinfo {year} {2021})}\BibitemShut {NoStop}%
\bibitem [{\citenamefont {Sanders}\ \emph {et~al.}(2022)\citenamefont
  {Sanders}, \citenamefont {Mole}, \citenamefont {Liu}, \citenamefont {Brown},
  \citenamefont {Yu}, \citenamefont {Ling},\ and\ \citenamefont
  {Rachel}}]{SandersArxiv2021}%
  \BibitemOpen
  \bibfield  {author} {\bibinfo {author} {\bibfnamefont {A.~L.}\ \bibnamefont
  {Sanders}}, \bibinfo {author} {\bibfnamefont {R.~A.}\ \bibnamefont {Mole}},
  \bibinfo {author} {\bibfnamefont {J.}~\bibnamefont {Liu}}, \bibinfo {author}
  {\bibfnamefont {A.~J.}\ \bibnamefont {Brown}}, \bibinfo {author}
  {\bibfnamefont {D.}~\bibnamefont {Yu}}, \bibinfo {author} {\bibfnamefont
  {C.~D.}\ \bibnamefont {Ling}},\ and\ \bibinfo {author} {\bibfnamefont
  {S.}~\bibnamefont {Rachel}},\ }\href
  {https://doi.org/10.1103/PhysRevB.106.014413} {\bibfield  {journal} {\bibinfo
   {journal} {Phys. Rev. B}\ }\textbf {\bibinfo {volume} {106}},\ \bibinfo
  {pages} {014413} (\bibinfo {year} {2022})}\BibitemShut {NoStop}%
\bibitem [{\citenamefont {Yao}\ \emph {et~al.}(2022)\citenamefont {Yao},
  \citenamefont {Iida}, \citenamefont {Kamazawa},\ and\ \citenamefont
  {Li}}]{YaoArxiv2022}%
  \BibitemOpen
  \bibfield  {author} {\bibinfo {author} {\bibfnamefont {W.}~\bibnamefont
  {Yao}}, \bibinfo {author} {\bibfnamefont {K.}~\bibnamefont {Iida}}, \bibinfo
  {author} {\bibfnamefont {K.}~\bibnamefont {Kamazawa}},\ and\ \bibinfo
  {author} {\bibfnamefont {Y.}~\bibnamefont {Li}},\ }\href
  {https://doi.org/10.1103/PhysRevLett.129.147202} {\bibfield  {journal}
  {\bibinfo  {journal} {Phys. Rev. Lett.}\ }\textbf {\bibinfo {volume} {129}},\
  \bibinfo {pages} {147202} (\bibinfo {year} {2022})}\BibitemShut {NoStop}%
\bibitem [{\citenamefont {Lampen-Kelley}\ \emph {et~al.}(2018)\citenamefont
  {Lampen-Kelley}, \citenamefont {Rachel}, \citenamefont {Reuther},
  \citenamefont {Yan}, \citenamefont {Banerjee}, \citenamefont {Bridges},
  \citenamefont {Cao}, \citenamefont {Nagler},\ and\ \citenamefont
  {Mandrus}}]{LampenPRB2018}%
  \BibitemOpen
  \bibfield  {author} {\bibinfo {author} {\bibfnamefont {P.}~\bibnamefont
  {Lampen-Kelley}}, \bibinfo {author} {\bibfnamefont {S.}~\bibnamefont
  {Rachel}}, \bibinfo {author} {\bibfnamefont {J.}~\bibnamefont {Reuther}},
  \bibinfo {author} {\bibfnamefont {J.-Q.}\ \bibnamefont {Yan}}, \bibinfo
  {author} {\bibfnamefont {A.}~\bibnamefont {Banerjee}}, \bibinfo {author}
  {\bibfnamefont {C.~A.}\ \bibnamefont {Bridges}}, \bibinfo {author}
  {\bibfnamefont {H.~B.}\ \bibnamefont {Cao}}, \bibinfo {author} {\bibfnamefont
  {S.~E.}\ \bibnamefont {Nagler}},\ and\ \bibinfo {author} {\bibfnamefont
  {D.}~\bibnamefont {Mandrus}},\ }\href
  {https://doi.org/10.1103/PhysRevB.98.100403} {\bibfield  {journal} {\bibinfo
  {journal} {Phys. Rev. B}\ }\textbf {\bibinfo {volume} {98}},\ \bibinfo
  {pages} {100403} (\bibinfo {year} {2018})}\BibitemShut {NoStop}%
\bibitem [{\citenamefont {Balz}\ \emph {et~al.}(2021)\citenamefont {Balz},
  \citenamefont {Janssen}, \citenamefont {Lampen-Kelley}, \citenamefont
  {Banerjee}, \citenamefont {Liu}, \citenamefont {Yan}, \citenamefont
  {Mandrus}, \citenamefont {Vojta},\ and\ \citenamefont
  {Nagler}}]{BalzPRB2021}%
  \BibitemOpen
  \bibfield  {author} {\bibinfo {author} {\bibfnamefont {C.}~\bibnamefont
  {Balz}}, \bibinfo {author} {\bibfnamefont {L.}~\bibnamefont {Janssen}},
  \bibinfo {author} {\bibfnamefont {P.}~\bibnamefont {Lampen-Kelley}}, \bibinfo
  {author} {\bibfnamefont {A.}~\bibnamefont {Banerjee}}, \bibinfo {author}
  {\bibfnamefont {Y.~H.}\ \bibnamefont {Liu}}, \bibinfo {author} {\bibfnamefont
  {J.-Q.}\ \bibnamefont {Yan}}, \bibinfo {author} {\bibfnamefont {D.~G.}\
  \bibnamefont {Mandrus}}, \bibinfo {author} {\bibfnamefont {M.}~\bibnamefont
  {Vojta}},\ and\ \bibinfo {author} {\bibfnamefont {S.~E.}\ \bibnamefont
  {Nagler}},\ }\href {https://doi.org/10.1103/PhysRevB.103.174417} {\bibfield
  {journal} {\bibinfo  {journal} {Phys. Rev. B}\ }\textbf {\bibinfo {volume}
  {103}},\ \bibinfo {pages} {174417} (\bibinfo {year} {2021})}\BibitemShut
  {NoStop}%
\bibitem [{\citenamefont {Kocsis}\ \emph {et~al.}(2022)\citenamefont {Kocsis},
  \citenamefont {Kaib}, \citenamefont {Riedl}, \citenamefont {Gass},
  \citenamefont {Lampen-Kelley}, \citenamefont {Mandrus}, \citenamefont
  {Nagler}, \citenamefont {P\'erez}, \citenamefont {Nielsch}, \citenamefont
  {B\"uchner}, \citenamefont {Wolter},\ and\ \citenamefont
  {Valent\'{\i}}}]{KocsisArxiv2022}%
  \BibitemOpen
  \bibfield  {author} {\bibinfo {author} {\bibfnamefont {V.}~\bibnamefont
  {Kocsis}}, \bibinfo {author} {\bibfnamefont {D.~A.~S.}\ \bibnamefont {Kaib}},
  \bibinfo {author} {\bibfnamefont {K.}~\bibnamefont {Riedl}}, \bibinfo
  {author} {\bibfnamefont {S.}~\bibnamefont {Gass}}, \bibinfo {author}
  {\bibfnamefont {P.}~\bibnamefont {Lampen-Kelley}}, \bibinfo {author}
  {\bibfnamefont {D.~G.}\ \bibnamefont {Mandrus}}, \bibinfo {author}
  {\bibfnamefont {S.~E.}\ \bibnamefont {Nagler}}, \bibinfo {author}
  {\bibfnamefont {N.}~\bibnamefont {P\'erez}}, \bibinfo {author} {\bibfnamefont
  {K.}~\bibnamefont {Nielsch}}, \bibinfo {author} {\bibfnamefont
  {B.}~\bibnamefont {B\"uchner}}, \bibinfo {author} {\bibfnamefont {A.~U.~B.}\
  \bibnamefont {Wolter}},\ and\ \bibinfo {author} {\bibfnamefont
  {R.}~\bibnamefont {Valent\'{\i}}},\ }\href
  {https://doi.org/10.1103/PhysRevB.105.094410} {\bibfield  {journal} {\bibinfo
   {journal} {Phys. Rev. B}\ }\textbf {\bibinfo {volume} {105}},\ \bibinfo
  {pages} {094410} (\bibinfo {year} {2022})}\BibitemShut {NoStop}%
\bibitem [{SM()}]{SM}%
  \BibitemOpen
  \bibinfo {note} {See Supplemental Material for additional methods, data, and
  analyses, which include additional Refs.
  \cite{WuJOI2016,BarryEPJ2015,KajimotoJPSJ2011,InamuraJPSJ2013,EwingsNI2016}.}\BibitemShut
  {Stop}%
\bibitem [{\citenamefont {Sears}\ \emph {et~al.}(2017)\citenamefont {Sears},
  \citenamefont {Zhao}, \citenamefont {Xu}, \citenamefont {Lynn},\ and\
  \citenamefont {Kim}}]{SearsPRB2017}%
  \BibitemOpen
  \bibfield  {author} {\bibinfo {author} {\bibfnamefont {J.~A.}\ \bibnamefont
  {Sears}}, \bibinfo {author} {\bibfnamefont {Y.}~\bibnamefont {Zhao}},
  \bibinfo {author} {\bibfnamefont {Z.}~\bibnamefont {Xu}}, \bibinfo {author}
  {\bibfnamefont {J.~W.}\ \bibnamefont {Lynn}},\ and\ \bibinfo {author}
  {\bibfnamefont {Y.-J.}\ \bibnamefont {Kim}},\ }\href
  {https://doi.org/10.1103/PhysRevB.95.180411} {\bibfield  {journal} {\bibinfo
  {journal} {Phys. Rev. B}\ }\textbf {\bibinfo {volume} {95}},\ \bibinfo
  {pages} {180411} (\bibinfo {year} {2017})}\BibitemShut {NoStop}%
\bibitem [{\citenamefont {Hardy}\ \emph {et~al.}(2004)\citenamefont {Hardy},
  \citenamefont {Lees}, \citenamefont {Petrenko}, \citenamefont {Paul},
  \citenamefont {Flahaut}, \citenamefont {H\'ebert},\ and\ \citenamefont
  {Maignan}}]{HardyPRB2004}%
  \BibitemOpen
  \bibfield  {author} {\bibinfo {author} {\bibfnamefont {V.}~\bibnamefont
  {Hardy}}, \bibinfo {author} {\bibfnamefont {M.~R.}\ \bibnamefont {Lees}},
  \bibinfo {author} {\bibfnamefont {O.~A.}\ \bibnamefont {Petrenko}}, \bibinfo
  {author} {\bibfnamefont {D.~M.}\ \bibnamefont {Paul}}, \bibinfo {author}
  {\bibfnamefont {D.}~\bibnamefont {Flahaut}}, \bibinfo {author} {\bibfnamefont
  {S.}~\bibnamefont {H\'ebert}},\ and\ \bibinfo {author} {\bibfnamefont
  {A.}~\bibnamefont {Maignan}},\ }\href
  {https://doi.org/10.1103/PhysRevB.70.064424} {\bibfield  {journal} {\bibinfo
  {journal} {Phys. Rev. B}\ }\textbf {\bibinfo {volume} {70}},\ \bibinfo
  {pages} {064424} (\bibinfo {year} {2004})}\BibitemShut {NoStop}%
\bibitem [{\citenamefont {Ueda}\ \emph {et~al.}(2005)\citenamefont {Ueda},
  \citenamefont {Katori}, \citenamefont {Mitamura}, \citenamefont {Goto},\ and\
  \citenamefont {Takagi}}]{UedaPRL2005}%
  \BibitemOpen
  \bibfield  {author} {\bibinfo {author} {\bibfnamefont {H.}~\bibnamefont
  {Ueda}}, \bibinfo {author} {\bibfnamefont {H.~A.}\ \bibnamefont {Katori}},
  \bibinfo {author} {\bibfnamefont {H.}~\bibnamefont {Mitamura}}, \bibinfo
  {author} {\bibfnamefont {T.}~\bibnamefont {Goto}},\ and\ \bibinfo {author}
  {\bibfnamefont {H.}~\bibnamefont {Takagi}},\ }\href
  {https://doi.org/10.1103/PhysRevLett.94.047202} {\bibfield  {journal}
  {\bibinfo  {journal} {Phys. Rev. Lett.}\ }\textbf {\bibinfo {volume} {94}},\
  \bibinfo {pages} {047202} (\bibinfo {year} {2005})}\BibitemShut {NoStop}%
\bibitem [{\citenamefont {Cao}\ \emph {et~al.}(2007)\citenamefont {Cao},
  \citenamefont {Durairaj}, \citenamefont {Chikara}, \citenamefont {Parkin},\
  and\ \citenamefont {Schlottmann}}]{CaoPRB2007}%
  \BibitemOpen
  \bibfield  {author} {\bibinfo {author} {\bibfnamefont {G.}~\bibnamefont
  {Cao}}, \bibinfo {author} {\bibfnamefont {V.}~\bibnamefont {Durairaj}},
  \bibinfo {author} {\bibfnamefont {S.}~\bibnamefont {Chikara}}, \bibinfo
  {author} {\bibfnamefont {S.}~\bibnamefont {Parkin}},\ and\ \bibinfo {author}
  {\bibfnamefont {P.}~\bibnamefont {Schlottmann}},\ }\href
  {https://doi.org/10.1103/PhysRevB.75.134402} {\bibfield  {journal} {\bibinfo
  {journal} {Phys. Rev. B}\ }\textbf {\bibinfo {volume} {75}},\ \bibinfo
  {pages} {134402} (\bibinfo {year} {2007})}\BibitemShut {NoStop}%
\bibitem [{\citenamefont {Jo}\ \emph {et~al.}(2009)\citenamefont {Jo},
  \citenamefont {Lee}, \citenamefont {Choi}, \citenamefont {Yi}, \citenamefont
  {Ratcliff}, \citenamefont {Choi}, \citenamefont {Kiryukhin}, \citenamefont
  {Cheong},\ and\ \citenamefont {Balicas}}]{JoPRB2009}%
  \BibitemOpen
  \bibfield  {author} {\bibinfo {author} {\bibfnamefont {Y.~J.}\ \bibnamefont
  {Jo}}, \bibinfo {author} {\bibfnamefont {S.}~\bibnamefont {Lee}}, \bibinfo
  {author} {\bibfnamefont {E.~S.}\ \bibnamefont {Choi}}, \bibinfo {author}
  {\bibfnamefont {H.~T.}\ \bibnamefont {Yi}}, \bibinfo {author} {\bibfnamefont
  {W.}~\bibnamefont {Ratcliff}}, \bibinfo {author} {\bibfnamefont {Y.~J.}\
  \bibnamefont {Choi}}, \bibinfo {author} {\bibfnamefont {V.}~\bibnamefont
  {Kiryukhin}}, \bibinfo {author} {\bibfnamefont {S.~W.}\ \bibnamefont
  {Cheong}},\ and\ \bibinfo {author} {\bibfnamefont {L.}~\bibnamefont
  {Balicas}},\ }\href {https://doi.org/10.1103/PhysRevB.79.012407} {\bibfield
  {journal} {\bibinfo  {journal} {Phys. Rev. B}\ }\textbf {\bibinfo {volume}
  {79}},\ \bibinfo {pages} {012407} (\bibinfo {year} {2009})}\BibitemShut
  {NoStop}%
\bibitem [{\citenamefont {Ross}\ \emph {et~al.}(2009)\citenamefont {Ross},
  \citenamefont {Ruff}, \citenamefont {Adams}, \citenamefont {Gardner},
  \citenamefont {Dabkowska}, \citenamefont {Qiu}, \citenamefont {Copley},\ and\
  \citenamefont {Gaulin}}]{RossPRL2009}%
  \BibitemOpen
  \bibfield  {author} {\bibinfo {author} {\bibfnamefont {K.~A.}\ \bibnamefont
  {Ross}}, \bibinfo {author} {\bibfnamefont {J.~P.~C.}\ \bibnamefont {Ruff}},
  \bibinfo {author} {\bibfnamefont {C.~P.}\ \bibnamefont {Adams}}, \bibinfo
  {author} {\bibfnamefont {J.~S.}\ \bibnamefont {Gardner}}, \bibinfo {author}
  {\bibfnamefont {H.~A.}\ \bibnamefont {Dabkowska}}, \bibinfo {author}
  {\bibfnamefont {Y.}~\bibnamefont {Qiu}}, \bibinfo {author} {\bibfnamefont
  {J.~R.~D.}\ \bibnamefont {Copley}},\ and\ \bibinfo {author} {\bibfnamefont
  {B.~D.}\ \bibnamefont {Gaulin}},\ }\href
  {https://doi.org/10.1103/PhysRevLett.103.227202} {\bibfield  {journal}
  {\bibinfo  {journal} {Phys. Rev. Lett.}\ }\textbf {\bibinfo {volume} {103}},\
  \bibinfo {pages} {227202} (\bibinfo {year} {2009})}\BibitemShut {NoStop}%
\bibitem [{\citenamefont {Thompson}\ \emph {et~al.}(2011)\citenamefont
  {Thompson}, \citenamefont {McClarty}, \citenamefont {R\o{}nnow},
  \citenamefont {Regnault}, \citenamefont {Sorge},\ and\ \citenamefont
  {Gingras}}]{ThompsonPRL2011}%
  \BibitemOpen
  \bibfield  {author} {\bibinfo {author} {\bibfnamefont {J.~D.}\ \bibnamefont
  {Thompson}}, \bibinfo {author} {\bibfnamefont {P.~A.}\ \bibnamefont
  {McClarty}}, \bibinfo {author} {\bibfnamefont {H.~M.}\ \bibnamefont
  {R\o{}nnow}}, \bibinfo {author} {\bibfnamefont {L.~P.}\ \bibnamefont
  {Regnault}}, \bibinfo {author} {\bibfnamefont {A.}~\bibnamefont {Sorge}},\
  and\ \bibinfo {author} {\bibfnamefont {M.~J.~P.}\ \bibnamefont {Gingras}},\
  }\href {https://doi.org/10.1103/PhysRevLett.106.187202} {\bibfield  {journal}
  {\bibinfo  {journal} {Phys. Rev. Lett.}\ }\textbf {\bibinfo {volume} {106}},\
  \bibinfo {pages} {187202} (\bibinfo {year} {2011})}\BibitemShut {NoStop}%
\bibitem [{\citenamefont {Scheie}\ \emph {et~al.}(2020)\citenamefont {Scheie},
  \citenamefont {Kindervater}, \citenamefont {Zhang}, \citenamefont
  {Changlani}, \citenamefont {Sala}, \citenamefont {Ehlers}, \citenamefont
  {Heinemann}, \citenamefont {Tucker}, \citenamefont {Koohpayeh},\ and\
  \citenamefont {Broholm}}]{ScheiePNAS2020}%
  \BibitemOpen
  \bibfield  {author} {\bibinfo {author} {\bibfnamefont {A.}~\bibnamefont
  {Scheie}}, \bibinfo {author} {\bibfnamefont {J.}~\bibnamefont {Kindervater}},
  \bibinfo {author} {\bibfnamefont {S.}~\bibnamefont {Zhang}}, \bibinfo
  {author} {\bibfnamefont {H.~J.}\ \bibnamefont {Changlani}}, \bibinfo {author}
  {\bibfnamefont {G.}~\bibnamefont {Sala}}, \bibinfo {author} {\bibfnamefont
  {G.}~\bibnamefont {Ehlers}}, \bibinfo {author} {\bibfnamefont
  {A.}~\bibnamefont {Heinemann}}, \bibinfo {author} {\bibfnamefont {G.~S.}\
  \bibnamefont {Tucker}}, \bibinfo {author} {\bibfnamefont {S.~M.}\
  \bibnamefont {Koohpayeh}},\ and\ \bibinfo {author} {\bibfnamefont
  {C.}~\bibnamefont {Broholm}},\ }\href
  {https://doi.org/10.1073/pnas.2008791117} {\bibfield  {journal} {\bibinfo
  {journal} {Proceedings of the National Academy of Sciences}\ }\textbf
  {\bibinfo {volume} {117}},\ \bibinfo {pages} {27245} (\bibinfo {year}
  {2020})}\BibitemShut {NoStop}%
\bibitem [{\citenamefont {Scheie}\ \emph {et~al.}(2022)\citenamefont {Scheie},
  \citenamefont {Benton}, \citenamefont {Taillefumier}, \citenamefont
  {Jaubert}, \citenamefont {Sala}, \citenamefont {Jalarvo}, \citenamefont
  {Koohpayeh},\ and\ \citenamefont {Shannon}}]{ScheieArxiv2022}%
  \BibitemOpen
  \bibfield  {author} {\bibinfo {author} {\bibfnamefont {A.}~\bibnamefont
  {Scheie}}, \bibinfo {author} {\bibfnamefont {O.}~\bibnamefont {Benton}},
  \bibinfo {author} {\bibfnamefont {M.}~\bibnamefont {Taillefumier}}, \bibinfo
  {author} {\bibfnamefont {L.~D.}\ \bibnamefont {Jaubert}}, \bibinfo {author}
  {\bibfnamefont {G.}~\bibnamefont {Sala}}, \bibinfo {author} {\bibfnamefont
  {N.}~\bibnamefont {Jalarvo}}, \bibinfo {author} {\bibfnamefont {S.~M.}\
  \bibnamefont {Koohpayeh}},\ and\ \bibinfo {author} {\bibfnamefont
  {N.}~\bibnamefont {Shannon}},\ }\bibfield  {journal} {\bibinfo  {journal}
  {arXiv preprint arXiv:2202.11085}\ }\href
  {https://doi.org/10.48550/arXiv.2202.11085} {10.48550/arXiv.2202.11085}
  (\bibinfo {year} {2022})\BibitemShut {NoStop}%
\bibitem [{\citenamefont {Zheng}\ \emph {et~al.}(2017)\citenamefont {Zheng},
  \citenamefont {Ran}, \citenamefont {Li}, \citenamefont {Wang}, \citenamefont
  {Wang}, \citenamefont {Liu}, \citenamefont {Liu}, \citenamefont {Normand},
  \citenamefont {Wen},\ and\ \citenamefont {Yu}}]{ZhengPRL2017}%
  \BibitemOpen
  \bibfield  {author} {\bibinfo {author} {\bibfnamefont {J.}~\bibnamefont
  {Zheng}}, \bibinfo {author} {\bibfnamefont {K.}~\bibnamefont {Ran}}, \bibinfo
  {author} {\bibfnamefont {T.}~\bibnamefont {Li}}, \bibinfo {author}
  {\bibfnamefont {J.}~\bibnamefont {Wang}}, \bibinfo {author} {\bibfnamefont
  {P.}~\bibnamefont {Wang}}, \bibinfo {author} {\bibfnamefont {B.}~\bibnamefont
  {Liu}}, \bibinfo {author} {\bibfnamefont {Z.-X.}\ \bibnamefont {Liu}},
  \bibinfo {author} {\bibfnamefont {B.}~\bibnamefont {Normand}}, \bibinfo
  {author} {\bibfnamefont {J.}~\bibnamefont {Wen}},\ and\ \bibinfo {author}
  {\bibfnamefont {W.}~\bibnamefont {Yu}},\ }\href
  {https://doi.org/10.1103/PhysRevLett.119.227208} {\bibfield  {journal}
  {\bibinfo  {journal} {Phys. Rev. Lett.}\ }\textbf {\bibinfo {volume} {119}},\
  \bibinfo {pages} {227208} (\bibinfo {year} {2017})}\BibitemShut {NoStop}%
\bibitem [{\citenamefont {Wolter}\ \emph {et~al.}(2017)\citenamefont {Wolter},
  \citenamefont {Corredor}, \citenamefont {Janssen}, \citenamefont {Nenkov},
  \citenamefont {Sch\"onecker}, \citenamefont {Do}, \citenamefont {Choi},
  \citenamefont {Albrecht}, \citenamefont {Hunger}, \citenamefont {Doert},
  \citenamefont {Vojta},\ and\ \citenamefont {B\"uchner}}]{WolterPRB2017}%
  \BibitemOpen
  \bibfield  {author} {\bibinfo {author} {\bibfnamefont {A.~U.~B.}\
  \bibnamefont {Wolter}}, \bibinfo {author} {\bibfnamefont {L.~T.}\
  \bibnamefont {Corredor}}, \bibinfo {author} {\bibfnamefont {L.}~\bibnamefont
  {Janssen}}, \bibinfo {author} {\bibfnamefont {K.}~\bibnamefont {Nenkov}},
  \bibinfo {author} {\bibfnamefont {S.}~\bibnamefont {Sch\"onecker}}, \bibinfo
  {author} {\bibfnamefont {S.-H.}\ \bibnamefont {Do}}, \bibinfo {author}
  {\bibfnamefont {K.-Y.}\ \bibnamefont {Choi}}, \bibinfo {author}
  {\bibfnamefont {R.}~\bibnamefont {Albrecht}}, \bibinfo {author}
  {\bibfnamefont {J.}~\bibnamefont {Hunger}}, \bibinfo {author} {\bibfnamefont
  {T.}~\bibnamefont {Doert}}, \bibinfo {author} {\bibfnamefont
  {M.}~\bibnamefont {Vojta}},\ and\ \bibinfo {author} {\bibfnamefont
  {B.}~\bibnamefont {B\"uchner}},\ }\href
  {https://doi.org/10.1103/PhysRevB.96.041405} {\bibfield  {journal} {\bibinfo
  {journal} {Phys. Rev. B}\ }\textbf {\bibinfo {volume} {96}},\ \bibinfo
  {pages} {041405} (\bibinfo {year} {2017})}\BibitemShut {NoStop}%
\bibitem [{\citenamefont {Leahy}\ \emph {et~al.}(2017)\citenamefont {Leahy},
  \citenamefont {Pocs}, \citenamefont {Siegfried}, \citenamefont {Graf},
  \citenamefont {Do}, \citenamefont {Choi}, \citenamefont {Normand},\ and\
  \citenamefont {Lee}}]{LeahyPRL2017}%
  \BibitemOpen
  \bibfield  {author} {\bibinfo {author} {\bibfnamefont {I.~A.}\ \bibnamefont
  {Leahy}}, \bibinfo {author} {\bibfnamefont {C.~A.}\ \bibnamefont {Pocs}},
  \bibinfo {author} {\bibfnamefont {P.~E.}\ \bibnamefont {Siegfried}}, \bibinfo
  {author} {\bibfnamefont {D.}~\bibnamefont {Graf}}, \bibinfo {author}
  {\bibfnamefont {S.-H.}\ \bibnamefont {Do}}, \bibinfo {author} {\bibfnamefont
  {K.-Y.}\ \bibnamefont {Choi}}, \bibinfo {author} {\bibfnamefont
  {B.}~\bibnamefont {Normand}},\ and\ \bibinfo {author} {\bibfnamefont
  {M.}~\bibnamefont {Lee}},\ }\href
  {https://doi.org/10.1103/PhysRevLett.118.187203} {\bibfield  {journal}
  {\bibinfo  {journal} {Phys. Rev. Lett.}\ }\textbf {\bibinfo {volume} {118}},\
  \bibinfo {pages} {187203} (\bibinfo {year} {2017})}\BibitemShut {NoStop}%
\bibitem [{\citenamefont {Baek}\ \emph {et~al.}(2017)\citenamefont {Baek},
  \citenamefont {Do}, \citenamefont {Choi}, \citenamefont {Kwon}, \citenamefont
  {Wolter}, \citenamefont {Nishimoto}, \citenamefont {van~den Brink},\ and\
  \citenamefont {B\"uchner}}]{BaekPRL2017}%
  \BibitemOpen
  \bibfield  {author} {\bibinfo {author} {\bibfnamefont {S.-H.}\ \bibnamefont
  {Baek}}, \bibinfo {author} {\bibfnamefont {S.-H.}\ \bibnamefont {Do}},
  \bibinfo {author} {\bibfnamefont {K.-Y.}\ \bibnamefont {Choi}}, \bibinfo
  {author} {\bibfnamefont {Y.~S.}\ \bibnamefont {Kwon}}, \bibinfo {author}
  {\bibfnamefont {A.~U.~B.}\ \bibnamefont {Wolter}}, \bibinfo {author}
  {\bibfnamefont {S.}~\bibnamefont {Nishimoto}}, \bibinfo {author}
  {\bibfnamefont {J.}~\bibnamefont {van~den Brink}},\ and\ \bibinfo {author}
  {\bibfnamefont {B.}~\bibnamefont {B\"uchner}},\ }\href
  {https://doi.org/10.1103/PhysRevLett.119.037201} {\bibfield  {journal}
  {\bibinfo  {journal} {Phys. Rev. Lett.}\ }\textbf {\bibinfo {volume} {119}},\
  \bibinfo {pages} {037201} (\bibinfo {year} {2017})}\BibitemShut {NoStop}%
\bibitem [{\citenamefont {Yu}\ \emph {et~al.}(2018)\citenamefont {Yu},
  \citenamefont {Xu}, \citenamefont {Ran}, \citenamefont {Ni}, \citenamefont
  {Huang}, \citenamefont {Wang}, \citenamefont {Wen},\ and\ \citenamefont
  {Li}}]{YuPRL2018}%
  \BibitemOpen
  \bibfield  {author} {\bibinfo {author} {\bibfnamefont {Y.~J.}\ \bibnamefont
  {Yu}}, \bibinfo {author} {\bibfnamefont {Y.}~\bibnamefont {Xu}}, \bibinfo
  {author} {\bibfnamefont {K.~J.}\ \bibnamefont {Ran}}, \bibinfo {author}
  {\bibfnamefont {J.~M.}\ \bibnamefont {Ni}}, \bibinfo {author} {\bibfnamefont
  {Y.~Y.}\ \bibnamefont {Huang}}, \bibinfo {author} {\bibfnamefont {J.~H.}\
  \bibnamefont {Wang}}, \bibinfo {author} {\bibfnamefont {J.~S.}\ \bibnamefont
  {Wen}},\ and\ \bibinfo {author} {\bibfnamefont {S.~Y.}\ \bibnamefont {Li}},\
  }\href {https://doi.org/10.1103/PhysRevLett.120.067202} {\bibfield  {journal}
  {\bibinfo  {journal} {Phys. Rev. Lett.}\ }\textbf {\bibinfo {volume} {120}},\
  \bibinfo {pages} {067202} (\bibinfo {year} {2018})}\BibitemShut {NoStop}%
\bibitem [{\citenamefont {Wellm}\ \emph {et~al.}(2018)\citenamefont {Wellm},
  \citenamefont {Zeisner}, \citenamefont {Alfonsov}, \citenamefont {Wolter},
  \citenamefont {Roslova}, \citenamefont {Isaeva}, \citenamefont {Doert},
  \citenamefont {Vojta}, \citenamefont {B\"uchner},\ and\ \citenamefont
  {Kataev}}]{WellmPRB2018}%
  \BibitemOpen
  \bibfield  {author} {\bibinfo {author} {\bibfnamefont {C.}~\bibnamefont
  {Wellm}}, \bibinfo {author} {\bibfnamefont {J.}~\bibnamefont {Zeisner}},
  \bibinfo {author} {\bibfnamefont {A.}~\bibnamefont {Alfonsov}}, \bibinfo
  {author} {\bibfnamefont {A.~U.~B.}\ \bibnamefont {Wolter}}, \bibinfo {author}
  {\bibfnamefont {M.}~\bibnamefont {Roslova}}, \bibinfo {author} {\bibfnamefont
  {A.}~\bibnamefont {Isaeva}}, \bibinfo {author} {\bibfnamefont
  {T.}~\bibnamefont {Doert}}, \bibinfo {author} {\bibfnamefont
  {M.}~\bibnamefont {Vojta}}, \bibinfo {author} {\bibfnamefont
  {B.}~\bibnamefont {B\"uchner}},\ and\ \bibinfo {author} {\bibfnamefont
  {V.}~\bibnamefont {Kataev}},\ }\href
  {https://doi.org/10.1103/PhysRevB.98.184408} {\bibfield  {journal} {\bibinfo
  {journal} {Phys. Rev. B}\ }\textbf {\bibinfo {volume} {98}},\ \bibinfo
  {pages} {184408} (\bibinfo {year} {2018})}\BibitemShut {NoStop}%
\bibitem [{\citenamefont {Kasahara}\ \emph {et~al.}(2018)\citenamefont
  {Kasahara}, \citenamefont {Ohnishi}, \citenamefont {Mizukami}, \citenamefont
  {Tanaka}, \citenamefont {Ma}, \citenamefont {Sugii}, \citenamefont {Kurita},
  \citenamefont {Tanaka}, \citenamefont {Nasu}, \citenamefont {Motome},
  \citenamefont {Shibauchi},\ and\ \citenamefont
  {Matsuda}}]{KasaharaNature2018}%
  \BibitemOpen
  \bibfield  {author} {\bibinfo {author} {\bibfnamefont {Y.}~\bibnamefont
  {Kasahara}}, \bibinfo {author} {\bibfnamefont {T.}~\bibnamefont {Ohnishi}},
  \bibinfo {author} {\bibfnamefont {Y.}~\bibnamefont {Mizukami}}, \bibinfo
  {author} {\bibfnamefont {O.}~\bibnamefont {Tanaka}}, \bibinfo {author}
  {\bibfnamefont {S.}~\bibnamefont {Ma}}, \bibinfo {author} {\bibfnamefont
  {K.}~\bibnamefont {Sugii}}, \bibinfo {author} {\bibfnamefont
  {N.}~\bibnamefont {Kurita}}, \bibinfo {author} {\bibfnamefont
  {H.}~\bibnamefont {Tanaka}}, \bibinfo {author} {\bibfnamefont
  {J.}~\bibnamefont {Nasu}}, \bibinfo {author} {\bibfnamefont {Y.}~\bibnamefont
  {Motome}}, \bibinfo {author} {\bibfnamefont {T.}~\bibnamefont {Shibauchi}},\
  and\ \bibinfo {author} {\bibfnamefont {Y.}~\bibnamefont {Matsuda}},\ }\href
  {https://doi.org/10.1038/s41586-018-0274-0} {\bibfield  {journal} {\bibinfo
  {journal} {Nature}\ }\textbf {\bibinfo {volume} {559}},\ \bibinfo {pages}
  {227} (\bibinfo {year} {2018})}\BibitemShut {NoStop}%
\bibitem [{\citenamefont {Sahasrabudhe}\ \emph {et~al.}(2020)\citenamefont
  {Sahasrabudhe}, \citenamefont {Kaib}, \citenamefont {Reschke}, \citenamefont
  {German}, \citenamefont {Koethe}, \citenamefont {Buhot}, \citenamefont
  {Kamenskyi}, \citenamefont {Hickey}, \citenamefont {Becker}, \citenamefont
  {Tsurkan}, \citenamefont {Loidl}, \citenamefont {Do}, \citenamefont {Choi},
  \citenamefont {Gr\"uninger}, \citenamefont {Winter}, \citenamefont {Wang},
  \citenamefont {Valent\'{\i}},\ and\ \citenamefont {van
  Loosdrecht}}]{SahasrabudhePRB2020}%
  \BibitemOpen
  \bibfield  {author} {\bibinfo {author} {\bibfnamefont {A.}~\bibnamefont
  {Sahasrabudhe}}, \bibinfo {author} {\bibfnamefont {D.~A.~S.}\ \bibnamefont
  {Kaib}}, \bibinfo {author} {\bibfnamefont {S.}~\bibnamefont {Reschke}},
  \bibinfo {author} {\bibfnamefont {R.}~\bibnamefont {German}}, \bibinfo
  {author} {\bibfnamefont {T.~C.}\ \bibnamefont {Koethe}}, \bibinfo {author}
  {\bibfnamefont {J.}~\bibnamefont {Buhot}}, \bibinfo {author} {\bibfnamefont
  {D.}~\bibnamefont {Kamenskyi}}, \bibinfo {author} {\bibfnamefont
  {C.}~\bibnamefont {Hickey}}, \bibinfo {author} {\bibfnamefont
  {P.}~\bibnamefont {Becker}}, \bibinfo {author} {\bibfnamefont
  {V.}~\bibnamefont {Tsurkan}}, \bibinfo {author} {\bibfnamefont
  {A.}~\bibnamefont {Loidl}}, \bibinfo {author} {\bibfnamefont {S.~H.}\
  \bibnamefont {Do}}, \bibinfo {author} {\bibfnamefont {K.~Y.}\ \bibnamefont
  {Choi}}, \bibinfo {author} {\bibfnamefont {M.}~\bibnamefont {Gr\"uninger}},
  \bibinfo {author} {\bibfnamefont {S.~M.}\ \bibnamefont {Winter}}, \bibinfo
  {author} {\bibfnamefont {Z.}~\bibnamefont {Wang}}, \bibinfo {author}
  {\bibfnamefont {R.}~\bibnamefont {Valent\'{\i}}},\ and\ \bibinfo {author}
  {\bibfnamefont {P.~H.~M.}\ \bibnamefont {van Loosdrecht}},\ }\href
  {https://doi.org/10.1103/PhysRevB.101.140410} {\bibfield  {journal} {\bibinfo
   {journal} {Phys. Rev. B}\ }\textbf {\bibinfo {volume} {101}},\ \bibinfo
  {pages} {140410} (\bibinfo {year} {2020})}\BibitemShut {NoStop}%
\bibitem [{\citenamefont {Yokoi}\ \emph {et~al.}(2021)\citenamefont {Yokoi},
  \citenamefont {Ma}, \citenamefont {Kasahara}, \citenamefont {Kasahara},
  \citenamefont {Shibauchi}, \citenamefont {Kurita}, \citenamefont {Tanaka},
  \citenamefont {Nasu}, \citenamefont {Motome}, \citenamefont {Hickey},
  \citenamefont {Trebst},\ and\ \citenamefont {Matsuda}}]{YokoiScience2021}%
  \BibitemOpen
  \bibfield  {author} {\bibinfo {author} {\bibfnamefont {T.}~\bibnamefont
  {Yokoi}}, \bibinfo {author} {\bibfnamefont {S.}~\bibnamefont {Ma}}, \bibinfo
  {author} {\bibfnamefont {Y.}~\bibnamefont {Kasahara}}, \bibinfo {author}
  {\bibfnamefont {S.}~\bibnamefont {Kasahara}}, \bibinfo {author}
  {\bibfnamefont {T.}~\bibnamefont {Shibauchi}}, \bibinfo {author}
  {\bibfnamefont {N.}~\bibnamefont {Kurita}}, \bibinfo {author} {\bibfnamefont
  {H.}~\bibnamefont {Tanaka}}, \bibinfo {author} {\bibfnamefont
  {J.}~\bibnamefont {Nasu}}, \bibinfo {author} {\bibfnamefont {Y.}~\bibnamefont
  {Motome}}, \bibinfo {author} {\bibfnamefont {C.}~\bibnamefont {Hickey}},
  \bibinfo {author} {\bibfnamefont {S.}~\bibnamefont {Trebst}},\ and\ \bibinfo
  {author} {\bibfnamefont {Y.}~\bibnamefont {Matsuda}},\ }\href
  {https://doi.org/10.1126/science.aay5551} {\bibfield  {journal} {\bibinfo
  {journal} {Science}\ }\textbf {\bibinfo {volume} {373}},\ \bibinfo {pages}
  {568} (\bibinfo {year} {2021})}\BibitemShut {NoStop}%
\bibitem [{\citenamefont {Samarakoon}\ \emph {et~al.}(2018)\citenamefont
  {Samarakoon}, \citenamefont {Wachtel}, \citenamefont {Yamaji}, \citenamefont
  {Tennant}, \citenamefont {Batista},\ and\ \citenamefont
  {Kim}}]{SamarakoonPRB2018}%
  \BibitemOpen
  \bibfield  {author} {\bibinfo {author} {\bibfnamefont {A.~M.}\ \bibnamefont
  {Samarakoon}}, \bibinfo {author} {\bibfnamefont {G.}~\bibnamefont {Wachtel}},
  \bibinfo {author} {\bibfnamefont {Y.}~\bibnamefont {Yamaji}}, \bibinfo
  {author} {\bibfnamefont {D.~A.}\ \bibnamefont {Tennant}}, \bibinfo {author}
  {\bibfnamefont {C.~D.}\ \bibnamefont {Batista}},\ and\ \bibinfo {author}
  {\bibfnamefont {Y.~B.}\ \bibnamefont {Kim}},\ }\href
  {https://doi.org/10.1103/PhysRevB.98.045121} {\bibfield  {journal} {\bibinfo
  {journal} {Phys. Rev. B}\ }\textbf {\bibinfo {volume} {98}},\ \bibinfo
  {pages} {045121} (\bibinfo {year} {2018})}\BibitemShut {NoStop}%
\bibitem [{\citenamefont {Saha}\ \emph {et~al.}(2019)\citenamefont {Saha},
  \citenamefont {Fan}, \citenamefont {Zhang},\ and\ \citenamefont
  {Chern}}]{SahaPRL2019}%
  \BibitemOpen
  \bibfield  {author} {\bibinfo {author} {\bibfnamefont {P.}~\bibnamefont
  {Saha}}, \bibinfo {author} {\bibfnamefont {Z.}~\bibnamefont {Fan}}, \bibinfo
  {author} {\bibfnamefont {D.}~\bibnamefont {Zhang}},\ and\ \bibinfo {author}
  {\bibfnamefont {G.-W.}\ \bibnamefont {Chern}},\ }\href
  {https://doi.org/10.1103/PhysRevLett.122.257204} {\bibfield  {journal}
  {\bibinfo  {journal} {Phys. Rev. Lett.}\ }\textbf {\bibinfo {volume} {122}},\
  \bibinfo {pages} {257204} (\bibinfo {year} {2019})}\BibitemShut {NoStop}%
\bibitem [{\citenamefont {Luo}\ \emph {et~al.}(2021)\citenamefont {Luo},
  \citenamefont {Zhao}, \citenamefont {Kee},\ and\ \citenamefont
  {Wang}}]{LuoNPJ2021}%
  \BibitemOpen
  \bibfield  {author} {\bibinfo {author} {\bibfnamefont {Q.}~\bibnamefont
  {Luo}}, \bibinfo {author} {\bibfnamefont {J.}~\bibnamefont {Zhao}}, \bibinfo
  {author} {\bibfnamefont {H.-Y.}\ \bibnamefont {Kee}},\ and\ \bibinfo {author}
  {\bibfnamefont {X.}~\bibnamefont {Wang}},\ }\href
  {https://doi.org/10.1038/s41535-021-00356-z} {\bibfield  {journal} {\bibinfo
  {journal} {npj Quantum Materials}\ }\textbf {\bibinfo {volume} {6}},\
  \bibinfo {pages} {57} (\bibinfo {year} {2021})}\BibitemShut {NoStop}%
\bibitem [{\citenamefont {Wu}\ \emph {et~al.}(2016)\citenamefont {Wu},
  \citenamefont {Deng}, \citenamefont {Gardner}, \citenamefont {Vorderwisch},
  \citenamefont {Li}, \citenamefont {Yano}, \citenamefont {Peng},\ and\
  \citenamefont {Imamovic}}]{WuJOI2016}%
  \BibitemOpen
  \bibfield  {author} {\bibinfo {author} {\bibfnamefont {C.-M.}\ \bibnamefont
  {Wu}}, \bibinfo {author} {\bibfnamefont {G.}~\bibnamefont {Deng}}, \bibinfo
  {author} {\bibfnamefont {J.}~\bibnamefont {Gardner}}, \bibinfo {author}
  {\bibfnamefont {P.}~\bibnamefont {Vorderwisch}}, \bibinfo {author}
  {\bibfnamefont {W.-H.}\ \bibnamefont {Li}}, \bibinfo {author} {\bibfnamefont
  {S.}~\bibnamefont {Yano}}, \bibinfo {author} {\bibfnamefont {J.-C.}\
  \bibnamefont {Peng}},\ and\ \bibinfo {author} {\bibfnamefont
  {E.}~\bibnamefont {Imamovic}},\ }\href
  {https://doi.org/10.1088/1748-0221/11/10/p10009} {\bibfield  {journal}
  {\bibinfo  {journal} {Journal of Instrumentation}\ }\textbf {\bibinfo
  {volume} {11}},\ \bibinfo {pages} {P10009 (2016)}}\BibitemShut {NoStop}%
\bibitem [{\citenamefont {{Winn, Barry}}\ \emph {et~al.}(2015)\citenamefont
  {{Winn, Barry}}, \citenamefont {{Filges, Uwe}}, \citenamefont {{Garlea, V.
  Ovidiu}}, \citenamefont {{Graves-Brook, Melissa}}, \citenamefont {{Hagen,
  Mark}}, \citenamefont {{Jiang, Chenyang}}, \citenamefont {{Kenzelmann,
  Michel}}, \citenamefont {{Passell, Larry}}, \citenamefont {{Shapiro, Stephen
  M.}}, \citenamefont {{Tong, Xin}},\ and\ \citenamefont {{Zaliznyak,
  Igor}}}]{BarryEPJ2015}%
  \BibitemOpen
  \bibfield  {author} {\bibinfo {author} {\bibnamefont {{Winn, Barry}}},
  \bibinfo {author} {\bibnamefont {{Filges, Uwe}}}, \bibinfo {author}
  {\bibnamefont {{Garlea, V. Ovidiu}}}, \bibinfo {author} {\bibnamefont
  {{Graves-Brook, Melissa}}}, \bibinfo {author} {\bibnamefont {{Hagen, Mark}}},
  \bibinfo {author} {\bibnamefont {{Jiang, Chenyang}}}, \bibinfo {author}
  {\bibnamefont {{Kenzelmann, Michel}}}, \bibinfo {author} {\bibnamefont
  {{Passell, Larry}}}, \bibinfo {author} {\bibnamefont {{Shapiro, Stephen
  M.}}}, \bibinfo {author} {\bibnamefont {{Tong, Xin}}},\ and\ \bibinfo
  {author} {\bibnamefont {{Zaliznyak, Igor}}},\ }\href
  {https://doi.org/10.1051/epjconf/20158303017} {\bibfield  {journal} {\bibinfo
   {journal} {EPJ Web of Conferences}\ }\textbf {\bibinfo {volume} {83}},\
  \bibinfo {pages} {03017} (\bibinfo {year} {2015})}\BibitemShut {NoStop}%
\bibitem [{\citenamefont {Kajimoto}\ \emph {et~al.}(2011)\citenamefont
  {Kajimoto}, \citenamefont {Nakamura}, \citenamefont {Inamura}, \citenamefont
  {Mizuno}, \citenamefont {Nakajima}, \citenamefont {Ohira-Kawamura},
  \citenamefont {Yokoo}, \citenamefont {Nakatani}, \citenamefont {Maruyama},
  \citenamefont {Soyama}, \citenamefont {Shibata}, \citenamefont {Suzuya},
  \citenamefont {Sato}, \citenamefont {Aizawa}, \citenamefont {Arai},
  \citenamefont {Wakimoto}, \citenamefont {Ishikado}, \citenamefont {ichi
  Shamoto}, \citenamefont {Fujita}, \citenamefont {Hiraka}, \citenamefont
  {Ohoyama}, \citenamefont {Yamada},\ and\ \citenamefont
  {Lee}}]{KajimotoJPSJ2011}%
  \BibitemOpen
  \bibfield  {author} {\bibinfo {author} {\bibfnamefont {R.}~\bibnamefont
  {Kajimoto}}, \bibinfo {author} {\bibfnamefont {M.}~\bibnamefont {Nakamura}},
  \bibinfo {author} {\bibfnamefont {Y.}~\bibnamefont {Inamura}}, \bibinfo
  {author} {\bibfnamefont {F.}~\bibnamefont {Mizuno}}, \bibinfo {author}
  {\bibfnamefont {K.}~\bibnamefont {Nakajima}}, \bibinfo {author}
  {\bibfnamefont {S.}~\bibnamefont {Ohira-Kawamura}}, \bibinfo {author}
  {\bibfnamefont {T.}~\bibnamefont {Yokoo}}, \bibinfo {author} {\bibfnamefont
  {T.}~\bibnamefont {Nakatani}}, \bibinfo {author} {\bibfnamefont
  {R.}~\bibnamefont {Maruyama}}, \bibinfo {author} {\bibfnamefont
  {K.}~\bibnamefont {Soyama}}, \bibinfo {author} {\bibfnamefont
  {K.}~\bibnamefont {Shibata}}, \bibinfo {author} {\bibfnamefont
  {K.}~\bibnamefont {Suzuya}}, \bibinfo {author} {\bibfnamefont
  {S.}~\bibnamefont {Sato}}, \bibinfo {author} {\bibfnamefont {K.}~\bibnamefont
  {Aizawa}}, \bibinfo {author} {\bibfnamefont {M.}~\bibnamefont {Arai}},
  \bibinfo {author} {\bibfnamefont {S.}~\bibnamefont {Wakimoto}}, \bibinfo
  {author} {\bibfnamefont {M.}~\bibnamefont {Ishikado}}, \bibinfo {author}
  {\bibfnamefont {S.}~\bibnamefont {ichi Shamoto}}, \bibinfo {author}
  {\bibfnamefont {M.}~\bibnamefont {Fujita}}, \bibinfo {author} {\bibfnamefont
  {H.}~\bibnamefont {Hiraka}}, \bibinfo {author} {\bibfnamefont
  {K.}~\bibnamefont {Ohoyama}}, \bibinfo {author} {\bibfnamefont
  {K.}~\bibnamefont {Yamada}},\ and\ \bibinfo {author} {\bibfnamefont {C.-H.}\
  \bibnamefont {Lee}},\ }\href {https://doi.org/10.1143/JPSJS.80SB.SB025}
  {\bibfield  {journal} {\bibinfo  {journal} {Journal of the Physical Society
  of Japan}\ }\textbf {\bibinfo {volume} {80}},\ \bibinfo {pages} {SB025}
  (\bibinfo {year} {2011})}\BibitemShut {NoStop}%
\bibitem [{\citenamefont {Inamura}\ \emph {et~al.}(2013)\citenamefont
  {Inamura}, \citenamefont {Nakatani}, \citenamefont {Suzuki},\ and\
  \citenamefont {Otomo}}]{InamuraJPSJ2013}%
  \BibitemOpen
  \bibfield  {author} {\bibinfo {author} {\bibfnamefont {Y.}~\bibnamefont
  {Inamura}}, \bibinfo {author} {\bibfnamefont {T.}~\bibnamefont {Nakatani}},
  \bibinfo {author} {\bibfnamefont {J.}~\bibnamefont {Suzuki}},\ and\ \bibinfo
  {author} {\bibfnamefont {T.}~\bibnamefont {Otomo}},\ }\href
  {https://doi.org/10.7566/JPSJS.82SA.SA031} {\bibfield  {journal} {\bibinfo
  {journal} {Journal of the Physical Society of Japan}\ }\textbf {\bibinfo
  {volume} {82}},\ \bibinfo {pages} {SA031} (\bibinfo {year}
  {2013})}\BibitemShut {NoStop}%
\bibitem [{\citenamefont {Ewings}\ \emph {et~al.}(2016)\citenamefont {Ewings},
  \citenamefont {Buts}, \citenamefont {Le}, \citenamefont {van Duijn},
  \citenamefont {Bustinduy},\ and\ \citenamefont {Perring}}]{EwingsNI2016}%
  \BibitemOpen
  \bibfield  {author} {\bibinfo {author} {\bibfnamefont {R.}~\bibnamefont
  {Ewings}}, \bibinfo {author} {\bibfnamefont {A.}~\bibnamefont {Buts}},
  \bibinfo {author} {\bibfnamefont {M.}~\bibnamefont {Le}}, \bibinfo {author}
  {\bibfnamefont {J.}~\bibnamefont {van Duijn}}, \bibinfo {author}
  {\bibfnamefont {I.}~\bibnamefont {Bustinduy}},\ and\ \bibinfo {author}
  {\bibfnamefont {T.}~\bibnamefont {Perring}},\ }\href
  {https://doi.org/https://doi.org/10.1016/j.nima.2016.07.036} {\bibfield
  {journal} {\bibinfo  {journal} {Nuclear Instruments and Methods in Physics
  Research Section A: Accelerators, Spectrometers, Detectors and Associated
  Equipment}\ }\textbf {\bibinfo {volume} {834}},\ \bibinfo {pages} {132 }
  (\bibinfo {year} {2016})}\BibitemShut {NoStop}%
\end{thebibliography}%

\pagebreak
\pagebreak

\widetext
\begin{center}
\textbf{\large Supplemental Material for \\``Giant magnetic in-plane anisotropy and competing instabilities in \ch{Na_3Co_2SbO_6}''}
\end{center}
\makeatletter
\renewcommand{\theequation}{S\arabic{equation}}
\renewcommand{\thetable}{S\arabic{table}}
\renewcommand{\thefigure}{S\arabic{figure}}

\section{Methods}

\begin{figure}[h]
\centering{\includegraphics[clip,width=10cm]{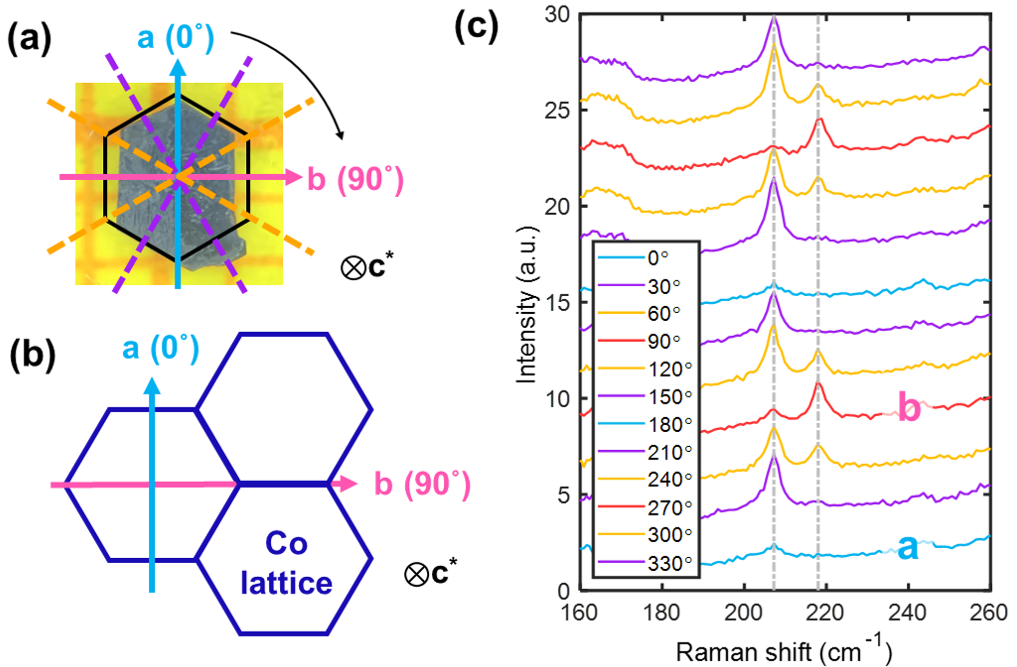}}
\caption{\textbf{Finding twin-free samples with Raman spectroscopy.} (a) Photograph of a twin-free single crystal. The $a$ and $b$-axis are indicated by solid arrows, whereas dashed lines indicate directions at 60$^\circ$ from the $a$ and $b$-axis. (b) The corresponding Co honeycomb sub-lattice. (c) Raman spectra obtained with different parallel polarization directions color-coded with (a). Given the space group  $C$2/m of \ch{Na_3Co_2SbO_6}, there are 15 Raman-active optical phonons, which can be labeled as irreducible representations of the C$_{\text{2h}}$ point group: $7A_\text{g}$ and $8B_\text{g}$ modes. The spectra in (c) reveal intensity variations of two $A_\text{g}$ modes at 207 cm$^{-1}$ and 218 cm$^{-1}$. This empirical knowledge can be used to identify twin-free crystals by scanning the laser spot over the top and bottom surfaces of the entire crystal.}
\label{figS1}
\end{figure}

\begin{figure}[h]
\centering{\includegraphics[clip,width=17cm]{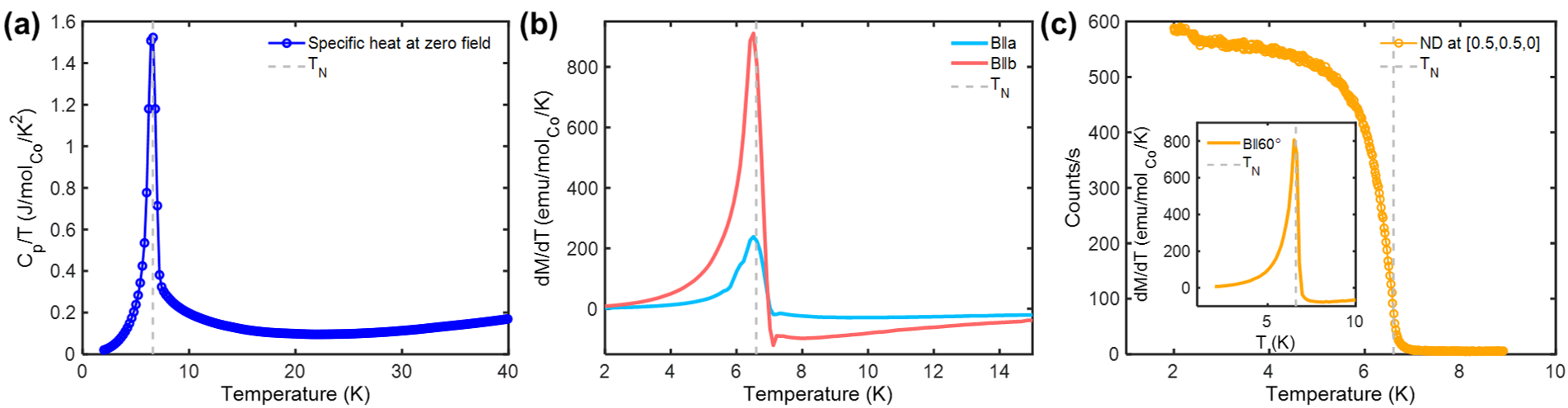}}
\caption{\textbf{A twin-free crystal's signatures of $T_\mathrm{N}$.} (a) Specific heat in zero field. (b) Derivatives of DC magnetic susceptibility measured with fields of 0.1 T along the $a$ and $b$-axes. The measurements were performed on a temperature ramp (+1~K/min). (c) AFM${\frac{1}{2}}$ reflection at [0.5$\mathbf{a^*}$, 0.5$\mathbf{b^*}$, 0] measured as a function of temperature. Inset: the derivatives of DC susceptibility with fields of 0.1 T along 60$^\circ$ from $\mathbf{a}$, which is in the same direction as in Fig.~\ref{figS5}. All three measurements in the figure were performed on the same twin-free crystal of 6.2 mg in mass, and $T_\mathrm{N}=6.6$~K is indicated by grey dashed lines in each panel.}
\label{figS2}
\end{figure}

\begin{figure}[h]
\centering{\includegraphics[clip,width=17cm]{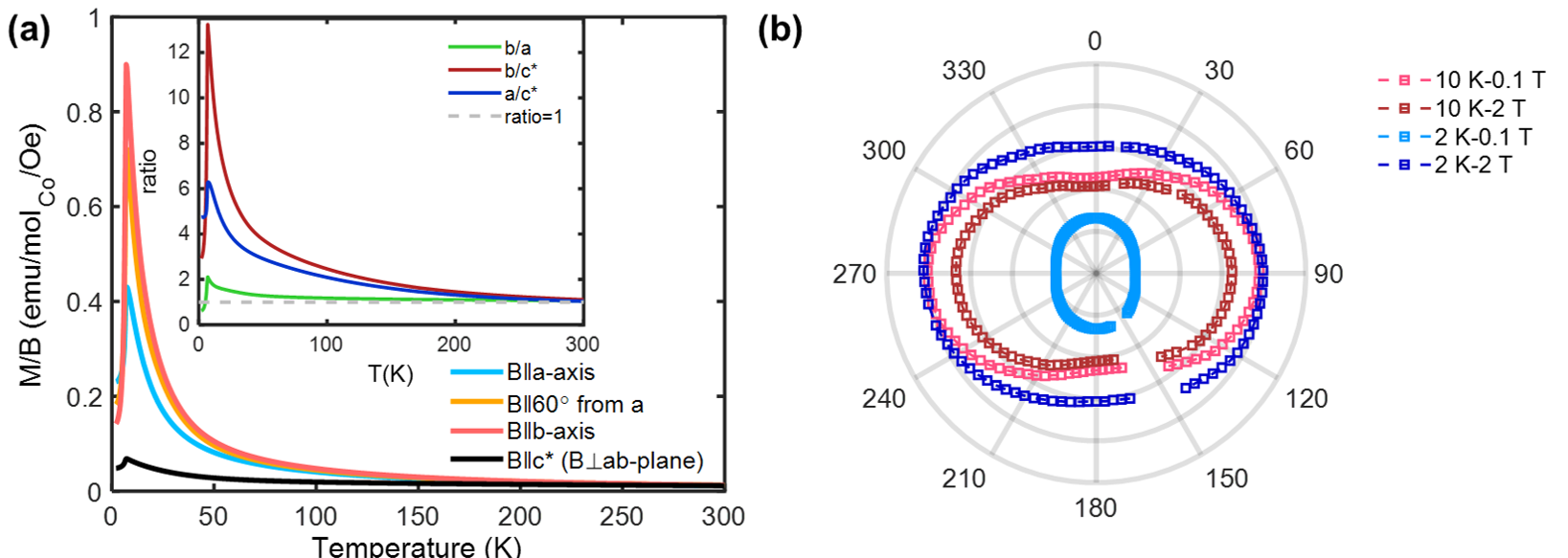}}
\caption{\textbf{Magnetic anisotropy at high temperature and high field.} (a) Susceptibility from 2 K to 300 K measured with $B = 0.1$ T applied along four different directions. Inset shows ratio between the directions. (b) In-plane angle-dependent susceptibility plotted in polar coordinates measured at different temperatures and fields. The outermost ring represents $M / B=1$ emu/mol$_\mathrm{Co}$/Oe. It is noteworthy that in a relatively large magnetic field of 2 T, the thermally disordered state at 10 K ($> T_\mathrm{N}$) has stronger anisotropy (by the $a/b$ ratio) than at 2 K. This indicates that fluctuations under anisotropic interactions contribute significantly to the observed anisotropy.}
\label{figS3}
\end{figure}

\begin{figure}[h]
\centering{\includegraphics[clip,width=17cm]{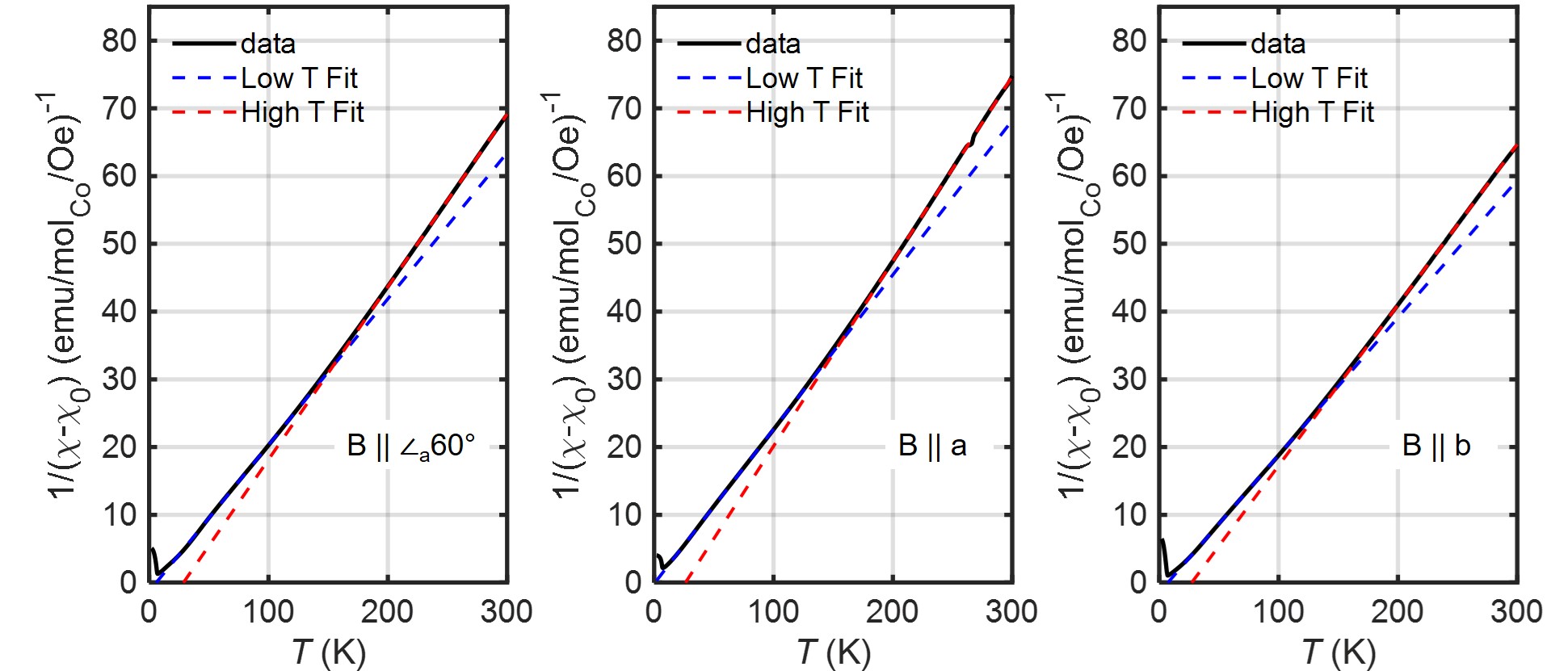}}
\caption{\textcolor[RGB]{0,0,0}{\textbf{Inverse magnetic susceptibility and Curie-Weiss fitting.} Same data as in Fig.~\ref{figS3}(a). A small background constant has been subtracted from the susceptibility (to account for contribution from the sample mount) before taking the inverse. The inverse susceptibility is found to be well approximated by linear functions of temperature over the 20 K $<T<$ 120 K low-$T$ range and the 200 K $<T<$ 300 K high-$T$ range, but not as well in between. The associated Curie-Weiss fit parameters of the two temperature ranges are somewhat different, as summarized in Table~\ref{tabS1}. This difference may arise from thermal activation to the $J_\mathrm{eff}=3/2$ electronic states in the high-$T$ range. Hence, results extracted from the low-$T$ range may better reflect the physics of the $J_\mathrm{eff}=1/2$ states.}}
\label{figS4}
\end{figure}

\begin{figure}[h]
\centering{\includegraphics[clip,scale=0.4]{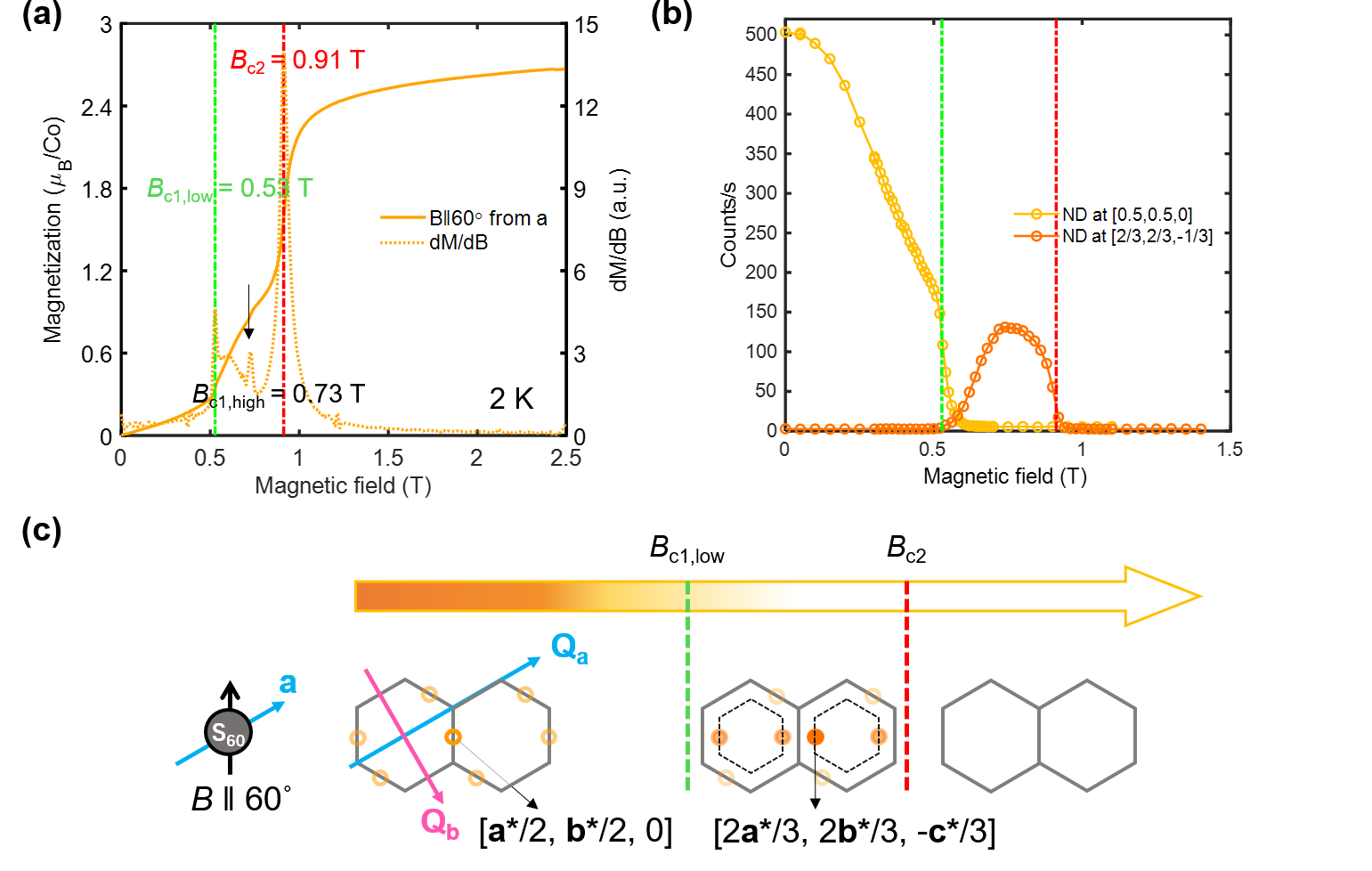}}
\caption{\textbf{AFM$\frac{1}{2}$ to AFM$\frac{1}{3}$ wave vector switch in a twin-free crystal.} (a) Magnetization versus field (solid, field-up) and its derivative (dashed) at $T=2$~K with field along $60^{\circ}$ from $a$-axis. In this direction, the phase transition from AFM${\frac{1}{2}}$ to AFM${\frac{1}{3}}$ occurs in two steps, at $B_{\mathrm{c1,low}}=0.53$~T and $B_{\mathrm{c1,high}}=0.73$~T. Above $B_{\mathrm{c}2}=0.91$~T, the system enters into a field-saturated state. (b) Field-evolution of magnetic diffraction at $[1/2\mathbf{a^*},\,1/2\mathbf{b^*},\,0]$ (light) and $[2/3\mathbf{a^*},\,2/3\mathbf{b^*},\,-1/3\mathbf{c^*}]$ (dark, which is $[1/3\mathbf{a^*},\,1/3\mathbf{b^*},\,1/3\mathbf{c^*}]$ subtracting $[0,\,0,\,\mathbf{c^*}]$). (c) Schematic of magnetic wave-vector switching according to (b). The indicated wave vectors are in the monoclinic notation.}
\label{figS5}
\end{figure}

\textbf{Single crystal growth.} Single crystals of \ch{Na_3Co_2SbO_6} were synthesized with a flux method similar to the method for growing \ch{Na_2Co_2TeO_6} \cite{YaoPRB2020}. Highest-quality crystals were of a dark purple color and a flaky hexagonal shape [Fig.~\ref{figS1}(a)]. The correspondence between sample shape and the cobalt honeycomb sub-lattice is shown in Fig.~\ref{figS1}.

\textbf{Single-crystal $X$-ray diffraction.} The measurements were performed with a custom-designed instrument equipped with a Xenocs Genix3D Mo K$_{\alpha}$ (17.48 keV) $X$-ray source, which produced a beam-spot diameter of 150 $\mu$m at the sample position with $2.5\times10^7$ photons/sec. The samples were mounted on a Huber four-circle diffractometer. A highly sensitive PILATUS3 R 1M solid-state pixel array detector (with $980\times1042$ pixels, 172 $\mu$m $\times172 \mu$m per pixel) was used to collect the diffraction signals. Three-dimensional mapping of the reciprocal space was achieved by taking diffraction images in 0.1$^\circ$ sample-rotation increments. Due to a small non-monochromaticity of the $X$-rays, each observed Bragg peak consisted of closely-spaced K$_{\alpha1}$ and K$_{\alpha2}$ components, and was further accompanied by a weak K$_\beta$ peak and a radial tail [red dashed lines in Fig.~1(c)].

\textbf{Raman spectroscopy screening for twin-free samples.} The measurements were performed under a microscope in a confocal backscattering geometry, using a Horiba Jobin Yvon LabRAM HR Evolution spectrometer equipped with 1800 gr/mm gratings and a liquid-nitrogen-cooled CCD detector. A He-Ne laser with $\lambda = 632.8$ nm was used for excitation (with 1.2 mW laser power). The measurements were performed at room temperature, in air, and on crystal surfaces parallel to the $ab$-plane. Linear polarizations of both incident and scattered photons were set parallel to each other and along twelve in-plane directions, as illustrated in Fig.~\ref{figS1}(a).

\textbf{Specific-heat measurement and magnetometry.} Specific-heat measurements were performed with a Quantum Design PPMS, using a relaxation method. DC magnetometry was performed with a Quantum Design MPMS equipped with a sample rotator. The angle dependence of magnetization in Fig.~2(c) was measured in a sweep-field mode (+3 Oe/sec) after initially cooling the sample in zero field. The field was reset to zero between adjacent orientations without thermal cycling. The systematic results in Fig.~2(c) indicate a lack of dependence on the field history.

\textbf{Neutron diffraction on a twin-free sample.} The experiment was performed on the SIKA cold neutron triple-axis spectrometer at the Australian Nuclear Science and Technology Organization (ANSTO) \cite{WuJOI2016}. A twin-free crystal was mounted with reciprocal vectors $[K\mathbf{a^*}, K\mathbf{b^*}, L\mathbf{c^*}]$ in the horizontal scattering plane, and the measurement was performed with $k_\mathrm{i}=k_\mathrm{f}=1.97$~\AA$^{-1}$ neutrons and a beryllium filter. In this experiment, only diffractions in the horizontal scattering plane could be accessed, due to the use of a cryomagnet which did not allow for sample tilting. The applied vertical fields were at 60$^\circ$ (anticlockwise) from the $\mathbf{a}$-axis [Fig.~\ref{figS5}(c)]. This geometry made the studied sample equivalent to S$_{60}$ discussed in main text, and allowed us to access the $[1/2\mathbf{a^*},\,1/2\mathbf{b^*},\,0]$ reflection in the AFM${\frac{1}{2}}$ state, as well as the $[2/3\mathbf{a^*},\,2/3\mathbf{b^*},\,-1/3\mathbf{c^*}]$ reflection in the AFM${\frac{1}{3}}$ state.

\textbf{Neutron diffraction on twinned samples.} The neutron diffraction experiments displayed in Figs.~3-6 of main text were performed with twinned crystal arrays, which were co-aligned in the $ab$-plane on aluminum plates with a hydrogen-free adhesive (Cytop). 1/3 of the sample mass belonged to $\mathrm{S}_a$ (see text), for which the vertical direction was parallel to the $\mathbf{a}$-axis, and the corresponding horizontal scattering plane was $[0,\,K\mathbf{b^*},\,L\mathbf{c^*}]$. The other 2/3 of the sample belonged to $\mathrm{S}_{60}$, for which the vertical direction was at 60$^\circ$ from the $a$-axis and the horizontal plane was $[K\mathbf{a^*},\,K\mathbf{b^*},\,L\mathbf{c^*}]$ (or symmetry equivalent). As explained in main text and below, we use $(Q_a,\,Q_b,\,Q_\perp)$ to denote wave vectors. $Q_a$ is along the vertical direction, and the horizontal plane is $(Q_b,\,Q_\perp)$. $(Q_a,\,Q_b)$ in Fig.~3(a), Fig.~4 and Fig.~5(c) of main text denotes 2D wave vectors in the vertical (real-space) $ab$-plane.

The diffraction experiment in magnetic fields (Fig.~4 of main text) was performed on the HYSPEC time-of-flight spectrometer at the SNS, Oak Ridge National Laboratory \cite{BarryEPJ2015}, using a helium-3 insert and a 14 T vertical-field cryomagnet as sample environment. Equipped with the incident beam-focusing optics, the narrow vertical opening angle of $\pm 7^\circ $ of the cryomagnet allowed us to detect the magnetic diffractions out of the horizontal scattering plane. We used a relatively high incident neutron energy $E_\mathrm{i} = 35$~meV to have the full out-of-plane access for acquiring the data in Figs.~4 and \ref{figS10}, and a relatively low incident neutron energy $E_\mathrm{i} = 15$~meV for accurate peak indexing Fig.~\ref{figS7}. The sample was 0.5 gram in total mass, and had a full mosaic spread of about 2.3$^\circ$. Diffraction data were acquired by rotating the sample over a 78$^\circ$ range in 0.5$^\circ$ steps, resulting in a three-dimensional data set, which we further symmetrize with $Q_b$ and $Q_\perp$ mirror operations for plotting.

\begin{figure}[h]
\centering{\includegraphics[clip,scale=0.45]{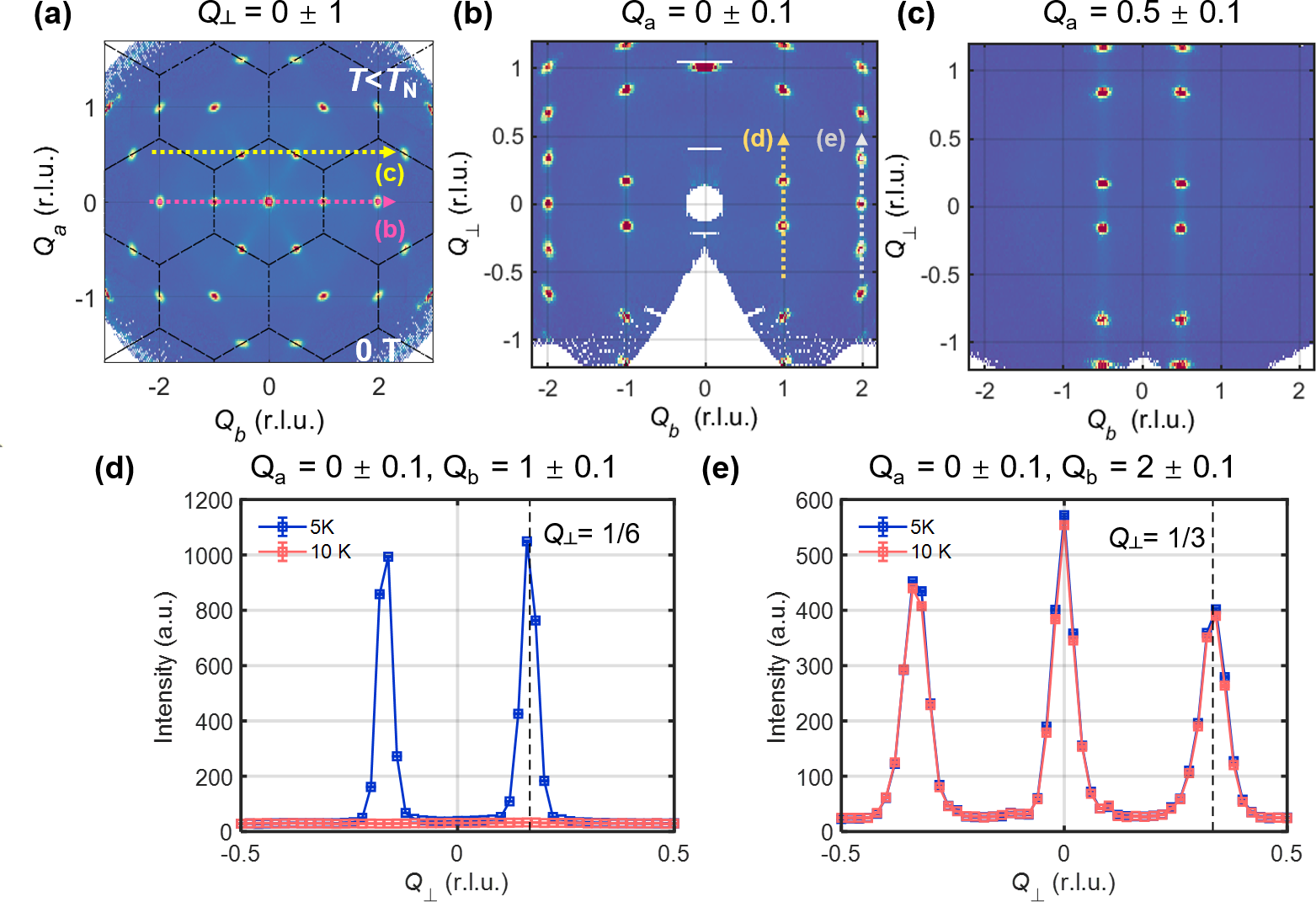}}
\caption{\textbf{Peak indexing in zero field for a twinned sample.} (a) Symmetrized ($Q_a$, $Q_b$) plane of neutron diffraction results with $Q_\perp$ integrated within 1 r.l.u. around zero. All observed peaks on the Brillouin zone boundaries (dashed-dotted lines) are due to the AFM${\frac{1}{2}}$ order. The measurement was performed in zero field, at a temperature of about 6 K (below $T_\text{N}$), using a twinned sample. Dashed arrows indicate $Q_a=0$ and $Q_a=0.5$ and are reference for (b) and (c). (b-c) Viewing the same data as in (a), but with $Q_\perp$ displayed as vertical axis. \textcolor[RGB]{0,0,0}{(d-e) Cuts through magnetic and structural peaks, respectively.}}
\label{figS6}
\end{figure}

\begin{figure}[h]
\centering{\includegraphics[clip,width=12cm]{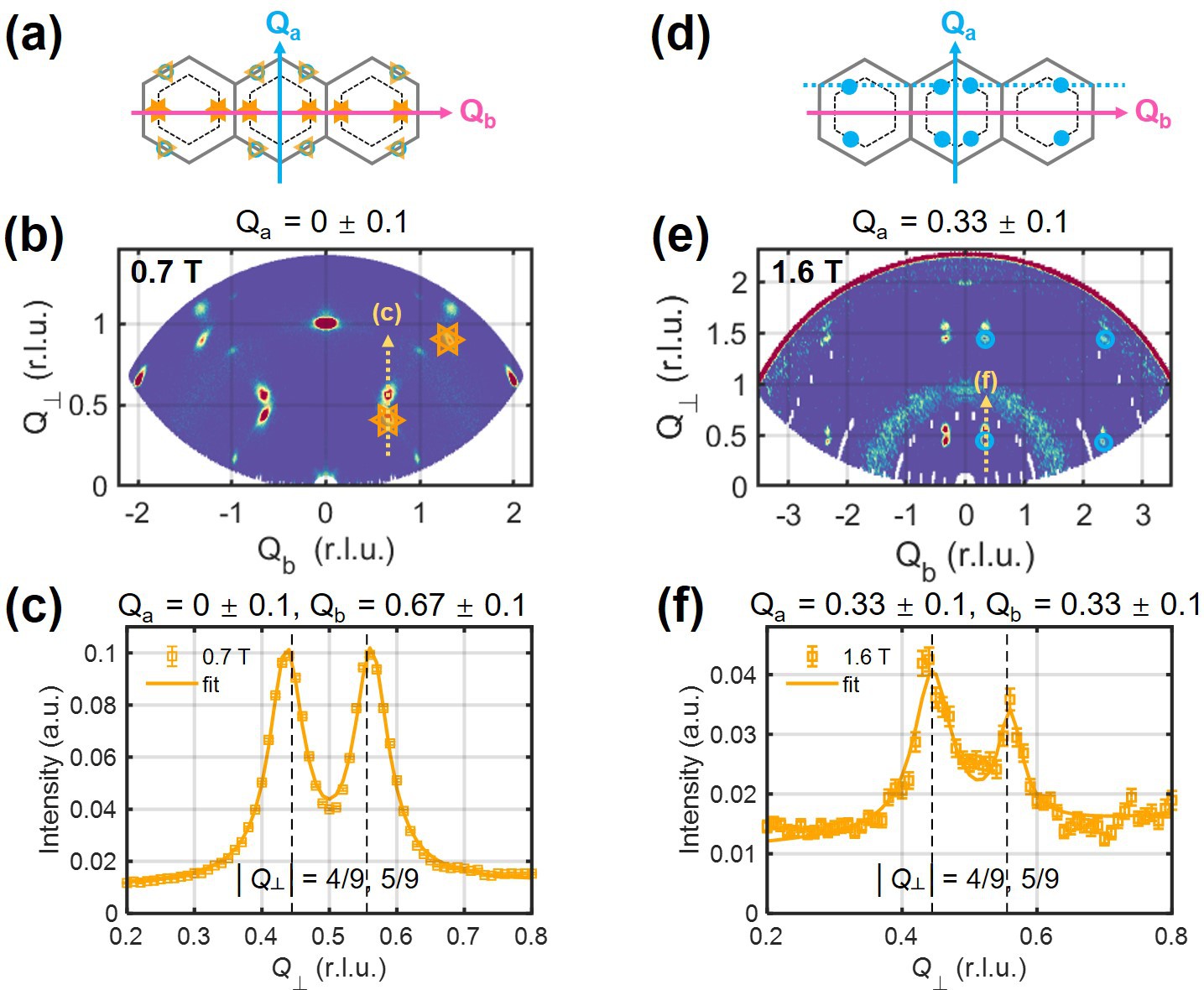}}
\caption{\textbf{Peak indexing in vertical fields for a twinned sample.} (a-b) Magnetic peaks in the ($Q_a$, $Q_b$) plane in a vertical field of 0.7~T and at $T = 250 mK$, together with their $Q_\perp$ locations to be checked against the attributions in Fig.~4 of main text. \textcolor[RGB]{0,0,0}{(c) Line-cut through the data in (b).} (d-e) Magnetic peaks in the ($Q_a$, $Q_b$) plane in a vertical field of 1.6~T, together with their $Q_\perp$ locations to be checked against the attributions in Fig.~4 of main text. \textcolor[RGB]{0,0,0}{(f) Line-cut through the data in (e).} The integrated energy window is $\pm 0.2$ meV. }
\label{figS7}
\end{figure}

\begin{figure}[h]
\centering{\includegraphics[clip,width=16cm]{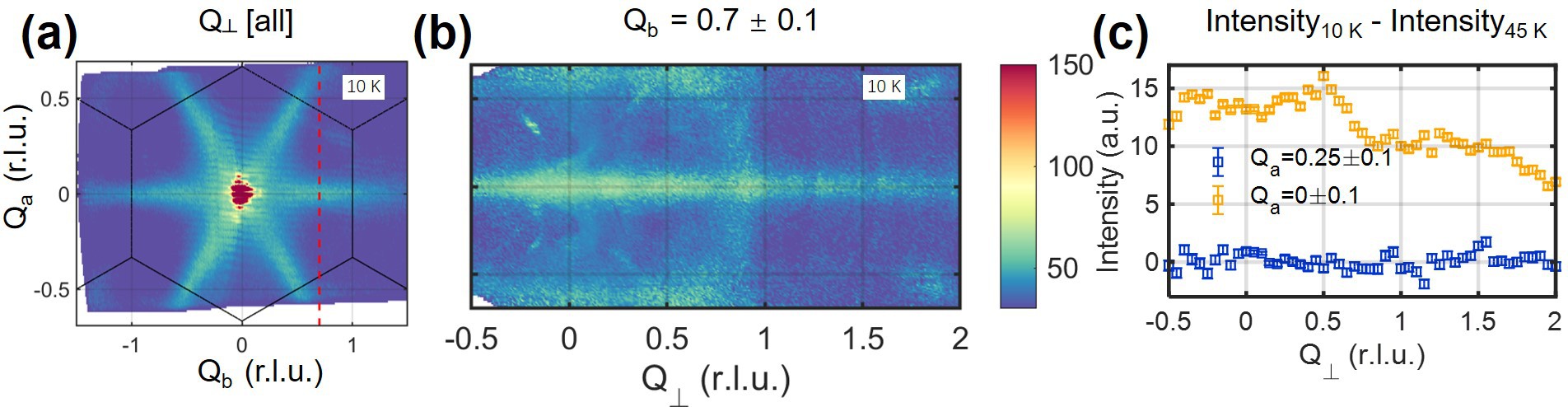}}
\caption{\textcolor[RGB]{0,0,0}{\textbf{Momentum structure of the diffuse scattering above $T_\mathrm{N}$.} (a) $Q_\perp$-integrated data obtained at $T = 10 K$, same as in main text Fig.~5. (b) Viewing the same data as in (a), but for the chosen $Q_b$ and with $Q_\perp$ displayed as horizontal axis. (c) Line-cuts at selected $Q_a$ positions in (b) after subtracting 45 K data as background. The magnetic diffuse scattering is seen to be only weakly structured along $Q_\perp$, where the intensity decrease with increasing $Q_\perp$ is due to the magnetic form factor.}}
\label{figS8}
\end{figure}

\begin{figure}[h]
\centering{\includegraphics[clip,width=9cm]{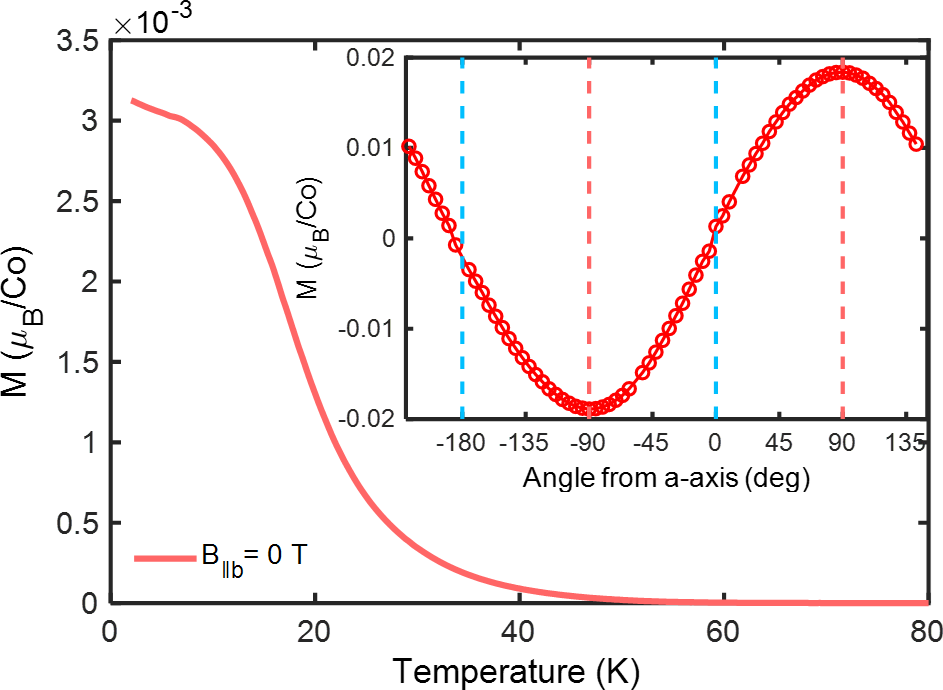}}
\caption{\textbf{Incipient ferromagnetism.} A weak moment in a twin-free sample, measured upon warm-up in zero field, after preparing the sample with $B_{\|b}=0.02$ T field-cooling to 2 K. Inset shows the angle dependence at 10 K after 0.4~T field-cooling, revealing a clear $C_1$ symmetry that is consistent with a frozen-in net ferromagnetic moment. The field 0.4 T is almost enough to ``saturate'' the moment, whereas the field of 0.02 T is not enough but it ensures that the apparatus has no remnant magnetic field. The weak ferromagnetism might originate from a small amount of disorder in the crystal (such as stacking faults) which stabilizes the ferromagnetic fluctuations discussed in main text.}
\label{figS9}
\end{figure}

The diffraction experiment at variable temperatures without magnetic field. (Figs.~3, 5, 6 of main text) was performed on the 4SEASONS time-of-flight spectrometer at the MLF, J-PARC, Japan \cite{KajimotoJPSJ2011}. The sample was 0.7 gram in total mass with a full mosaic spread of about 2.7$^\circ $. The data in Figs.~3(a), 5(c) 6(c) and \ref{figS6} were collected at fixed temperatures by rotating the sample over an 85$^\circ$ range in 1$^\circ$ steps. The data in Fig.~5(a-b) were acquired with a ``sit-and-count'' method, \textit{i.e.}, the sample was rotated to fixed orientations, and counting was continuously performed during a slow warm-up of the sample at a rate of about 0.022 K/min. Data were reduced and analyzed with the Utsusemi \cite{InamuraJPSJ2013} and Horace \cite{EwingsNI2016} software packages.

The time-of-flight scattering experiments generated high-dimensional data sets, which were sliced into lower dimensions for plotting. Detailed measurement conditions and slicing restrictions can be found in Table~\ref{tabS2}.

\textbf{Conversion between coordinate systems for the reciprocal space}. A fully-twinned sample contains all $C_6$-related counterparts of any chosen monoclinic domain, where the $C_6$ rotation is about the normal direction of the $ab$-plane, or $\mathbf{c^*}$. It is straightforward to show that, if a monoclinic domain has a physical wave vector of $[H\mathbf{a^*},\,K\mathbf{b^*},\,L\mathbf{c^*}]$, the wave vector will be observed at a total of six positions in the $(Q_a,\,Q_b,\,Q_\perp)$ coordinate system: $(H,\,K,\,L+H/3)$, $(H/2+K/2,\,-3H/2+K/2,\,L+H/3)$, $(-H/2+K/2,\,-3H/2-K/2,\,L+H/3)$, $(-H,\,-K,\,L+H/3)$, $(-H/2-K/2,\,3H/2-K/2,\,L+H/3)$, $(H/2-K/2,\,3H/2+K/2,\,L+H/3)$. Note that they share the same $Q_\perp$. It is straightforward to verify that all magnetic peaks in Fig.~\ref{figS6} in zero field are consistent with $\mathbf{Q}= [\frac{1}{2}\mathbf{a^*},\frac{1}{2}\mathbf{b^*},0] + \mathbf{G}$ and symmetry equivalent, and that all magnetic peaks in Fig.~\ref{figS7} above $B_\mathrm{c,1}$ are consistent with $\mathbf{Q}= [\frac{1}{3}\mathbf{a^*},\frac{1}{3}\mathbf{b^*},\frac{1}{3}\mathbf{c^*}] + \mathbf{G}$ and symmetry equivalent, where $\mathbf{G}$ is a reciprocal lattice vector. A reference table for the conversion between the coordinate systems can be found in Table~\ref{tabS3}.

\textbf{Discussion of magnetic structure}. Recent indications of triple-$\mathbf{q}$ order \cite{ChenPRB2021,LeePRB2021} in the sister compound \ch{Na_2Co_2TeO_6} motivate us to consider a multi-$\mathbf{q}$ possibility here as well. In this regard, the two-step transitions of $B_{\mathrm{c}1}$ (\textit{e.g.}, in $\mathrm{S}_{60}$) are intriguing. In zero field, the magnetic Hamiltonian has $C_2$ symmetry around the $b$-axis. In the single-$\mathbf{q}$ zigzag scenario, the symmetry of magnetic ground state is spontaneously lowered to $C_1$, and $C_2$-symmetry-related magnetic domains are formed, featuring FM zigzag chains running in directions related to each other by $C_2$ rotation around the $b$-axis. However, in an external magnetic field applied \textit{not} along the $b$-axis (or the $a$-axis), as is the case for $\mathrm{S}_{60}$, the magnetization energy of the $C_2$-related magnetic domains will no longer be degenerate. In other words, considering the combination of the crystallographic structure \textit{and} the field direction, the Hamiltonian's symmetry is lowered to \textit{none} ($C_1$) from the first place. This low symmetry is at the origin of the split between $B_\mathrm{c1,low}$ and $B_\mathrm{c1,high}$ for essentially all field directions between the $a$ and $b$-axes.

At $B_\mathrm{c1,low}$ and $B_\mathrm{c1,high}$, since the wave vectors switch only along the same $\Gamma$-M line, the resultant AFM${\frac{1}{3}}$ order would continue to have their FM ribbons/chains running in inequivalent directions, \textit{i.e.}, same as in the original zigzag domains. Then, by the same symmetry argument, it is natural to expect the two types of AFM${\frac{1}{3}}$ domains to have their next transitions occur at somewhat different $B_\mathrm{c2}$. However, only a single transition is experimentally observed. Under the zigzag scenario, this empirical ``simplicity'' can only be attributed to a coincidence not required by symmetry, yet it holds over a wide angular range -- magnetometry always sees a single $B_\mathrm{c2}$ transition [Fig.~2(c-d)]. Moreover, the two-step transitions between AFM${\frac{1}{2}}$ and AFM${\frac{1}{3}}$ are consistently observed independent of field history [Fig.~2(c)], which means that the above two types of domains always reappear after the symmetry-lowering field is removed. Such lack of domain repopulation by the fields is somewhat difficult to understand if the zero-field magnetic ground state's symmetry is indeed spontaneously lowered to $C_1$ (as for zigzag). Alternatively, these remaining puzzles for the zigzag scenario could imply that the ``domains'' are not macroscopically separated, but instead, the two sets of wave vectors arise from the same part of the sample, hinting at a multi-$\mathbf{q}$ scenario for at least some of the magnetic orders. We therefore believe that the precise nature of the AFM phases in \ch{Na_3Co_2SbO_6} is still open for further research.

\pagebreak	

\section{Supplementary Figures}

\pagebreak

\pagebreak

\pagebreak

\pagebreak

\pagebreak

\section{Supplementary Tables}

\begin{table}[h]
\begin{ruledtabular}
\renewcommand{\arraystretch}{1.3}
\begin{tabular}{l|ccc|cccc}
 & \multicolumn{3}{c|}{low-T} & \multicolumn{4}{c}{high-$T$} \\
\hline
 & $B \parallel \mathbf{a}$ & $B \parallel 60^{\circ}$ & $B \parallel \mathbf{b}$ & $B \parallel \mathbf{a}$ & $B \parallel 60^{\circ}$ & $B \parallel \mathbf{b}$ & $B \perp ab$ \\
\hline
$\mu_{\text{eff}}$ $(\mu_\mathrm{B}/\mathrm{Co}^{2+})$ & 5.9 & 6.1 & 6.3 & 5.4 &  5.8 & 5.8 & 6.2\\
$\theta$ (K) & 1.0 & 5.7 & 6.8 & 26.2 & 22.4 & 27.6 & $-220$ \\
$g$ & 6.8 & 7.1 & 7.3 & 6.3 & 6.7 & 6.7 & 7.2 \\
\end{tabular}	
\end{ruledtabular}
\caption{\textcolor[RGB]{0,0,0}{$g$-factor anisotropy revealed by magnetic susceptibility  [Figs.~\ref{figS3}(a) and \ref{figS4}] along four different directions. The effective moments $\mu_{\text{eff}}$ and Weiss temperature $\theta$ are obtained from Curie-Weiss fitting $\chi=\chi_{0}+C /(T-\theta)$ in two temperature ranges: 20 K $<T<$ 120 K (low-$T$), and 200 K $<T<$ 300 K (high-$T$).} $\chi_0$ accounts for a small background from the holder and glue, and $C=N\mu_0\mu_\mathrm{eff}^2/(3k_\mathrm{B})$. The Land\'{e} $g$-factor is related to $\mu_\mathrm{eff}$ as $\mu_\mathrm{eff}=g\sqrt{J(J+1)}\mu_\mathrm{B}$ with $J=1/2$. }
\label{tabS1}
\end{table}

\begin{table*}[!h]
\begin{ruledtabular}
\renewcommand{\arraystretch}{1.3}
\begin{tabular}{lrrrrrr}
Data & $Q_a$ range (r.l.u.) & $Q_b$ range (r.l.u.) & $Q_\perp$ range (r.l.u.) & $E_\mathrm{i}$ (meV)\footnote{Incident neutron energy.} & chopper frequency (Hz) &$\Delta E$ range (meV)\footnote{Neutron energy transfer.}\\
\hline
Fig. 3(a) & - & - & [-1,~1] & 16.8 &150 & [-0.1,~0.1]\\
Fig. 3(b) blue& [-0.1,~0.1] & [1.9,~2.1] & - & 16.8 &150 & [-0.1,~0.1]\\
Fig. 3(b) orange & [0.4,~0.6] & [0.4,~0.6] & - & 16.8 &150 & [-0.1,~0.1]\\
Fig. 3(c) & - & [0.4,~0.6] & [-5,~5] & 5.6 &150 & [-0.1,~0.1]\\
Fig. 3(d) & [0.4,~0.6] & - & [-5,~5] & 5.6 &150 & [-0.1,~0.1]\\
Fig. 4 left & - & - & [-0.2,~0.2] & 35 &120 & [-0.2,~0.2]\\
Fig. 4 right & - & - & [0.3,~0.7] & 35 &120 & [-0.2,~0.2]\\
Fig. 5(a) & - & [-0.2,~0.2] & [-2.2,~2.2] & 5.6 &150 & [-0.2,~0.2]\\
Fig. 5(b) & [-0.2,~0.2] & - & [-2.2,~2.2] & 5.6 &150 & [-0.2,~0.2]\\
Fig. 5(c) & - & - & [-2.2,~2.2] & 5.6 &150 & [-0.2,~0.2]\\
Fig. 5(c) & - & - & [-2.2,~2.2] & 16.8 &150 & [-0.1,~0.1]\\
Fig. 6(c) & - & - & [-1,~1] & 16.8 &150 & [-0.1,~0.1]\\
Fig. \ref{figS6}(a) & - & - & [-1,~1] & 16.8 &150 & [-0.1,~0.1]\\
Fig. \ref{figS6}(b) & [-0.1,~0.1] & - & - & 16.8 &150 & [-0.1,~0.1]\\
Fig. \ref{figS6}(c) & [0.4,~0.6] & - & - & 16.8 &150 & [-0.1,~0.1]\\
Fig. \ref{figS6}(d] & [-0.1,~0.1] & [0.9,~1.1] & - & 16.8 &150 & [-0.1,~0.1]\\
Fig. \ref{figS6}(e) & [-0.1,~0.1] & [1.9,~2.1] & - & 16.8 &150 & [-0.1,~0.1]\\
Fig. \ref{figS7}(b) & [-0.1,~0.1] & - & - & 5 &360 & [-0.2,~0.2]\\
Fig. \ref{figS7}(c) & [-0.1,~0.1] &[0.567,~0.767] & - & 5 &360 & [-0.2,~0.2]\\
Fig. \ref{figS7}(e) & [0.2,~0.285] & - & - & 15 &360 & [-0.2,~0.2]\\
Fig. \ref{figS7}(f) & [0.2,~0.285] & [0.233,~0.433] & - & 15 &360 & [-0.2,~0.2]\\
Fig. \ref{figS8}(a) & - & - & [-2.2,~2.2] & 5.6 &150 & [-0.2,~0.2]\\
Fig. \ref{figS8}(b) & - & [0.6,~0.8] & - & 5.6 &150 & [-0.2,~0.2]\\
Fig. \ref{figS8}(c) & - & [0.6,~0.8] & - & 5.6 &150 & [-0.2,~0.2]\\
\end{tabular}
\end{ruledtabular}
\caption{\textcolor[RGB]{0,0,0}{Detailed measurement and time-of-flight data-reduction conditions used in figures.}}
\label{tabS2}
\end{table*}

\pagebreak
\begin{table*}[h]
\renewcommand{\arraystretch}{1.3}
\begin{tabular}{|c|c|c|c|c|c|c|}
\hline
\begin{minipage}[b]{0.11\columnwidth}
		\centering
		\raisebox{-.4\height}{\includegraphics[width=16mm]{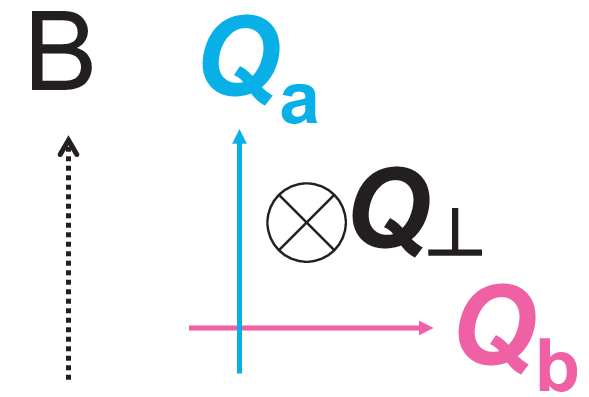}}
	\end{minipage} &  \boldmath{$B||a$} & \boldmath{$60^\circ$} \textbf{from} \boldmath{$a$} & \boldmath{$120^\circ$} \textbf{from} \boldmath{$a$} & \boldmath{$180^\circ$} \textbf{from} \boldmath{$a$} & \boldmath{$240^\circ$} \textbf{from} \boldmath{$a$} & \boldmath{$300^\circ$} \textbf{from} \boldmath{$a$} \\
\hline
\tabincell{c}{\textbf{AFM}\boldmath{$_{\frac{1}{2}}$}\\
\boldmath{$(H, K, L)$} \textbf{index}\\
\boldmath{\textcolor{red}{$\pm(\frac{1}{2}, \frac{1}{2}, 0)$}}\\\boldmath{\textcolor[RGB]{0,139,0}{$\pm(\frac{1}{2}, -\frac{1}{2}, 0)$}}} & \begin{minipage}[b]{0.11\columnwidth}
		\centering
		\raisebox{-.45\height}{\includegraphics[scale=0.6]{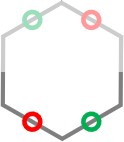}}
	\end{minipage} & \begin{minipage}[b]{0.11\columnwidth}
		\centering
		\raisebox{-.45\height}{\includegraphics[scale=0.6]{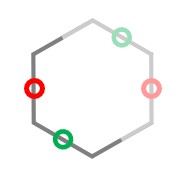}}
	\end{minipage} & \begin{minipage}[b]{0.11\columnwidth}
		\centering
		\raisebox{-.45\height}{\includegraphics[scale=0.6]{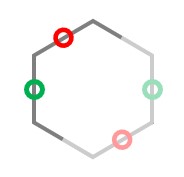}}
	\end{minipage} & \begin{minipage}[b]{0.11\columnwidth}
		\centering
		\raisebox{-.45\height}{\includegraphics[scale=0.6]{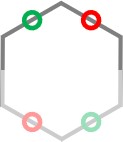}}
	\end{minipage} & \begin{minipage}[b]{0.11\columnwidth}
		\centering
		\raisebox{-.45\height}{\includegraphics[scale=0.6]{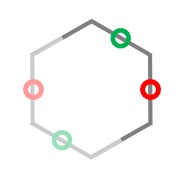}}
	\end{minipage} & \begin{minipage}[b]{0.11\columnwidth}
		\centering
		\raisebox{-.45\height}{\includegraphics[scale=0.6]{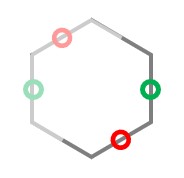}}
	\end{minipage} \\
\hline
\boldmath{$(Q_{a}, Q_{b}, Q_{\perp})$} & \tabincell{c}{\textcolor{red}{$\pm(\frac{1}{2}, \frac{1}{2}, \frac{1}{6})$}\\\textcolor[RGB]{0,139,0}{$\pm(\frac{1}{2}, -\frac{1}{2}, \frac{1}{6})$}} & \tabincell{c}{\textcolor{red}{$\pm (0, 1, \frac{1}{6})$}\\\textcolor[RGB]{0,139,0}{$\pm (\frac{1}{2}, \frac{1}{2}, \frac{1}{6})$}} & \tabincell{c}{\textcolor{red}{$\pm (-\frac{1}{2}, \frac{1}{2}, \frac{1}{6})$}\\\textcolor[RGB]{0,139,0}{$\pm (0, 1, \frac{1}{6})$}} & \tabincell{c}{\textcolor{red}{$\pm (-\frac{1}{2}, -\frac{1}{2}, \frac{1}{6})$}\\\textcolor[RGB]{0,139,0}{$\pm (\frac{1}{2}, \frac{1}{2}, -\frac{1}{6})$}} & \tabincell{c}{\textcolor{red}{$\pm (0, -1, \frac{1}{6})$}\\\textcolor[RGB]{0,139,0}{$\pm (-\frac{1}{2}, -\frac{1}{2}, \frac{1}{6})$}} & \tabincell{c}{\textcolor{red}{$\pm (\frac{1}{2}, -\frac{1}{2}, \frac{1}{6})$}\\\textcolor[RGB]{0,139,0}{$\pm (0, -1, \frac{1}{6})$}}
\\\hline
\tabincell{c}{\textbf{AFM}\boldmath{$_{\frac{1}{3}}$}\\
\boldmath{$(H, K, L)$} \textbf{index}\\
\boldmath{\textcolor{red}{$\pm(\frac{1}{3}, \frac{1}{3}, \frac{1}{3})$}}\\\boldmath{\textcolor[RGB]{0,139,0}{$\pm(\frac{1}{3}, -\frac{1}{3}, \frac{1}{3})$}}} & \begin{minipage}[b]{0.11\columnwidth}
		\centering
		\raisebox{-.45\height}{\includegraphics[scale=0.6]{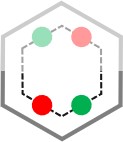}}
	\end{minipage} & \begin{minipage}[b]{0.11\columnwidth}
		\centering
		\raisebox{-.45\height}{\includegraphics[scale=0.6]{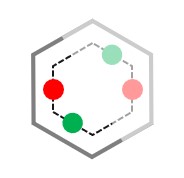}}
	\end{minipage} & \begin{minipage}[b]{0.11\columnwidth}
		\centering
		\raisebox{-.45\height}{\includegraphics[scale=0.6]{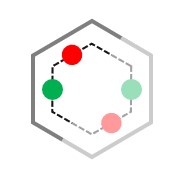}}
	\end{minipage} & \begin{minipage}[b]{0.11\columnwidth}
		\centering
		\raisebox{-.45\height}{\includegraphics[scale=0.6]{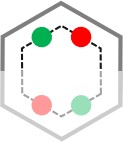}}
	\end{minipage} & \begin{minipage}[b]{0.11\columnwidth}
		\centering
		\raisebox{-.45\height}{\includegraphics[scale=0.6]{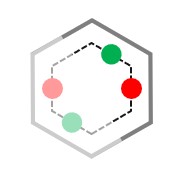}}
	\end{minipage} & \begin{minipage}[b]{0.11\columnwidth}
		\centering
		\raisebox{-.45\height}{\includegraphics[scale=0.6]{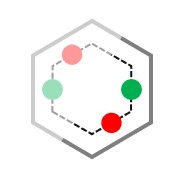}}
	\end{minipage} \\\hline
\boldmath{$(Q_{a}, Q_{b}, Q_{\perp})$} & \tabincell{c}{\textcolor{red}{$\pm(\frac{1}{3}, \frac{1}{3}, \frac{4}{9})$}\\\textcolor[RGB]{0,139,0}{$\pm(\frac{1}{3}, -\frac{1}{3}, \frac{4}{9})$}} & \tabincell{c}{\textcolor{red}{$\pm (0, \frac{2}{3}, \frac{4}{9})$}\\\textcolor[RGB]{0,139,0}{$\pm (\frac{1}{3}, \frac{1}{3}, \frac{4}{9})$}} & \tabincell{c}{\textcolor{red}{$\pm (-\frac{1}{3}, \frac{1}{3}, \frac{4}{9})$}\\\textcolor[RGB]{0,139,0}{$\pm (0, \frac{2}{3}, \frac{4}{9})$}} & \tabincell{c}{\textcolor{red}{$\pm (-\frac{1}{3}, -\frac{1}{3}, \frac{4}{9})$}\\\textcolor[RGB]{0,139,0}{$\pm (-\frac{1}{3}, \frac{1}{3}, \frac{4}{9})$}} & \tabincell{c}{\textcolor{red}{$\pm (0, -\frac{2}{3}, \frac{4}{9})$}\\\textcolor[RGB]{0,139,0}{$\pm (-\frac{1}{3}, -\frac{1}{3}, \frac{4}{9})$}} & \tabincell{c}{\textcolor{red}{$\pm (\frac{1}{3}, -\frac{1}{3}, \frac{4}{9})$}\\\textcolor[RGB]{0,139,0}{$\pm (0, -\frac{2}{3}, \frac{4}{9})$}} \\
\hline
\end{tabular}
\caption{\textcolor[RGB]{0,0,0}{Index conversion between the physical coordinate system, $(H,\,K,\,L)$, and the hybrid coordinate system, $(Q_a,\,Q_b,\,Q_\perp)$, for different crystallographic orientations with respect to the vertical magnetic field $B$. Magnetic diffraction wave vectors in the first 2D Brillouin zone in the two AFM phases are illustrated by circles color-coded with their indices in the table. Solid and faded-out halves of the illustrations indicate negative and positive $Q_\perp$ components, respectively.}}
\label{tabS3}
\end{table*}

\pagebreak

\section{Supplementary Animation}

\begin{figure}[h]
\centering{\animategraphics[height=15cm, label=Dmap, controls]{2}{Dmap_rev2_}{01}{14}}
\caption{\textbf{Full data behind Fig.~4 of main text.} (Animated) Symmetrized diffraction maps measured as a function of vertical field, presented in three constant-$Q_\perp$ slices \textcolor[RGB]{0,0,0}{with $Q_\perp$ integrated over the ranges indicated in the panel titles. Figure 3 of main text presents an overview of these data.}  $Q_a=0$ corresponds to the horizontal scattering plane.}
\label{figS10}
\end{figure}

\end{document}